\documentclass[%
 reprint,
 amsmath,amssymb,
 aps,
pre
]{revtex4-2}
\usepackage{graphicx}
\usepackage{wrapfig}
\usepackage{enumitem}
\usepackage{amsmath,mathtools}
\usepackage{amssymb}
\usepackage{marginnote}
\usepackage{comment}
\usepackage[dvipsnames]{xcolor}
\usepackage[colorlinks, allcolors=NavyBlue]{hyperref}
\usepackage{url}
\usepackage{esint}
\usepackage{multirow}
\usepackage{comment}
\usepackage{tikz}
\usetikzlibrary{arrows.meta}

\usepackage{listings}
\usepackage{xcolor}
\usepackage{pgfplots}
\pgfplotsset{compat=1.9, , every axis/.append style={font=\footnotesize}}
\usepgfplotslibrary{external}
\usepackage{xspace}
\usepackage{tensor}
\usepackage{physics}

\definecolor{codegray}{rgb}{0.5,0.5,0.5}
\definecolor{codered}{rgb}{0.75,0,0}
\definecolor{backcolour}{rgb}{0.95,0.95,0.92}
\lstdefinestyle{mystyle}{
    backgroundcolor=\color{backcolour},
    keywordstyle=\color{blue},
    numberstyle=\tiny\color{codegray},
    stringstyle=\color{codered},
    basicstyle=\ttfamily\footnotesize,
    breakatwhitespace=false,         
    captionpos=b,                    
    keepspaces=true,                 
    numbers=left,                    
    numbersep=5pt,                  
    showspaces=false,                
    showstringspaces=false,
    showtabs=false,                  
    tabsize=2
}
\newcommand{\mb}{\mathbf}
\newcommand{\bs}{\boldsymbol}
\newcommand{\X}{\mathcal{X}}

\newcommand{\C}{\mathcal{C}}

\newcommand{\G}{\mathcal{G}}
\newcommand{\M}{\mathcal{M}}
\newcommand{\Z}{\mathbb{Z}}
\newcommand{\T}{\mathcal{T}}
\newcommand{\ov}[1]{\overline{#1}}
\renewcommand{\H}{H^{\rm eff}}

\DeclareMathOperator*{\argmin}{argmin}

\begin{document}

\title{Frustrated neurons: Energy landscapes and relaxation dynamics \\ in repulsive phase oscillators}

\author{Brandon B. Le}
\email{sxh3qf@virginia.edu}
\affiliation{Department of Physics, University of Virginia, Charlottesville, Virginia 22904-4714, USA}

\date{\today}

\begin{abstract}
Geometrical frustration, a central paradigm in condensed matter physics, provides a unifying language for systems in which locally preferred interactions cannot be made globally compatible. Here, we use this language to formulate a minimal theory of frustrated neural timing, mapping repulsively coupled rhythmic units onto antiferromagnetic $XY$ models. In this formulation, the condensed-matter concepts of local constraints, degenerate ground-state manifolds, metastability, and quench dynamics are used to guide our analysis of structured neural phase dynamics. This paper analyzes a hierarchy of geometries, beginning with a triangle as the minimal frustrated motif with two chiral $120^\circ$ timing states, then a tetrahedron whose reduced ground-state manifold consists of intersecting continuous branches associated with antipodal pairings, and finally a kagome lattice on which local constraints define a constrained three-coloring manifold. The kagome lattice reveals that zero-temperature relaxation suppresses global synchrony but typically selects low-energy metastable torque-balanced states rather than exact ground states. We then show how the phase theory can be carried back towards biophysical neural models by treating the minimal framework as an effective-interaction target, where geometrical timing frustration arises from preferred phase lags that become incompatible around closed motifs. Our results suggest that weak global coherence in neural systems doesn't necessarily signal disordered activity, but can instead reflect structured local timing order within a frustrated dynamical landscape.
\end{abstract}

\maketitle

\tableofcontents

\section{Introduction}

Many interacting systems are governed by local preferences that cannot all be satisfied at once. In condensed matter physics, this incompatibility appears as geometrical frustration, where the geometry of the lattice forces a collective compromise among competing local constraints~\cite{toulouse1987theory, sadoc1999geometrical, diep2013frustrated, ronceray2019range}. This simple mechanism gives rise to a broad range of emergent phenomena, such as degeneracy, slow dynamics, emergent gauge structure, and unconventional phases of matter~\cite{ramirez1994strongly, moessner2006geometrical, lacroix2011introduction, balents2010spin, broholm2020quantum}. The underlying incompatibility is already visible in an antiferromagnetic triangle, where pairwise antiparallel alignment is impossible on all three bonds, and it recurs in extended form on triangular, kagome, and pyrochlore lattices~\cite{wannier1950antiferromagnetism, anderson1956ordering, chalker1992hidden, moessner1998low, gardner2010magnetic}.  The system must therefore compromise by replacing the perfect satisfaction of each bond with constrained patterns that distribute the mismatch across the motif or lattice~\cite{huse1992classical, reimers1992absence, ortiz2019colloquium}.

This leads to one of the most important consequences of frustration: the low-energy problem changes from the selection of a single ordered configuration to the organization of a structured landscape of degenerate or nearly degenerate states~\cite{henley1989ordering, henley2010coulomb, rau2019frustrated, pitts2022order, le2026energy}. Accordingly, frustrated systems are rarely characterized by a single global order parameter alone. Rather, their low-energy behavior is often organized by local constraint satisfaction, correlations within the allowed manifold, and the rearrangements or barriers that connect distinct low-energy configurations~\cite{chern2013dipolar, taillefumier2014semiclassical, wan2016color}. In constrained classical systems, this viewpoint is often expressed through local satisfaction rules and their violations, such as the Bernal--Fowler rules of water ice and the two-in/two-out rule and monopole-like defects of spin ice~\cite{bernal1933theory, pauling1935structure, harris1997geometrical, bramwell2001spin, castelnovo2008magnetic}. In noncollinear antiferromagnets, the local compromise can carry an additional handedness degree of freedom, making chirality a natural diagnostic of frustrated order alongside energy and spin correlations~\cite{kawamura1984phase, kawamura2001spin, seabra2011phase}. Spin glasses provide a useful precedent for comparing many competing outcomes, as the low-energy landscape contains numerous states. Then, the similarity between configurations becomes a natural quantity to examine, motivating overlap-based descriptions of phase space~\cite{edwards1975theory, parisi1983order, binder1986spin, dahlberg2025spin}. More broadly, spin ice and quantum spin liquids illustrate how frustrated constraints can produce residual entropy, emergent gauge structure, and excitations beyond conventional symmetry-breaking magnetic order~\cite{nisoli2013colloquium, gingras2014quantum, savary2017quantum, zhou2017quantum, knolle2019field}. For the purposes of this paper, the key point is that frustration organizes low-energy behavior through local constraints, degenerate manifolds, chiral structure, and the relations among competing states.

Frustration is also intrinsically tied to dynamics, since the same landscape that underlies the low-energy states shapes how the system relaxes from generic initial conditions. Starting far from equilibrium, relaxation can sample this landscape in several ways, such as ending in one region of a degenerate manifold, passing between locally compatible configurations through collective rearrangements~\cite{von1993spin, cepas2011resonating, le2026phase}, or freezing in metastable states when barriers divide the accessible phase space~\cite{klich2014glassiness,samarakoon2016aging,kim2018metastable}. Indeed, this dynamical viewpoint appears across frustrated magnetism. For example, order-by-disorder selects particular states from degenerate manifolds~\cite{savary2012order, ross2014order, oitmaa2013phase, rau2018pseudo}; spin-ice dynamics are controlled by the creation and motion of monopole-like defects~\cite{snyder2004low, jaubert2009signature, paulsen2014far, stoter2020extremely}; and clean frustrated magnets can exhibit glassy slowing, jamming, and metastable landscape structure~\cite{bilitewski2019dynamics, mitsumoto2023replica, mraz2023manipulation}. Across these examples, geometrical frustration fundamentally shapes both structure and dynamics by organizing the low-energy phase space and constraining the pathways by which the system evolves.

This viewpoint suggests a broader question: whenever a network contains locally preferred relations that cannot be made globally compatible, can the same language of frustration be useful outside magnetism? Many systems suggest that the answer is yes. Generalizing from the specific case of magnetic moments and exchange interactions, the essential ingredients become interacting degrees of freedom, locally preferred pairwise relations, and a network geometry that makes those preferences mutually incompatible. Geometrical frustration then provides a general language for understanding how local incompatibility can produce multiple collective states, partially satisfied constraints, nontrivial order parameters, and complex relaxation behavior in systems far removed from magnetic materials. 

The present work develops this broader viewpoint in the context of neuroscience. Neural systems are natural candidates for frustration because they are networks of interacting dynamical units whose collective function depends on timing, coupling type, and competition between local and global organization. Complete neural synchrony is generally not a signature of healthy brain dynamics; in fact, excessive synchronization is associated with pathological states such as seizures~\cite{jiruska2013synchronization, ren2021connectivity}. Instead, brain dynamics are organized through processes such as oscillatory coherence, phase relations, inhibition, modular coupling, and transient metastable patterns~\cite{buzsaki2004neuronal, fries2005mechanism, buzsaki2006rhythms, breakspear2017dynamic}. This makes frustration a potentially useful language for neural systems, since local interactions may favor particular timing relations, but the architecture of the network may make those relations mutually incompatible. Before formulating the specific geometrical analogy developed here, it is therefore important to clarify the diverse ways in which the word ``frustration'' has already been used in the neuroscience and brain-network theory literature.

One prominent usage comes from signed-network and structural-balance theory, where positive and negative functional connections define balanced and imbalanced arrangements, and frustration measures the extent to which the signed relations cannot all be made mutually consistent~\cite{saberi2021requirement, saberi2022pattern, moradimanesh2021altered, soleymani2023impact}. Another usage appears in synchronization-based models of brain and neural dynamics.  In cortical-motif models, frustration refers to closed-loop architectures that disrupt simple zero-lag synchrony and allow multiple metastable phase-synchronization patterns to coexist ~\cite{gollo2014frustrated, gollo2014mechanisms}. In connectome-based Kuramoto models, hierarchical modularity and intrinsic frequency heterogeneity can produce frustrated synchronization, metastability, and chimera-like states between the synchronized and incoherent regimes~\cite{villegas2014frustrated}. Related oscillator approaches emphasize other routes to obstructed global coherence, including phase frustration or phase lags in Kuramoto-type models of cortical activity~\cite{breakspear2010generative, nicosia2013remote, vuksanovic2014functional, caprioglio2024emergence} and network-geometric effects in brain-inspired oscillator networks~\cite{millan2018complex}. More recently, micropatterned biological neuron networks have been shown to exhibit dynamic
frustration, where increased gap-junction-mediated connectivity suppresses calcium synchronization under slow periodic ATP driving, consistent with coupling-induced anti-phase tendencies~\cite{li2025collective}. All these uses of the word ``frustration'' share the broad idea of competing tendencies in a network but differ in the mechanisms that generate the conflict. This variety makes it important to separate the general neuroscience ideas of frustration from the more specific condensed-matter notion of geometrical frustration.

With this distinction in mind, we turn our attention towards a specific geometrical form of neural frustration. Specifically, we consider rhythmic neural units at the phase level, where the relevant variable is the timing phase of an ongoing oscillation~\cite{ermentrout1991multiple, hoppensteadt1997weakly, izhikevich1999weakly, brown2004phase, ashwin2016mathematical}. The key interaction is anti-synchronizing, i.e., two coupled units locally prefer to fire out of phase. This is the direct phase-oscillator analog of antiferromagnetic coupling in an $XY$ magnet, where a planar spin is described by an angle and antiferromagnetic bonds favor opposite orientations~\cite{miyashita1984nature, teitel1983phase, bach2021antiferromagnetic}. While a single pair can satisfy this antiphase timing preference exactly, a non-bipartite motif cannot. On a triangle, for example, every edge locally favors a phase difference of $\pi$, but no assignment of three phases can satisfy all three edges simultaneously. This local incompatibility is the minimal frustrated-neuron mechanism developed in this work.

Before formulating this mechanism explicitly, it is useful to place it within the broader hierarchy of neuron models. At the most detailed level, conductance-based models such as the famous Hodgkin--Huxley model describe membrane excitability through voltage-dependent ionic currents~\cite{hodgkin1952quantitative}.  Reduced excitable-cell models, such as the FitzHugh--Nagumo~\cite{fitzhugh1961impulses, nagumo1962active} and Izhikevich~\cite{izhikevich2003simple} models, retain nonlinear spiking and bursting dynamics while compressing the underlying ionic variables into a smaller dynamical system. Map-based models such as the Chialvo~\cite{chialvo1995generic} and Rulkov maps~\cite{rulkov2002modeling, rulkov2001regularization} push this reduction further, replacing continuous-time membrane dynamics by low-dimensional discrete-time rules that still reproduce excitable, spiking, and bursting behavior. Phase reductions are the final step in this hierarchy. For rhythmic units with a stable oscillation, they retain only the timing variable and describe interactions through the evolution of oscillator phases~\cite{ermentrout1991multiple, brown2004phase, ashwin2016mathematical}.

Going all the way to the phase level is important for a first theory because it separates timing frustration from the additional mechanisms present in more biophysically detailed models. Excitable-cell equations and spiking maps can already generate multistability, complex attractor structure, final-state sensitivity, and coupling-dependent synchronization through their intrinsic nonlinear firing dynamics~\cite{izhikevich2007dynamical, xu2023extreme, bashkirtseva2023multistability, le2025chaotic, lamb2025final, le2026hyperchaos}. A phase description removes these additional sources of complexity while retaining the continuous timing variable, so the central question becomes clearer: what collective states arise from anti-synchronizing timing constraints alone? For identical oscillators with symmetric and repulsive sinusoidal coupling, the rotating-frame dynamics form a gradient flow of an antiferromagnetic $XY$ energy~\cite{kuramoto1984chemical, jadbabaie2004stability, acebron2005kuramoto, ha2013formation}. Energy-landscape descriptions also have an established precedent in Hopfield neural networks, whose stable states represent stored activity patterns~\cite{hopfield1982neural, hopfield1984neurons}. Here, the landscape follows from oscillator phase dynamics, and its minima represent phase-locked timing configurations. The phase model therefore provides the natural baseline for translating the condensed-matter language of frustrated magnetism into a controlled theory of neural timing.

At the phase level, the present work connects to a broad oscillator literature in which ``frustration'' has been implemented in several distinct ways. In Kuramoto--Sakaguchi models, frustration usually denotes a phase-lag parameter in the coupling function, which changes the synchronization problem and can support nontrivial order parameter dynamics or chimera-like states~\cite{sakaguchi1986soluble, brede2016frustration, botha2018analysis, yue2020model}. Other studies introduce frustration through random phase shifts, random interactions, heterogeneous frequencies, or mixtures of attractive and repulsive couplings, leading to oscillator-glass behavior, slow relaxation, contrarian synchronization, or desynchronization transitions~\cite{daido1992quasientrainment, zanette2005synchronization, hong2011kuramoto, iatsenko2014glassy}. Closest to the present construction are studies of repulsive or antiphase-coupled oscillator systems, including triangular chemical oscillators~\cite{yoshimoto1993coupling}, phase-repulsive networks with topology-dependent link frustration and final phase-locked patterns~\cite{levnajic2011emergent}, evolutionary design of non-frustrated phase-repulsive networks~\cite{tsimring2005repulsive, levnajic2012evolutionary}, geometrically frustrated rings of relaxation oscillators~\cite{goldstein2015synchronization}, and time-delay-coupled oscillators on frustrated motifs~\cite{thakur2017collective}. Related experimental oscillator platforms, such as negatively coupled laser arrays on kagome lattices, have also realized geometrical frustration and direct mappings to $XY$-spin energies~\cite{nixon2013observing}. These works show that antiphase coupling can realize frustration in oscillator systems, but their emphasis is typically on synchronization patterns, link-frustration measures, network design, chemical-oscillator states, or delay-controlled collective dynamics.

In this work, we take the minimal geometrical limit of this broader literature, namely, identical oscillators, symmetric sinusoidal coupling, no delays, no phase lags, and no mixture of attractive and repulsive links. In this limit, the single source of frustration is uniform repulsive timing constraints placed on non-bipartite motifs and lattices. Since the rotating-frame dynamics are exactly the antiferromagnetic $XY$ phase model, the object of study is shifted from synchronization failure to the frustrated-matter structure of ground-state manifolds, chiral sectors, metastable phase-locked attractors, basin weights, overlaps, and relaxation pathways. The result is a minimal framework in which geometrical frustration organizes neural timing states.

This paper is the first in a series on ``frustrated neurons,'' a broader program aimed at developing a controlled theoretical framework for geometrical frustration in neural timing. The present work establishes the minimal phase-oscillator foundation of that framework, where the frustrated-timing mechanism can be isolated before adding further neural complexity. Later papers will extend the same frustrated-timing principle to more biophysically detailed neural models, heterogeneous populations, stochastic dynamics, and larger neural architectures. These extensions will then be used to determine which conclusions of the minimal phase theory are robust and where they break down.

The remainder of this paper is organized as follows. Section~\ref{sec:frust_mag_neural_timing} reviews the condensed-matter ingredients needed for the analogy, introducing geometrical frustration in antiferromagnetic $XY$ systems, translating neural timing into a frustrated phase problem, and formulating a ``dictionary'' between frustrated magnets and neural oscillators. Section~\ref{sec:minimal_phase_theory} develops the minimal phase-oscillator theory, deriving the rotating-frame dynamics and the Lyapunov energy landscape, then establishing the exact mapping to the antiferromagnetic $XY$ phase model and introducing the diagnostics used throughout. Section~\ref{sec:oscillator_dyn_energy_land} is the main analysis section, studying the dynamics and energy landscapes across a hierarchy of frustrated geometries. It begins with the two-oscillator unfrustrated baseline, then proceeds to the triangle as the minimal frustrated motif, the tetrahedron as the first case with a continuous ground-state manifold, and a small kagome lattice as an extended network of corner-sharing triangles. Section~\ref{sec:biophys_real} interprets the phase-level results as a guide for biophysical realization, identifying the effective coupling mechanisms, observable diagnostics, and limitations that must be addressed in more detailed neural models. Section~\ref{sec:conclusions} summarizes the results and outlines future directions, including how subsequent papers in the frustrated-neurons series will extend the framework beyond the minimal phase description.

\section{Frustrated magnetism and neural timing}
\label{sec:frust_mag_neural_timing}

In this section, we make the analogy between geometrically frustrated magnets and phase-repulsive oscillators precise. To do this, we extract the important ideas from the condensed-matter framework and translate them into the language of neural timing. We begin by introducing the main condensed-matter reference system for this work, the antiferromagnetic $XY$ model, in which planar spins interact through bonds that locally prefer antiparallel alignment. We then formulate the corresponding neural timing problem, simplifying the relevant variable from a voltage amplitude to the phase of a neural oscillator. Finally, we flesh out the analogy in a ``dictionary'' that relates the central concepts of frustrated magnetism to the phase-oscillator diagnostics used in the rest of the paper.

\subsection{\texorpdfstring{Geometrical frustration in antiferromagnetic $XY$ systems}{Geometrical frustration in antiferromagnetic XY systems}}
\label{subsec:frustration_in_xy}

\begin{figure}[tbp!]
    \centering
    \includegraphics[width=0.7\linewidth]{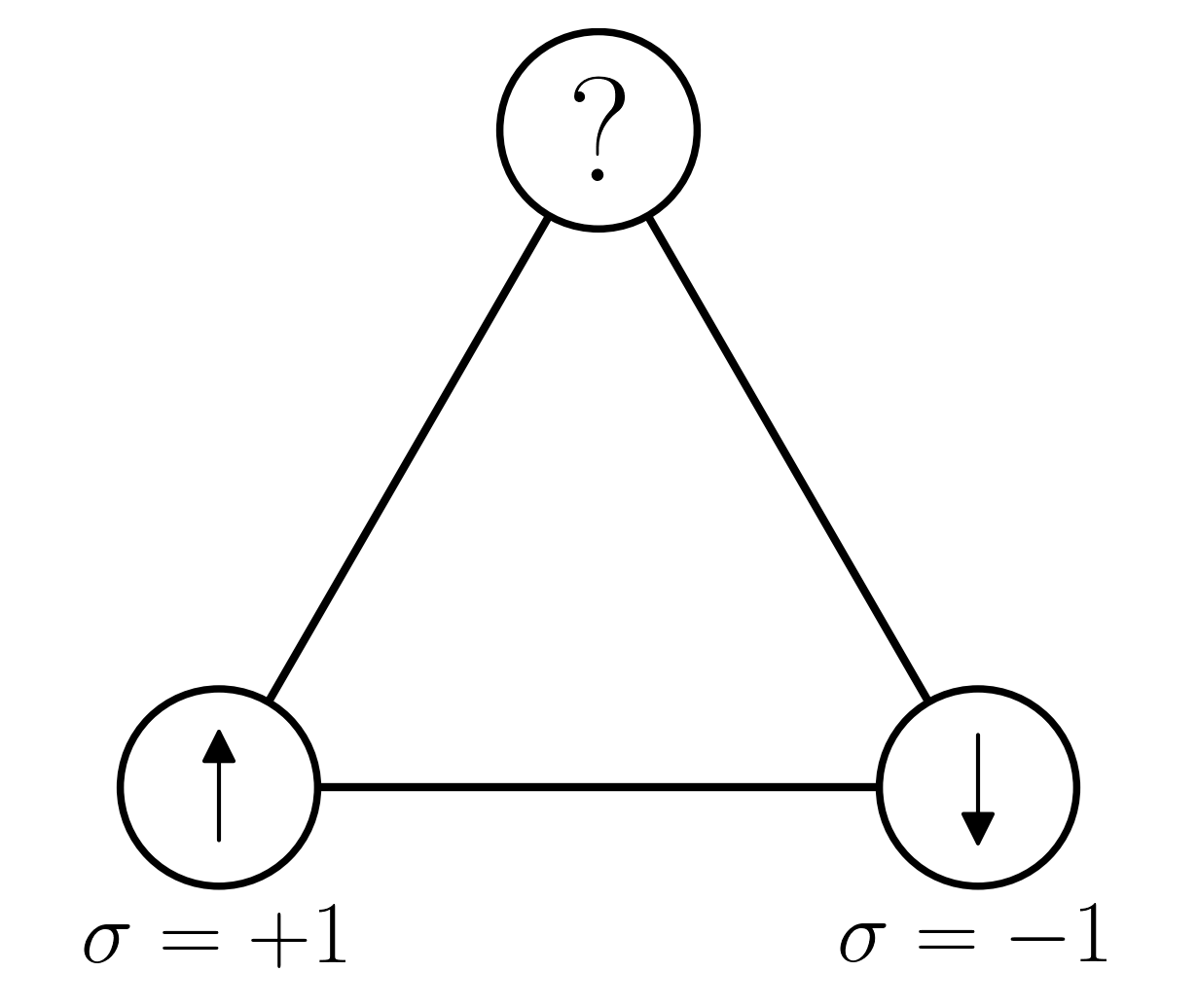}
    \caption{Antiferromagnetic Ising frustration on a triangle. Fixing two neighboring spins to satisfy one bond leaves the third spin unable to be antiparallel to both, so at least one bond must remain unsatisfied.}
    \label{fig:ising_triangle}
\end{figure}

The simplest setting in which geometrical frustration appears in condensed-matter physics is the antiferromagnetic Ising model. On a graph $G=(V,\mathcal{E})$, the Ising variables are discrete spins $\sigma_i = \pm 1$, and the antiferromagnetic Hamiltonian is
\begin{equation}
    \mathcal{H}_{\rm Ising} = J\sum_{\ev{ij}}\sigma_i\sigma_j,\quad J>0,
\end{equation}
where the sum is over all edges of the graph. The energy of each bond is minimized when neighboring spins are opposite each other: $\sigma_i\sigma_j = -1$. On a bipartite graph, this local rule can be satisfied globally by assigning opposite spins to the two sublattices. On a triangle, however, one bond is necessarily unsatisfied if the other two are satisfied (see Fig.~\ref{fig:ising_triangle}). Therefore, even the elementary triangular motif shows the basic mechanism of geometrical frustration, where locally preferred pairwise relations cannot all be satisfied simultaneously due to the geometry of the interaction graph~\cite{toulouse1987theory,sadoc1999geometrical,diep2013frustrated,ronceray2019range}. This phenomenon underlies the triangular antiferromagnet as well as more complex structures like kagome and pyrochlore lattices~\cite{wannier1950antiferromagnetism,anderson1956ordering,chalker1992hidden,moessner1998low,gardner2010magnetic}.

While the Ising example is useful because it exhibits frustration in its simplest discrete form, this paper is concerned with continuous timing variables. Therefore, the more useful magnetic analog is obtained by replacing the discrete Ising variable with a continuous spin direction. The Heisenberg model gives full orientational freedom by allowing each spin to point in any direction in three-dimensional space, but for reasons that will be made explicit in Sec.~\ref{sec:minimal_phase_theory}, the natural magnetic reference model is the $XY$ model, in which each spin is restricted to a plane. Then, each spin $\mb{S}_i$ is described by a single angle $\theta_i$:
\begin{equation}
    \mb{S}_i = (\cos\theta_i,\sin\theta_i),\quad|\mb{S}_i| = 1.
\end{equation}
The antiferromagnetic $XY$ Hamiltonian is
\begin{equation}
    \mathcal{H}_{XY} = J\sum_{\ev{ij}}\mb{S}_i\cdot\mb{S}_j = J\sum_{\ev{ij}}\cos(\theta_i - \theta_j),\quad J>0,
\end{equation}
so the energy of a single antiferromagnetic bond is minimized by antiparallel alignment:
\begin{equation}
    \theta_i - \theta_j\pmod{2\pi}.
\end{equation}
Hence, the local constraint is again that neighboring spins prefer to point in opposite directions. The difference from the Ising case is that the system can now compromise continuously. On a triangle, for example, the three bonds cannot all be antiparallel, but the energy can be minimized by distributing the frustration evenly, giving a $120^\circ$ pattern [see Fig.~\ref{fig:phase_oscillator_schematic}(a)].

This frustration becomes richer on larger motifs and extended lattices. On the tetrahedron, i.e., the complete graph $K_4$, every spin is coupled to every other spin, so pairwise antiparallel alignment is again impossible. The $XY$ antiferromagnet instead minimizes its energy by arranging the four planar spins in such a way that their vector sum vanishes, which produces a continuous ground-state manifold. The kagome lattice extends the frustrated triangle from an isolated motif to a two-dimensional network of corner-sharing triangles. Each triangle still favors a $120^\circ$ spin pattern, but globally, these local constraints can be satisfied in many compatible ways, producing a highly structured manifold of low-energy states~\cite{huse1992classical, chalker1992hidden, cepas2011resonating, le2026phase}. Built from corner-sharing tetrahedra, the pyrochlore lattice is the analogous extension of the tetrahedral motif, providing a natural three-dimensional counterpart to the kagome lattice. In this paper, we focus on the triangle, tetrahedron, and kagome lattice, leaving the pyrochlore extension for future work. For the purposes of this work, the main point is that frustration changes the object of study from a single ordered state to a manifold of degenerate or nearly degenerate configurations generated by incompatible local constraints~\cite{henley1989ordering,henley2010coulomb,rau2019frustrated}. We now translate this constraint-based viewpoint into neural timing language.

\subsection{Neural timing as a frustrated phase problem}
\label{subsec:neural_timing_frustrated_phase}

Many neuron models exhibit self-sustained rhythmic activity, where the state of an individual neuron repeatedly traverses a stable cycle after transients decay. At the biophysical level, this state may be high-dimensional, involving membrane voltage, recovery variables, synaptic currents, and ionic conductances (see Sec.~\ref{sec:biophys_real}). However, when the oscillation is stable and the coupling between neurons is sufficiently weak, the dominant variable is the position of the system along its cycle. The dynamics of neuron $i$ can then be described in terms of a single phase variable
\begin{equation}
    \theta_i \in [0,2\pi),
\end{equation}
where one full rotation corresponds to one period of neural activity. This phase-reduction viewpoint is standard in the theory of weakly coupled neural oscillators and provides the natural phase-level description of rhythmic neural units~\cite{ermentrout1991multiple, hoppensteadt1997weakly, izhikevich1999weakly, brown2004phase, ashwin2016mathematical}.

\begin{table*}[t]
    \caption{\label{tab:dictionary} Dictionary between frustrated-magnet concepts and their neural-timing analogs.}
    \renewcommand{\arraystretch}{1.3}
    \begin{ruledtabular}
    \begin{tabular}{cc}
    \textbf{Frustrated-magnet concept} & \textbf{Neural-timing analog} \\
    \hline
    Antiferromagnetic exchange interaction & Anti-synchronizing/phase-repulsive coupling \\
    $XY$ spin angle & Neural oscillation phase \\
    Magnetic Hamiltonian & Effective phase energy serving as a Lyapunov function \\
    Local constraint & Preferred antiphase timing relation on an edge \\
    Unsatisfied bond & Residual bond-frustration cost \\
    Geometrical frustration & Incompatible local antiphase timing constraints \\
    Ground state & Global-minimum phase-locked timing pattern \\
    Ground-state degeneracy & Multiple equal-cost phase-locked timing patterns \\
    Local order & Motif-level phase organization without global synchrony \\
    Chirality & Handedness of local phase winding \\
    Order parameter & Global phase coherence, measured by the Kuramoto order parameter \\
    Energy landscape & Effective dynamical landscape for phase configurations \\
    Metastable state & Stable non-ground-state phase-locked attractor \\
    Quench dynamics & Relaxation from random initial phases \\
    Overlap distribution & Pairwise similarity distribution of final phase-locked patterns \\
    Temperature & Noise strength or stochastic neural variability \\
    \end{tabular}
    \end{ruledtabular}
\end{table*}

We emphasize that the phase variable should not be identified with membrane voltage. Instead, $\theta_i$ labels the position of oscillator $i$ along its rhythmic trajectory, with spikes, bursts, or activity maxima occurring at particular phases. In the absence of coupling, the dynamics of each neuron obey
\begin{equation}
    \dot{\theta}_i = \omega_i,
\end{equation}
where $\omega_i$ is the neuron's intrinsic angular frequency. Coupling changes this phase velocity according to the relative timing of neighboring units. In the weak-coupling limit, this leads naturally to phase equations of the general form
\begin{equation}
    \dot{\theta}_i = \omega_i + \sum_jA_{ij}\Gamma(\theta_j-\theta_i),
\end{equation}
where $A_{ij}$ is the adjacency matrix of the interaction graph and $\Gamma$ is an effective phase-coupling function. The Kuramoto model~\cite{kuramoto1984chemical, acebron2005kuramoto} corresponds to the simplest sinusoidal choice of this interaction function, which the minimal model developed in Sec.~\ref{sec:minimal_phase_theory} will implement.

This phase description makes it possible to formulate a direct neural analog of antiferromagnetic frustration. In the magnetic system, the relevant variable is the angle of an $XY$ spin; in the neural system, it is the timing phase of an oscillator. Synchronizing interactions favor small phase differences, while anti-synchronizing interactions favor antiphase timing:
\begin{equation}
    \theta_i-\theta_j = \pi \pmod{2\pi}.
\end{equation}
For an isolated pair of oscillators, this local preference can be satisfied exactly when the two units occupy opposite points of their cycles.

Geometrical frustration appears when this same local timing preference is imposed on a network whose geometry prevents all preferred relations from being satisfied at once. On a triangle, for example, three pairwise antiphase relations cannot all hold simultaneously, so the network must compromise, producing a collective phase pattern that partially satisfies the incompatible local timing constraints. This is the neural-timing analog of the antiferromagnetic $XY$ frustration described above.

For now, we keep this formulation general. The essential ingredient is a local tendency toward antiphase timing, together with a network geometry that can make those local timing preferences mutually incompatible. The biological origin of such effective phase preferences is discussed in Sec.~\ref{sec:biophys_real}, where they are related to the phase-response and coupling mechanisms induced by more detailed voltage- and synapse-level dynamics. The next subsection organizes this analogy as a dictionary between frustrated magnetism and neural timing, then Sec.~\ref{sec:minimal_phase_theory} specializes it to the concrete phase-oscillator model used in the rest of the paper.

\subsection{Dictionary between frustrated magnets and neural oscillators}

We have now introduced the two sides of the analogy: antiferromagnetic $XY$ systems, where planar spins are subject to incompatible local alignment constraints, and neural phase-oscillator systems, where rhythmic units are subject to incompatible local timing constraints. We now make this correspondence explicit by using the language of frustrated magnetism as an organizing framework for neural systems.

Table~\ref{tab:dictionary} summarizes the main dictionary used throughout the paper. The most basic entries are the phase variable and the interaction. In the magnetic problem, each $XY$ spin is described by an angle, and antiferromagnetic interactions favor antiparallel orientations. In the neural-timing problem, the corresponding variable is the oscillation phase, and the corresponding interaction is anti-synchronizing phase coupling. In other words, the magnetic preference for antiparallel spins becomes a timing preference for antiphase oscillation.

The local-constraint language is especially important. A single antiferromagnetic bond can be satisfied by making two spins antiparallel, just as a single anti-synchronizing connection can be satisfied by placing two oscillators in antiphase. Geometrical frustration appears when the network makes these local preferences mutually incompatible. The neural analog of an unsatisfied bond is therefore a residual bond-frustration cost that results from a local timing relation that deviates from perfect antiphase.

The ground-state language carries over with one important caveat. A magnetic ground state maps to a global-minimum phase-locked state in the phase energy landscape, while ground-state degeneracy becomes the existence of multiple equal-energy phase-locked states. Here, the ``phase energy'' should be understood operationally, not literally. It assigns a cost to phase configurations and organizes the resulting landscape, rather than representing a thermodynamic energy of the neural system. In Sec.~\ref{sec:minimal_phase_theory}, this interpretation is made precise by showing that the symmetric phase dynamics possess a Lyapunov function.

The remaining entries in the table describe how this landscape is diagnosed and sampled. Local order refers to motif-level phase organization that can persist even when global synchrony is suppressed. Chirality measures the handedness of local phase winding on triangular motifs. Metastable states are stable non-ground-state phase-locked attractors, while quench dynamics refers to relaxation from randomly chosen initial phases. Overlap distributions compare the similarity between final phase-locked patterns reached from different initial conditions. Finally, the analog of temperature is noise strength or stochastic neural variability. The present work focuses on the zero-noise, or zero-temperature, limit, and the corresponding quench protocol is defined explicitly in Sec.~\ref{sec:minimal_phase_theory}. Together, these concepts provide the vocabulary for the minimal phase-oscillator theory and the hierarchy of frustrated geometries analyzed below.

\section{Minimal phase-oscillator theory}
\label{sec:minimal_phase_theory}

The preceding section established the general analogy between frustrated magnetic interactions and phase-repulsive neural timing. We now specialize to the minimal dynamical model used throughout the rest of the paper. The goal of this model is to isolate the phase degree of freedom of a stable neural unit and study how repulsive timing interactions behave on frustrated graphs. As we do not aim to represent the full biophysical state of the neuron at this level, spike-shape variation, synaptic delays, heterogeneity, and noise are omitted from this first reference model. Sec.~\ref{sec:biophys_real} explains how these ingredients re-enter as biophysical extensions, and subsequent work will study them explicitly. What remains for now is a controlled phase-only limit in which frustration, degeneracy, metastability, and basin structure can be defined sharply.

Let $G = (V,\mathcal{E})$ be an interaction graph with $N=|V|$ oscillators and adjacency matrix $A_{ij}$. Here, $G$ specifies the interaction topology, not a physical embedding, so the motifs and lattices considered below represent patterns of connectivity and shouldn't necessarily be interpreted as anatomical arrangements. We take identical phase oscillators with common angular frequency $\omega$ and repulsive sinusoidal coupling, i.e., a negative-coupling Kuramoto-type interaction~\cite{kuramoto1984chemical,acebron2005kuramoto,tsimring2005repulsive}:
\begin{equation}
    \dot{\theta}_i = \omega - K\sum_j A_{ij}\sin(\theta_j-\theta_i),\quad K>0.
    \label{eq:repulsive_kuramoto}
\end{equation}
Here, $\theta_i\in[0,2\pi)$ is the phase of oscillator $i$. The sign convention in Eq.~\eqref{eq:repulsive_kuramoto} is chosen so that a single isolated bond favors antiphase locking. Indeed, for two coupled oscillators, the relative phase $\delta = \theta_1-\theta_2$ obeys
\begin{equation}
    \dot{\delta}=2K\sin\delta,
\end{equation}
so $\delta = 0$ is unstable and $\delta = \pi$ is stable. Therefore, each edge of the graph locally prefers
\begin{equation}
    \theta_i-\theta_j = \pi \pmod{2\pi}.
\end{equation}
Geometrical frustration arises when the graph prevents these local antiphase preferences from being satisfied simultaneously.

Because all oscillators have the same intrinsic frequency, the uniform drift can be removed by passing to a co-rotating frame,
\begin{equation}
    \phi_i(t) = \theta_i(t)-\omega t.
\end{equation}
The dynamics then become
\begin{equation}
    \dot{\phi}_i = -K \sum_j A_{ij}\sin(\phi_j-\phi_i),
    \label{eq:rotating_frame_phase_model}
\end{equation}
where the variables $\phi_i$ describe relative timing patterns. A phase-locked state in the laboratory variables therefore corresponds to a fixed point of Eq.~\eqref{eq:rotating_frame_phase_model}. Since the model is invariant under global $U(1)$ phase rotations
\begin{equation}
    \phi_i\mapsto \phi_i+\Phi,
    \qquad \Phi\in\mathbb{R}/2\pi\mathbb{Z},
\end{equation}
only phase differences are physically meaningful. Here, $\mathbb{R}/2\pi\mathbb{Z}$ denotes the quotient group of phases modulo $2\pi$, which is isomorphic to $U(1)$. In what follows, we use the representative interval $[0,2\pi)$ when explicit phase values are needed.

For the symmetric networks $A_{ij} = A_{ji}$ considered in this paper, Eq.~\eqref{eq:rotating_frame_phase_model} has a gradient-flow structure. Define
\begin{equation}
    E[\{\phi_i\}] = K \sum_{\langle ij\rangle}\cos(\phi_i-\phi_j),
    \label{eq:phase_energy}
\end{equation}
where the sum is over undirected edges of the graph, so each interacting pair is counted once. Then,
\begin{equation}
    \dot{\phi}_i = -\pdv{E}{\phi_i}.
    \label{eq:gradient_flow}
\end{equation}
Consequently,
\begin{equation}
    \dv{E}{t} = \sum_i\pdv{E}{\phi_i}\dot{\phi}_i = -\sum_i \dot{\phi}_i^2 \leq 0,
\end{equation}
so $E$ is a Lyapunov function for the symmetric phase-only dynamics~\cite{vanhemmen1993lyapunov}. 

This deterministic model can be viewed as the zero-temperature limit of overdamped Langevin dynamics on the same phase landscape~\cite{risken1996fokker,gardiner2009stochastic,acebron2005kuramoto}. A finite-temperature extension takes the form
\begin{equation}
    \dot{\phi}_i = -K\sum_j A_{ij}\sin(\phi_j-\phi_i) + \sqrt{2T}\,\eta_i(t),
\end{equation}
where the noise has zero mean and correlations
\begin{equation}
    \ev{\eta_i(t)} = 0,\quad \ev{\eta_i(t)\eta_j(t')} = \delta_{ij}\delta(t-t').
\end{equation}
In the magnetic language, $T$ plays the role of temperature, setting the strength of fluctuations that allow the system to wander through the energy landscape rather than follow a strictly downhill trajectory. In the neural-timing interpretation, the same parameter should be understood as an effective phase-noise strength, representing stochastic timing variability instead of a literal thermodynamic temperature. In this paper, we set $T=0$, so the dynamics reduce to deterministic relaxation in the phase-energy landscape.

The system evolves downhill in the effective landscape until it reaches a stationary phase-locked state: 
\begin{equation}
    \pdv{E}{\phi_i}=0
    \label{eq:stationary_condition}
\end{equation}
for all $i$. Global minima are ground states of the phase model, while stable nonground local minima correspond to metastable phase-locked attractors. Saddles and maxima are unstable under the gradient flow, but their stable manifolds organize the basin boundaries between the stable long-time outcomes. Importantly, the gradient flow does not search globally for the absolute minimum. It follows the downhill path selected by the initial condition and can therefore terminate in a nonground local minimum whenever such a minimum exists. In the kagome lattice discussed in Subsec.~\ref{subsec:kagome}, this same stationarity condition will be interpreted locally as a torque-balance condition.

The connection to the antiferromagnetic $XY$ model introduced in Subsec.~\ref{subsec:frustration_in_xy} is obtained by introducing planar unit spins
\begin{equation}
    \mathbf S_i=(\cos\phi_i,\sin\phi_i).
    \label{eq:spin_rep}
\end{equation}
Then, Eq.~\eqref{eq:phase_energy} becomes
\begin{equation}
    E = K \sum_{\langle ij\rangle} \mathbf{S}_i\cdot\mathbf{S}_j.
    \label{eq:xy_energy}
\end{equation}
Because each oscillator contributes one phase $\phi_i\in S^1$, this mapping is specifically to the $XY$ model rather than to the Heisenberg model discussed in Subsec.~\ref{subsec:frustration_in_xy}, whose spins have two independent degrees of freedom in $S^2$. Heisenberg systems can therefore access noncoplanar ground states that don't appear in our phase model, so the manifolds derived in Sec.~\ref{sec:oscillator_dyn_energy_land} are specifically $XY$ results.

In Eq.~\eqref{eq:xy_energy}, since $K>0$, each bond is minimized when neighboring spins are antiparallel. The phase-oscillator dynamics therefore describe overdamped relaxation in the energy landscape of a classical antiferromagnetic $XY$ model on the same graph. We note that this relation is the sign-reversed counterpart of the standard correspondence between attractive Kuramoto synchronization and ferromagnetic $XY$ order~\cite{kuramoto1984chemical, vanhemmen1993lyapunov, jadbabaie2004stability, acebron2005kuramoto, mehta2015algebraic, ling2019landscape}. Indeed, reversing the sign of the coupling in Eq.~\eqref{eq:rotating_frame_phase_model} gives
\begin{equation}
    \dot\phi_i = +K\sum_j A_{ij}\sin(\phi_j - \phi_i) = \pdv{E_{\rm F}}{\phi_i},
\end{equation}
with ferromagnetic $XY$ energy
\begin{equation}
    E_{\rm F}[\{\phi_i\}] = -K\sum_{\ev{ij}}\cos(\phi_i-\phi_j).
\end{equation}
Under the spin representation in Eq.~\eqref{eq:spin_rep}, phase coherence is equivalent to planar magnetization, and complete synchrony is equivalent to a ferromagnetic ground state. Since $E = -E_{\rm F}$ at the same phase configuration, the attractive and repulsive systems have the same stationary configurations with minima and maxima switched.

On a bipartite graph, the attractive and repulsive cases are related even more directly. Letting $\beta_i = 0$ on one sublattice and $\beta_i = 1$ on the other, we can define the staggered phases $\tilde\phi_i = \phi_i + \pi\beta_i$. Since every edge joins opposite sublattices, $\cos(\phi_i-\phi_j) = -\cos(\tilde\phi_i - \tilde\phi_j)$, so
\begin{equation}
    E[\{\phi_i\}] = E_{\rm F}[\{\tilde\phi_i\}].
\end{equation}
Hence, the repulsive dynamics in the original phases become attractive in the staggered phases, so every antiphase bond can be satisfied simultaneously on a bipartite graph, as discussed in Subsec.~\ref{subsec:frustration_in_xy}. However, an odd cycle obstructs this transformation and is therefore the elementary topological source of geometrical frustration~\cite{zanette2005synchronization, tsimring2005repulsive, levnajic2011emergent, chowdhury2020effect}. In the present neural-timing interpretation, however, $E$ should be understood as an effective dynamical landscape, not a literal thermodynamic energy of the system.

We now define the observables used to diagnose the relaxation dynamics. The first is the energy itself, usually reported either in dimensionless form $E/K$ for small motifs or as the energy density $E/KN$ for extended networks. A closely related local measure is the normalized bond-frustration cost
\begin{equation}
    f_{ij} = \frac{1+\cos(\phi_i-\phi_j)}{2}.
    \label{eq:bond_frustration}
\end{equation}
This satisfies $f_{ij} = 0$ for a perfectly satisfied antiphase bond and $f_{ij} = 1$ for a maximally unsatisfied in-phase bond. The mean bond frustration is
\begin{equation}
    \bar f = \frac{1}{N_b}\sum_{\langle ij\rangle} f_{ij},
    \label{eq:mean_bond_frustration}
\end{equation}
where $N_b = |\mathcal{E}|$ is the number of bonds. Using Eq.~\eqref{eq:phase_energy}, we have the general relation
\begin{equation}
    \bar f = \frac{1}{2} + \frac{E}{2KN_b},
    \label{eq:fbar_energy_relation_general}
\end{equation}
so $\bar f$ carries the same energetic information as $E$ but with the direct interpretation of an average local bond cost.

Global phase coherence is measured by the Kuramoto order parameter~\cite{kuramoto1984chemical, acebron2005kuramoto}
\begin{equation}
    R = \qty|\frac{1}{N}\sum_{j=1}^{N} e^{i\phi_j}|.
    \label{eq:kuramoto_order_parameter}
\end{equation}
A large $R$ indicates global synchrony, while a small $R$ indicates suppressed global coherence. In a frustrated antiphase network, however, small $R$ should not automatically be interpreted as either disorder or local frustrated order. It may reflect random phase cancellation, but it may also arise from organized cancellation imposed by local antiphase constraints. For this reason, $R$ must be supplemented by local diagnostics.

On graphs containing elementary triangles, we use two main local triangular observables. For an unoriented triangle $\triangle = (i,j,k)$, define the triangle moment
\begin{equation}
    m_\triangle =\qty|e^{i\phi_i} + e^{i\phi_j} + e^{i\phi_k}|,
    \label{eq:triangle_moment}
\end{equation}
with $0\leq m_\triangle\leq 3$. This quantity vanishes when the three phases form an exact $120^\circ$ pattern, since then
\begin{equation}
    e^{i\phi_i}+e^{i\phi_j}+e^{i\phi_k}=0.
\end{equation}
For a network of triangles $\T$, we define the average triangle moment
\begin{equation}
    \ov{m}_\triangle =\frac{1}{N_\triangle}\sum_{\triangle\in\mathcal T} m_\triangle,\quad N_\triangle=|\mathcal T|,
    \label{eq:mean_triangle_moment}
\end{equation}
which measures the average violation of the local $120^\circ$ constraint.

For an oriented triangle $\triangle=(i,j,k)$, we also define the local chirality
\begin{equation}
    \chi_\triangle = \frac{2}{3\sqrt{3}}\qty[\sin(\phi_j-\phi_i) + \sin(\phi_k-\phi_j) + \sin(\phi_i-\phi_k)].
    \label{eq:triangle_chirality}
\end{equation}
The two ideal $120^\circ$ states have $\chi_\triangle = \pm 1$, corresponding to opposite handedness of the phase winding around the triangle. In an extended triangular or kagome network, a useful coarse measure of chirality imbalance is the net chirality magnitude:
\begin{equation}
    |m_\chi| = \qty|\frac{1}{N_\triangle}\sum_{\triangle\in\mathcal T}\chi_\triangle|.
    \label{eq:net_chirality_magnitude}
\end{equation}
Small $|m_\chi|$ indicates approximate compensation between positive and negative local chiralities, while larger values indicate a net handedness imbalance.

Finally, we characterize the landscape dynamically using a zero-temperature quench protocol. Initial phases are sampled independently and uniformly from $[0,2\pi)$, and the deterministic gradient flow in Eq.~\eqref{eq:rotating_frame_phase_model} is integrated until a stationary state is reached. The phrase ``zero-temperature'' emphasizes that there is no noise or thermal activation; the trajectory can only move downhill in $E$. Therefore, the final state is determined by the basin of attraction~\cite{nusse1996basins, wagemakers2025basins} containing the initial condition.

If $\mathcal{A}$ denotes a final-state class, such as a single attractor, a branch of the degenerate ground-state manifold, or the exact ground-state sector, its basin probability or basin stability~\cite{menck2013basin, leng2016basin, schultz2017potentials} is estimated by
\begin{equation}
    p_{\mathcal A} = \frac{N_{\mathcal A}}{N_{\rm init}},
    \label{eq:basin_probability_general}
\end{equation}
where $N_{\mathcal A}$ is the number of trajectories ending in class $\mathcal A$ and $N_{\rm init}$ is the total number of initial conditions. When the relevant final states form a degenerate manifold rather than isolated points, the class label must be supplemented by coordinates along that manifold. This is the situation for the tetrahedral motif, where a trajectory selects both a discrete antipodal-pairing branch and a continuous internal angle along that branch.

\begin{figure*}
    \centering
    \includegraphics[width=\linewidth]{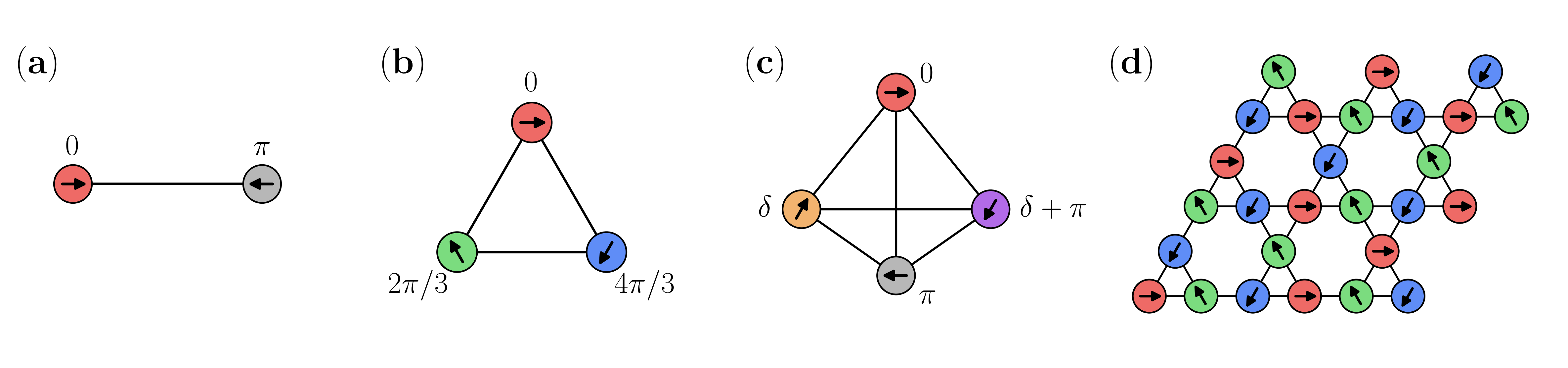}
    \caption{Schematic hierarchy of frustrated phase-oscillator geometries studied in Sec.~\ref{sec:oscillator_dyn_energy_land}. (a)~A two-oscillator pair provides the unfrustrated baseline, where the single repulsive bond can be satisfied exactly by an antiphase relation. (b)~The triangular motif is the minimal geometrically frustrated unit, where three repulsive bonds cannot all be satisfied simultaneously, so the reduced ground states form two chiral $120^\circ$ configurations. (c)~The tetrahedral motif introduces a continuous reduced ground-state manifold, organized by antipodal pairings of planar spins. (d)~A small kagome lattice extends the triangular constraint to a network of corner-sharing triangles, producing a many-state constrained landscape with richer basin and overlap structure.}
    \label{fig:phase_oscillator_schematic}
\end{figure*}

Finally, to compare two final phase patterns labeled by trajectory indices $\alpha$ and $\beta$, we use the phase overlap
\begin{equation}
    q_{\alpha\beta} = \qty|\frac{1}{N}\sum_{j=1}^{N}e^{i(\phi_j^{(\alpha)}-\phi_j^{(\beta)})}|,
    \label{eq:phase_overlap}
\end{equation}
which is invariant under independent global rotations of the two configurations. This quantity is the phase-oscillator analog of the replica overlap used in spin-glass theory~\cite{parisi1983order} and measures trial-to-trial similarity between final timing patterns. It satisfies $0\leq q_{\alpha\beta}\leq1$, with $q_{\alpha\beta}=1$ when the two final states agree up to a global phase shift. Broad overlap distributions indicate that the quench dynamics reach many inequivalent timing patterns, while distributions concentrated near large $q_{\alpha\beta}$ indicate a more restricted attractor set.

The minimal phase-oscillator theory developed in this section gives a common language for the examples below. In the rotating frame, the symmetric repulsive phase dynamics is gradient descent in an effective antiferromagnetic $XY$ landscape, so ground states, metastable attractors, and basin structure can be defined directly from the energy. The accompanying diagnostics separate different features of the relaxation: $E$ and $\bar f$ measure residual bond cost, $R$ measures global coherence, $m_\triangle$, $\ov{m}_\triangle$, and $\chi_\triangle$ probe local triangular order, while basin probabilities and overlaps describe the ensemble of final phase-locked states. The next section uses these quantities to analyze how local antiphase preferences are expressed on increasingly frustrated geometries, producing chirality, continuous degeneracy, and metastable attractor structure.

\section{Frustrated phase-oscillator dynamics and energy landscapes}
\label{sec:oscillator_dyn_energy_land}

The minimal phase-oscillator theory developed in Sec.~\ref{sec:minimal_phase_theory} turns geometrical frustration into an explicit dynamical problem, as the same interaction graph determines both the ground-state constraints and the relaxation pathways through the reduced energy landscape. In the rotating frame, the common drift is removed, and the remaining dynamics relax downhill in the interaction-energy landscape. In this section, we apply this framework to a sequence of interaction graphs that move from an exactly satisfiable repulsive bond to increasingly constrained frustrated networks.

The two-oscillator problem serves as the unfrustrated reference case because a single repulsive interaction can be satisfied exactly. By contrast, since three pairwise antiphase preferences cannot be satisfied simultaneously, the triangle is the first frustrated motif. After quotienting by the neutral global $U(1)$ rotation, its reduced ground-state structure consists of two chiral $120^\circ$ states. The tetrahedron introduces a qualitatively different type of degeneracy, in which the ground-state manifold contains both discrete pairing choices and a residual continuous coordinate. Finally, the kagome lattice extends the triangular constraint to an array of corner-sharing triangles, providing the first extended-network example in which frustrated constraints generate a many-state metastable landscape.

This hierarchy is summarized schematically in Fig.~\ref{fig:phase_oscillator_schematic}. The analysis below therefore has two aims: first, to identify the ground-state structure of each system, and second, to determine how zero-temperature relaxation partitions phase space among ground-state and metastable phase-locked attractors. The finite motifs isolate the basic mechanisms of antiphase locking, chirality, and degeneracy, while the kagome lattice provides the first extended-network setting in which local frustrated constraints generate a richer landscape of final states.

\subsection{Two-oscillator baseline}
\label{subsec:two_oscillator}

We first consider a two-oscillator system, which provides a useful baseline since the single repulsive bond can be satisfied exactly. In the rotating frame of the common intrinsic frequency $\omega$, by Eq.~\eqref{eq:rotating_frame_phase_model}, the phase dynamics are
\begin{equation}
    \begin{split}
        \dot{\phi}_1 &= -K\sin(\phi_2-\phi_1), \\
        \dot{\phi}_2 &= -K\sin(\phi_1-\phi_2).
    \end{split}
\end{equation}
Introducing the relative phase $\delta = \phi_1-\phi_2$, the dynamics reduce to
\begin{equation}
    \dot{\delta} = 2K\sin\delta.
    \label{eq:two_oscillator_relative_phase}
\end{equation}
The in-phase state $\delta=0$ is therefore unstable, while the antiphase state $\delta=\pi$ is stable.

\begin{figure}[tbp!]
    \centering
    \includegraphics[width=\linewidth]{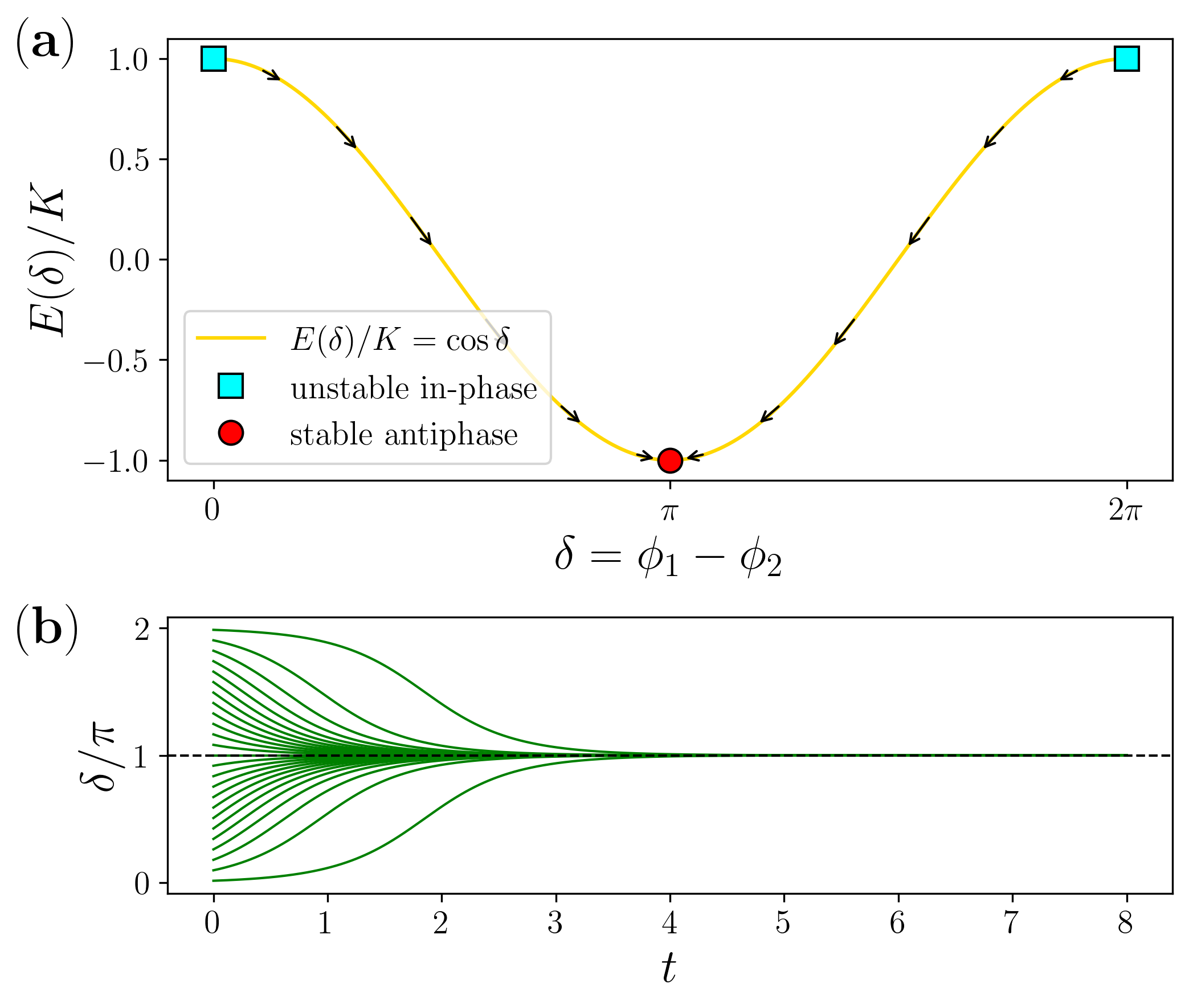}
    \caption{Reduced energy-landscape description of the two-oscillator phase-repulsive bond. (a)~The one-bond energy $E(\delta)=K\cos\delta$ has unstable in-phase maxima at $\delta=0$ and $2\pi$, and a stable antiphase minimum at $\delta=\pi$. Arrows indicate the direction of the reduced gradient flow in the relative phase $\delta$. (b)~Relative-phase trajectories from many initial conditions converge to the antiphase state $\delta=\pi$. The exactly in-phase state is unstable and remains fixed only for the fine-tuned initial condition $\delta=0 \mod 2\pi$.}
    \label{fig:two_oscillator_landscape}
\end{figure}

Specializing the gradient-flow structure of Sec.~\ref{sec:minimal_phase_theory} to the single relative phase $\delta$, the effective or reduced one-dimensional landscape is
\begin{equation}
    E(\delta) = K\cos\delta,
\end{equation}
since Eq.~\eqref{eq:two_oscillator_relative_phase} can be written as $\dot{\delta} = -2\,\mathrm{d}E/\mathrm{d}\delta$. Then,
\begin{equation}
    \dv{E}{t} = \dv{E}{\delta}\dot{\delta} = -2K^2\sin^2\delta\leq 0,
\end{equation}
so the dynamics monotonically decrease the energy until the system reaches the antiphase minimum at $\delta = \pi$. This reduced landscape is shown in Fig.~\ref{fig:two_oscillator_landscape}(a), where the in-phase state $\delta=0$ is the maximum with $E(0)=K$, and the antiphase state $\delta=\pi$ is the minimum with $E(\pi)=-K$. The arrows indicate the direction of the reduced flow. Figure~\ref{fig:two_oscillator_landscape}(b) shows the corresponding phase dynamics from many initial values of $\delta$, demonstrating that all initial conditions flow away from the exactly in-phase unstable fixed point toward $\delta=\pi$.

\begin{figure}[tbp!]
    \centering
    \includegraphics[width=\linewidth]{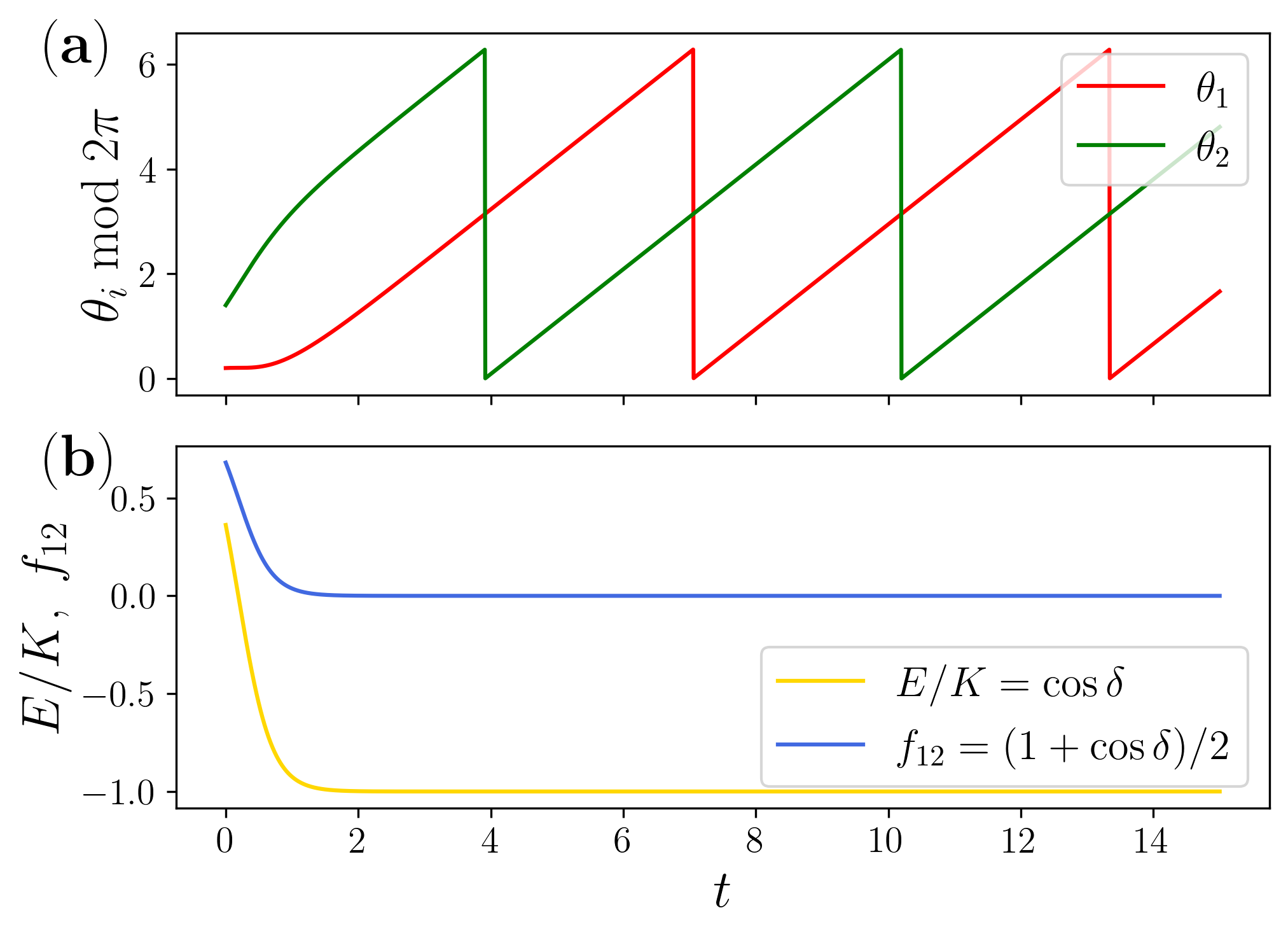}
    \caption{Two-oscillator baseline for a single phase-repulsive bond. (a)~The laboratory-frame phases $\theta_i$, shown modulo $2\pi$, continue to rotate while approaching a fixed antiphase separation. Passing to the rotating-frame variables $\phi_i=\theta_i-\omega t$ removes the common drift without changing the relative phase, $\delta=\theta_1-\theta_2=\phi_1-\phi_2$. (b)~The one-bond energy $E=K\cos\delta$ decreases toward its minimum value at $\delta=\pi$, while the bond-frustration measure $f_{12}=(1+\cos\delta)/2$ relaxes to zero, indicating a fully satisfied antiphase bond.}
    \label{fig:two_oscillator_time_series}
\end{figure}

The same relaxation is visible in the full oscillator dynamics. Figure~\ref{fig:two_oscillator_time_series}(a) shows the original, nonrotating phases $\theta_i$ modulo $2\pi$. Although the common rotation remains present in these variables, the phase separation approaches an antiphase relation. Since $\theta_1-\theta_2=\phi_1-\phi_2$, the energy landscape and frustration can be evaluated equivalently using either the rotating-frame or nonrotating phases. Figure~\ref{fig:two_oscillator_time_series}(b) shows the corresponding one-bond energy decreasing toward its minimum. Recalling the bond-frustration measure $f_{ij}=\qty(1+\cos(\phi_i-\phi_j))/2$ [Eq.~\eqref{eq:bond_frustration}] introduced in Sec.~\ref{sec:minimal_phase_theory},
the two-oscillator frustration $f_{12}$ simultaneously relaxes to zero, confirming that the single repulsive bond becomes perfectly satisfied.

Thus, the two-oscillator system is unfrustrated in the strict sense that its preferred antiphase constraint can be achieved exactly. It therefore provides the reference behavior for the frustrated motifs studied below, where the same local antiphase preference cannot be simultaneously satisfied on all bonds.

\subsection{Triangle motif: Minimal geometrical frustration}
\label{subsec:triangle}

We now consider dynamics on the smallest motif that exhibits geometrical frustration: a triangle. The two-oscillator system provides an unfrustrated baseline because its single repulsive bond can be satisfied exactly by an antiphase state. However, on a triangle, each bond still prefers antiphase alignment, but the three pairwise antiphase constraints are mutually incompatible around a closed odd cycle. It is therefore the minimal phase-oscillator analog of an antiferromagnetic triangular plaquette, the canonical example of geometrical frustration in condensed-matter physics.

The adjacency matrix on the three-node motif is
\begin{equation}
    A_\triangle = \mqty(0 & 1 & 1 \\ 1 & 0 & 1 \\ 1 & 1 & 0).
\end{equation}
Substituting this graph into the rotating-frame dynamics [Eq.~\eqref{eq:rotating_frame_phase_model}] gives
\begin{equation}
    \begin{split}
        \dot{\phi}_1 &= -K[\sin(\phi_2-\phi_1) + \sin(\phi_3-\phi_1)], \\
        \dot{\phi}_2 &= -K[\sin(\phi_1-\phi_2) + \sin(\phi_3-\phi_2)], \\
        \dot{\phi}_3 &= -K[\sin(\phi_1-\phi_3) + \sin(\phi_2-\phi_3)]. \\
    \end{split}
\end{equation}
As in the general symmetric model of Sec.~\ref{sec:minimal_phase_theory}, these equations have the gradient-flow form $\dot{\phi}_i = -\partial E/\partial\phi_i$ with triangular energy
\begin{equation}
\begin{split}
    E(\phi_1,\phi_2,\phi_3) &= K[\cos(\phi_1-\phi_2) \\ &\qquad + \cos(\phi_2-\phi_3) + \cos(\phi_3-\phi_1)].
    \label{eq:triangle_energy}
\end{split}
\end{equation}
It is straightforward to see that this energy landscape cannot be minimized by satisfying every bond independently. Each term in Eq.~\eqref{eq:triangle_energy} is minimized when the corresponding phase difference is $\pi$, but the three conditions
\begin{equation}
    \phi_1-\phi_2 = \pi,\quad \phi_2-\phi_3 = \pi,\quad \phi_3-\phi_1 = \pi\pmod{2\pi}
\end{equation}
cannot all hold simultaneously since adding them gives $0 = 3\pi\pmod{2\pi}$, a contradiction. The triangle therefore forces a compromise state. Because the three local antiphase preferences are mutually incompatible, the lowest-energy configuration must share the unavoidable residual bond frustration across the motif.

\subsubsection{Ground-state degeneracy and structure}

The ground-state structure of the triangular motif can be derived in direct analogy with the classical antiferromagnetic $XY$ model. Defining the unit vectors $\mb{S}_i = (\cos\phi_i,\sin\phi_i)$ as in Sec.~\ref{sec:minimal_phase_theory}, the triangular energy can be written as
\begin{equation}
    E = K\sum_{1\leq i<j\leq 3}\mb{S}_i\cdot\mb{S}_j = \frac{K}{2}\qty(|\mb{S}_1 + \mb{S}_2 + \mb{S}_3|^2 - 3).
\end{equation}
The energy is minimized when
\begin{equation}
    \mb{S}_1 + \mb{S}_2 + \mb{S}_3 = 0.
\end{equation}
For three unit vectors in the plane, this condition is satisfied when the phases are separated by $2\pi/3$. Hence, the frustrated triangular motif compromises by forming a $120^\circ$ phase pattern. This is directly analogous to the ground state of the classical antiferromagnetic $XY$ model on a triangular plaquette, where each pair of neighboring spins prefers to be antiparallel, but the closed triangular geometry makes it impossible to satisfy all three antiferromagnetic bonds simultaneously.

The minimum energy is
\begin{equation}
    E_{\text{min}} = \frac{K}{2}(0-3) = -\frac{3K}{2},
\end{equation}
or equivalently, in a $120^\circ$ state, each bond has $\cos(\phi_i - \phi_j) = -1/2$, so each of the three bonds contributes $-K/2$ to the total energy. In other words, the frustration is distributed equally over the triangle, rather than being localized to a single unsatisfied bond. This can also be seen using the bond-frustration measure defined in Eq.~\eqref{eq:bond_frustration}, where each bond has
\begin{equation}
    f_{ij} = \frac{1 + \cos(\phi_i-\phi_j)}{2} = \frac{1}{4},
\end{equation}
meaning each bond is equally frustrated. More generally, on the triangle, the mean bond frustration is directly related to the energy by Eq.~\eqref{eq:fbar_energy_relation_general}. Explicitly, since
\begin{equation}
    \bar f = \frac{1}{3}\sum_{i<j} f_{ij} = \frac{1}{3}\sum_{i<j}\frac{1+\cos(\phi_i-\phi_j)}{2},
\end{equation}
we have
\begin{equation}
    \bar f = \frac{1}{2} + \frac{1}{6}\sum_{i<j}\cos(\phi_i-\phi_j) = \frac{1}{2} + \frac{E}{6K},
    \label{eq:triangle_fbar_energy_relation}
\end{equation}
so the ground-state value $E_{\min}=-3K/2$ is equivalently $\bar f_{\min}=1/4$.

Because the energy depends only on phase differences, a uniform rotation $\phi_i\mapsto\phi_i + \Phi$ leaves the energy unchanged, meaning the absolute orientation of the phase pattern is arbitrary. Therefore, the physically meaningful information is contained in the relative phases, while the common phase $\Phi$ labels a continuous global $U(1)$ symmetry.

Before fixing the global phase, the full set of ground states can be written as $\G_\triangle = \G_+ \cup \G_-$, where
\begin{equation}
    \G_\sigma = \qty{\qty(\Phi + \sigma\frac{2\pi}{3},\Phi - \sigma\frac{2\pi}{3},\Phi)\bmod{2\pi}\,\Big|\,\Phi\in[0,2\pi)}
    \label{eq:trianglar_ground_states}
\end{equation}
and $\sigma = \pm 1$. Each branch $\G_\sigma$ is a continuous circle generated by the global phase rotation, and the two branches are related by reversing the cyclic ordering of the phases. Hence, the triangular ground-state manifold has the structure
\begin{equation}
    \G_\triangle\cong U(1)\times\mathbb{Z}_2,
\end{equation}
where the $U(1)$ factor is the continuous global phase degeneracy and the $\mathbb{Z}_2$ factor is the discrete chirality degeneracy. This is the phase-oscillator version of the degeneracy of the antiferromagnetic $XY$ model on a triangular plaquette.

\subsubsection{Energy landscape and phase-space geometry}

Since the global phase is dynamically neutral, it is useful to pass to relative coordinates:
\begin{equation}
    x = \phi_1-\phi_3,\quad y = \phi_2-\phi_3.
\end{equation}
This choice removes the continuous $U(1)$ freedom by measuring all phases relative to oscillator 3. In these coordinates, the energy becomes
\begin{equation}
    E(x,y) = K[\cos(x-y) + \cos x + \cos y].
    \label{eq:triangular_reduced_landscape}
\end{equation}
The two branches in Eq.~\eqref{eq:trianglar_ground_states} then reduce to the two isolated points
\begin{equation}
    \G_\triangle/U(1) = \qty{\qty(\frac{2\pi}{3},\frac{4\pi}{3}),\qty(\frac{4\pi}{3},\frac{2\pi}{3})},
    \label{eq:triangular_two_minima}
\end{equation}
so the two continuous families of energetically equivalent ground states in the full phase space appear as two degenerate minima in the reduced energy landscape.

\begin{figure*}[tbp!]
    \centering
    \includegraphics[width=0.95\linewidth]{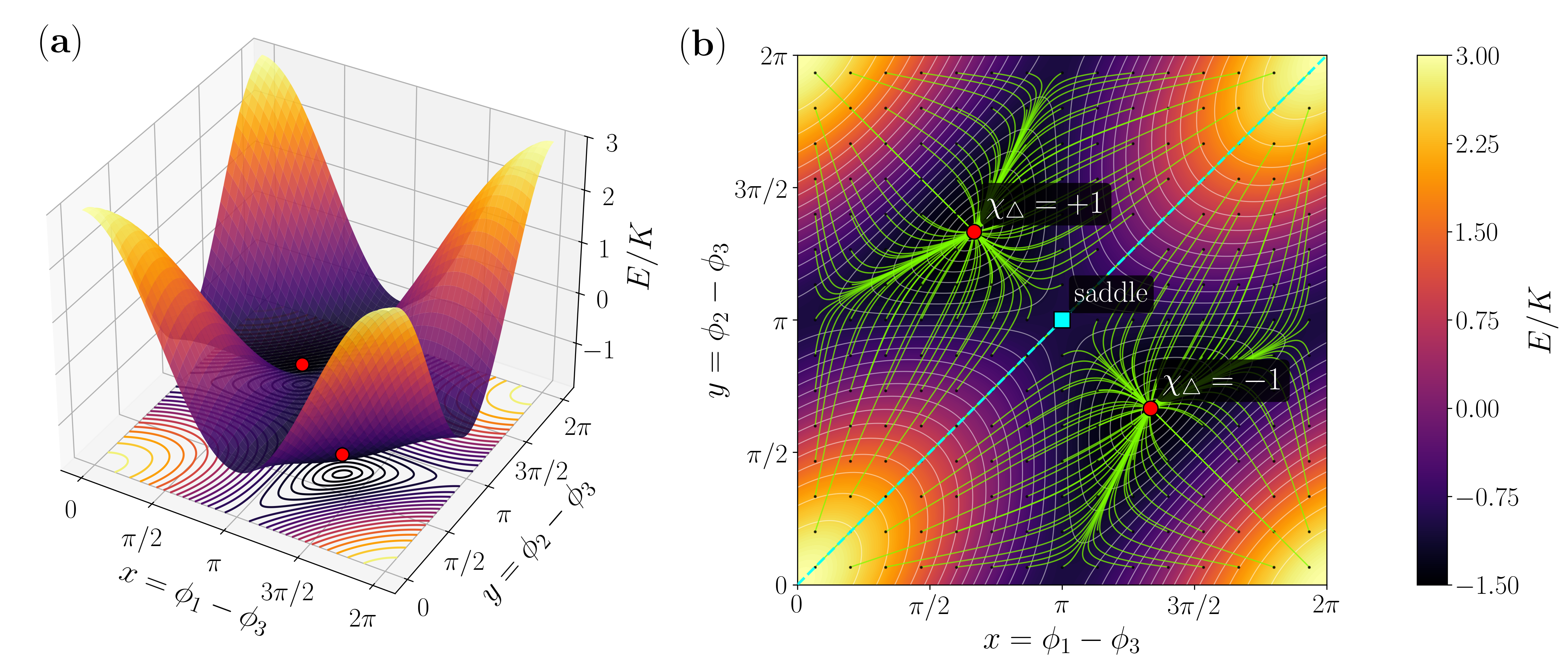}
    \caption{Reduced energy landscape and gradient-flow dynamics of the frustrated triangular motif. The global phase degree of freedom is removed by working in the relative rotating-frame coordinates $(x,y) = (\phi_1-\phi_3,\phi_2-\phi_3)$. (a)~Three-dimensional surface plot of the energy $E/K = \cos(\phi_1-\phi_2)+\cos(\phi_2-\phi_3)+\cos(\phi_3-\phi_1)$ expressed on the relative-phase plane. The two red points mark the degenerate $120^\circ$ minima, corresponding to the two possible chiral arrangements of the phases. (b)~Contour plot of the same landscape with white energy contours and green gradient-flow trajectories from many initial conditions. The flow relaxes toward one of the two chiral minima, labeled by $\chi_\triangle=\pm1$. The cyan square marks the saddle at $(\pi,\pi)$, while the dashed cyan line marks its stable manifold $x=y$. Because the plotted curves are the rotating-frame dynamics expressed in relative coordinates, they move downhill in the energy landscape but need not appear perpendicular to the contour lines.
}
    \label{fig:triangular_landscape}
\end{figure*}

To distinguish the two reduced minima, we recall the triangular chirality $\chi_\triangle$ introduced in Sec.~\ref{sec:minimal_phase_theory}. For planar phase vectors $\mb{S}_i = (\cos\phi_i,\sin\phi_i,0)$, the vector chirality points perpendicular to the phase plane. We use its signed out-of-plane component
\begin{equation}
    \chi_\triangle = \frac{2}{3\sqrt{3}}[\sin(\phi_2-\phi_1) + \sin(\phi_3-\phi_2) + \sin(\phi_1-\phi_3)],
\end{equation}
which measures the handedness of the phase winding around the triangle. For the ground states written in Eq.~\eqref{eq:trianglar_ground_states}, we have $\chi_\triangle = \sigma$ for $(\phi_1,\phi_2,\phi_3)\in\G_\sigma$. Therefore, the two ground-state branches have opposite chirality but identical energy $E[\G_+] = E[\G_-] = -3K/2$.

In the reduced coordinates, the chirality takes the form
\begin{equation}
    \chi_\triangle(x,y) = \frac{2}{3\sqrt{3}}[\sin(y-x) - \sin y + \sin x],
\end{equation}
so the $(2\pi/3,4\pi/3)$ state has $\chi_\triangle = +1$, and the $(4\pi/3,2\pi/3)$ state has $\chi_\triangle = -1$. The labels $\chi_\triangle = \pm 1$ used in Fig.~\ref{fig:triangular_landscape} identify these two energetically equivalent chiral ground states. 

Figure~\ref{fig:triangular_landscape} shows the reduced energy landscape and its associated gradient-flow dynamics. In Fig.~\ref{fig:triangular_landscape}(a), the surface defined by Eq.~\eqref{eq:triangular_reduced_landscape} is plotted over one periodic unit cell in $(x,y)$. The two low-energy points are precisely the reduced ground states in Eq.~\eqref{eq:triangular_two_minima}, i.e., the two degenerate realizations of the same frustrated compromise distinguished by chirality. The fully in-phase configuration at $x=y=0$ is the high-symmetry maximum of the landscape, where all three repulsive bonds are maximally unsatisfied. By contrast, the two $120^\circ$ minima distribute the bond frustration equally among the three edges.

Figure~\ref{fig:triangular_landscape}(b) shows trajectories of the gradient flow projected onto the same reduced coordinates. Since the rotating-frame dynamics obey $\dot{\phi}_i = -\partial E/\partial\phi_i$, the motion is downhill in the full energy landscape $E(\phi_1,\phi_2,\phi_3)$ [Eq.~\eqref{eq:triangle_energy}]. After passing to the relative coordinates $x,y$, the projected dynamics are still downhill in the reduced energy landscape. However, the projection modifies the Euclidean gradient flow $-(\partial_xE,\partial_yE)$ by an induced metric factor, giving
\begin{equation}
    \mqty(\dot{x} \\ \dot{y}) = -\mqty(2 & 1 \\ 1 & 2)\nabla_{(x,y)}E.
\end{equation}
The dynamics in the reduced landscape $E(x,y)$ [Eq.~\eqref{eq:triangular_reduced_landscape}] are therefore a metric-weighted gradient flow. Consequently, the green trajectories in Fig.~\ref{fig:triangular_landscape}(b) do not appear exactly perpendicular to the white contour lines in the plotted $(x,y)$ coordinates, even though they monotonically decrease the same energy:
\begin{equation}
    \dv{E}{t} = -(\nabla_{(x,y)}E)^\intercal\mqty(2 & 1 \\ 1 & 2)\nabla_{(x,y)}E\leq 0.
\end{equation}
Generic initial conditions relax to one of the two $120^\circ$ minima, and the labels $\chi_\triangle = \pm1$ identify the chirality selected by states in each basin of attraction.

The point $(x,y)=(\pi,\pi)$ provides a useful comparison with the Ising antiferromagnet on a triangle. At this point, oscillators 1 and 2 are synchronized with each other and antiphase to oscillator 3. Thus, two repulsive bonds are satisfied, while the 1--2 bond is maximally unsatisfied:
\begin{equation}
    f_{12}=1,\qquad f_{13}=f_{23}=0,\qquad \bar f=\frac{1}{3}.
\end{equation}
Equivalently,
\begin{equation}
    E(\pi,\pi)=K[\cos0+\cos\pi+\cos\pi]=-K.
\end{equation}
This is the direct analog of the antiferromagnetic Ising compromise on a triangle (see Fig.~\ref{fig:ising_triangle}). In the continuous phase model, however, this collinear compromise is not a ground state. Since $-K>-3K/2$, the system lowers its energy by rotating into one of the two $120^\circ$ configurations, where $f_{ij}=1/4$ for every bond and $\bar f=1/4$. Therefore, $(\pi,\pi)$ is a saddle point of the reduced landscape. The diagonal $x=y$ is its stable manifold, while arbitrarily small transverse perturbations select one of the two chiral $120^\circ$ minima.

Using the basin-probability language of Sec.~\ref{sec:minimal_phase_theory}, the same reduced landscape gives the simplest basin picture. Since the only stable reduced fixed points are the two chiral $120^\circ$ minima, a generic initial condition relaxes to either the $\chi_\triangle=+1$ or $\chi_\triangle=-1$ state. By symmetry, the two basin probabilities are equal,
\begin{equation}
    p_{\chi_\triangle=+1}=p_{\chi_\triangle=-1}=\frac{1}{2},
\end{equation}
for uniformly sampled initial phases. The basin boundary is the stable manifold of the saddle at $(x,y)=(\pi,\pi)$, given by the diagonal $x=y$. The triangle therefore provides the simplest example of frustration-driven selection. From a continuum of initial phases, the dynamics select one of two energetically equivalent chiral ground states.

\begin{figure*}
    \centering
    \includegraphics[width=\linewidth]{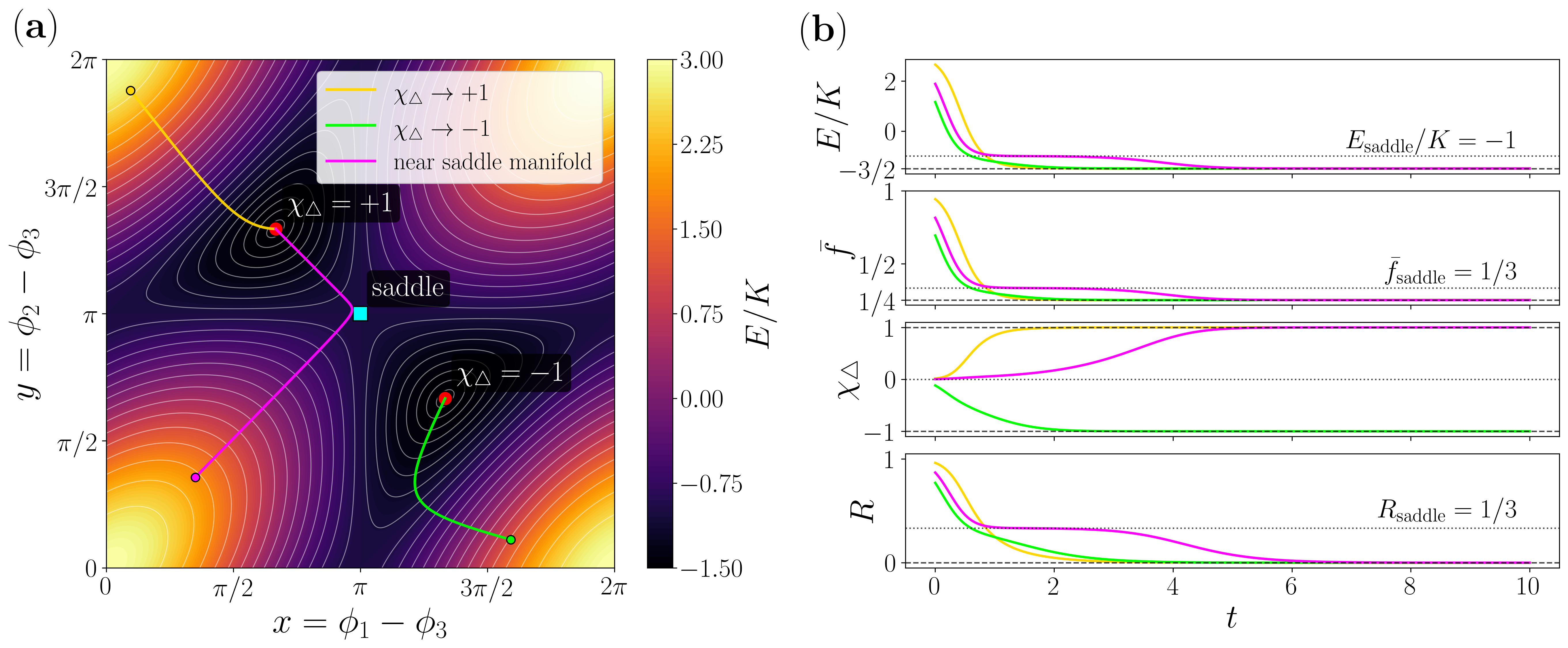}
    \caption{Representative reduced-phase trajectories and associated diagnostics for the frustrated triangular motif. (a)~Three representative trajectories in the reduced coordinates $x=\phi_1-\phi_3$ and $y=\phi_2-\phi_3$, plotted on top of the reduced energy landscape for reference. The background color denotes $E(x,y)/K$, and the white curves are energy contours. Red circles mark the two degenerate $120^\circ$ minima with opposite triangular chirality $\chi_\triangle=\pm1$, and the cyan square marks the collinear saddle at $(x,y)=(\pi,\pi)$. The yellow trajectory reaches the $\chi_\triangle=+1$ minimum, the green trajectory reaches the $\chi_\triangle=-1$ minimum, and the magenta trajectory is initialized near the stable manifold of the saddle. (b)~Time evolution of diagnostic quantities for the same three trajectories, with colors matching (a). From top to bottom, the plotted quantities are the energy $E/K$, mean bond frustration $\bar f$, triangular chirality $\chi_\triangle$, and Kuramoto order parameter $R$. Horizontal reference lines indicate the minimum values $E_{\min}/K=-3/2$, $\bar f_{\min}=1/4$, and $R_{\min} = 0$, the saddle values $E_{\rm saddle}/K=-1$, $\bar f_{\rm saddle}=1/3$, and $R_{\rm saddle}=1/3$, and the chirality values $\chi_\triangle=\pm1$.}
    \label{fig:triangle_representative_trajectories}
\end{figure*}

\subsubsection{Dynamical content and diagnostics}

The dynamical content of this basin-selection picture is illustrated in Fig.~\ref{fig:triangle_representative_trajectories}. In Fig.~\ref{fig:triangle_representative_trajectories}(a), we use the reduced energy surface from Fig.~\ref{fig:triangular_landscape} as a background for representative trajectories. The yellow and green trajectories begin on opposite sides of the saddle separatrix and relax to opposite chiral minima, while the magenta trajectory is initialized close to the stable manifold of the saddle point at $(x,y)=(\pi,\pi)$. Together, the three trajectories visualize the two generic outcomes of the frustrated triangle, as well as the special role played by the collinear saddle in separating them.

The magenta trajectory shows the effect of starting close to the basin boundary. Since it lies near the stable manifold of the saddle, it first approaches the Ising-like saddle state at $(x,y)=(\pi,\pi)$. To understand the behavior of the system during this transient, write
\begin{equation}
    x=\pi+u,\quad y=\pi+v,
    \label{eq:uv_def}
\end{equation}
and introduce coordinates parallel and transverse to the diagonal stable manifold,
\begin{equation}
    s=\frac{u+v}{2},\quad q=\frac{v-u}{2}.
    \label{eq:sq_def}
\end{equation}
Here, $q=0$ is the diagonal $x=y$, and $q$ measures displacement away from the basin boundary. Expanding the reduced energy near the saddle gives
\begin{equation}
    \frac{E}{K} = \cos(x-y)+\cos x+\cos y \simeq -1+s^2-q^2,
    \label{eq:energy_sq}
\end{equation}
so the energy differs from the saddle value only at quadratic order in the distance from $(\pi,\pi)$. By Eq.~\eqref{eq:triangle_fbar_energy_relation}, we have
\begin{equation}
    \bar f = \frac{1}{2}+\frac{E}{6K}
\end{equation}
for the triangle, so the mean bond frustration is also only quadratically displaced from its saddle value $\bar f_{\rm saddle}=1/3$. By contrast, the chirality is linear in the transverse coordinate:
\begin{equation}
    \chi_\triangle = \frac{2}{3\sqrt{3}}[\sin(y-x)-\sin y+\sin x] \simeq \frac{8}{3\sqrt{3}}q.
\end{equation}
Therefore, as the magenta trajectory begins to peel away from the stable manifold, Fig.~\ref{fig:triangle_representative_trajectories}(b) shows that $\chi_\triangle$ grows visibly while $E$ and $\bar f$ remain close to the saddle values $E_{\rm saddle}=-K$ and $\bar f_{\rm saddle}=1/3$. The chirality is therefore the more sensitive diagnostic of the unstable direction that determines the eventual basin choice. Eventually, this transverse displacement becomes large, and the trajectory relaxes to one of the two chiral $120^\circ$ minima.

The bottom panel of Fig.~\ref{fig:triangle_representative_trajectories}(b) shows the Kuramoto order parameter
\begin{equation}
    R = \left|\frac{1}{3}\sum_{j=1}^{3} e^{i\phi_j}\right|
\end{equation}
introduced in Sec.~\ref{sec:minimal_phase_theory}, which measures global phase coherence. For the triangular motif, $R$ is directly related to the energy. Since
\begin{equation}
    \left|\sum_{j=1}^{3}e^{i\phi_j}\right|^2 = 3+2\sum_{i<j}\cos(\phi_i-\phi_j),
\end{equation}
we have
\begin{equation}
    R^2 = \frac{1}{9}\left(3+\frac{2E}{K}\right).
\end{equation}
Therefore, the $120^\circ$ minima have $R_{\min}=0$ since $E/K=-3/2$, while the Ising-like saddle has
\begin{equation}
    R_{\rm saddle} = \sqrt{\frac{1}{9}(3-2)} = \frac{1}{3}.
\end{equation}
Then, using the local coordinates $s,q$ near the saddle, it follows from Eq.~\eqref{eq:energy_sq} that
\begin{equation}
    R \simeq \frac{1}{3}(1 + s^2 - q^2).
\end{equation}
Hence, $R$, like $E$ and $\bar f$, changes only quadratically near the saddle. This explains why the magenta trajectory shows a visible increase in $\chi_\triangle$ while $R$ remains close to the saddle value $1/3$. Only after the trajectory moves farther along the unstable direction does the global coherence decrease toward the $120^\circ$ value $R=0$.

The triangular motif therefore contains, in its simplest possible form, the mechanism by which repulsive phase interactions generate geometrical frustration. Unlike the two-oscillator case, the preferred antiphase relation cannot be imposed on every bond simultaneously. The system resolves this incompatibility by forming a $120^\circ$ phase pattern, distributing the residual frustration uniformly across the three bonds rather than localizing it on a single edge. After quotienting out the neutral global phase, the $U(1)\times\mathbb{Z}_2$ ground-state manifold reduces to two isolated minima distinguished by their chirality. The collinear Ising-like state at $(\phi_1-\phi_3,\phi_2-\phi_3)=(\pi,\pi)$ appears as a saddle whose stable manifold separates the two chiral basins. Importantly, the frustrated triangle suppresses global synchrony while leaving a discrete dynamical choice between two degenerate chiral phase-locked states. In the larger frustrated network considered in Subsec.~\ref{subsec:kagome}, the same local incompatibility is repeated across many motifs, producing a richer ground-state structure and more intricate phase-space geometry. The triangular plaquette is therefore the minimal setting in which the link between geometrical frustration and dynamical phase selection is fully visible.

\subsection{Tetrahedral motif}

We next consider the tetrahedral motif, whose interaction graph is the complete graph $K_4$. The tetrahedron contains four mutually coupled phase oscillators, so the adjacency matrix is
\begin{equation}
    A_{\rm tet} = \mqty(0 & 1 & 1 & 1 \\ 1 & 0 & 1 & 1 \\ 1 & 1 & 0 & 1 \\ 1 & 1 & 1 & 0).
\end{equation}
Then, in the rotating frame, the equations of motion are
\begin{equation}
    \dot{\phi}_i = -K\sum_{j\neq i}\sin(\phi_j-\phi_i) = -\pdv{E}{\phi_i},
\end{equation}
where the tetrahedral energy is
\begin{equation}
    E = K\sum_{1\leq i<j\leq 4}\cos(\phi_i-\phi_j).
\end{equation}
Using the planar-spin representation $\mb{S}_i = (\cos\phi_i,\sin\phi_i)$, we again obtain that the minimum-energy configurations satisfy
\begin{equation}
    \sum_{i=1}^4\mb{S}_i = 0
    \label{eq:tet_zero_sum_condition}
\end{equation}
with minimum energy
\begin{equation}
    E_{\rm min} = -2K.
\end{equation}
Since the energy and dynamics are invariant under a global phase shift $\phi_i\mapsto\phi_i+\Phi$, we can once again introduce reduced coordinates
\begin{equation}
    x = \phi_1-\phi_4,\quad y = \phi_2-\phi_4,\quad z = \phi_3-\phi_4,
    \label{eq:tet_reduced_coords}
\end{equation}
in which the energy takes the form
\begin{equation}
\begin{split}
    E(x,y,z) &= K[\cos(x-y)+\cos(x-z) \\ &\quad +\cos(y-z) +\cos x+\cos y+\cos z].
    \label{eq:tet_energy_landscape}
\end{split}
\end{equation}
Thus, after removing the global phase, the reduced tetrahedral energy landscape lives on a three-dimensional torus $(x,y,z)\in T^3$. This description will be used below to visualize the ground-state manifold, energy landscape, and basin structure of the tetrahedral motif.

\begin{figure*}[tbp!]
    \centering
    \includegraphics[width=\linewidth]{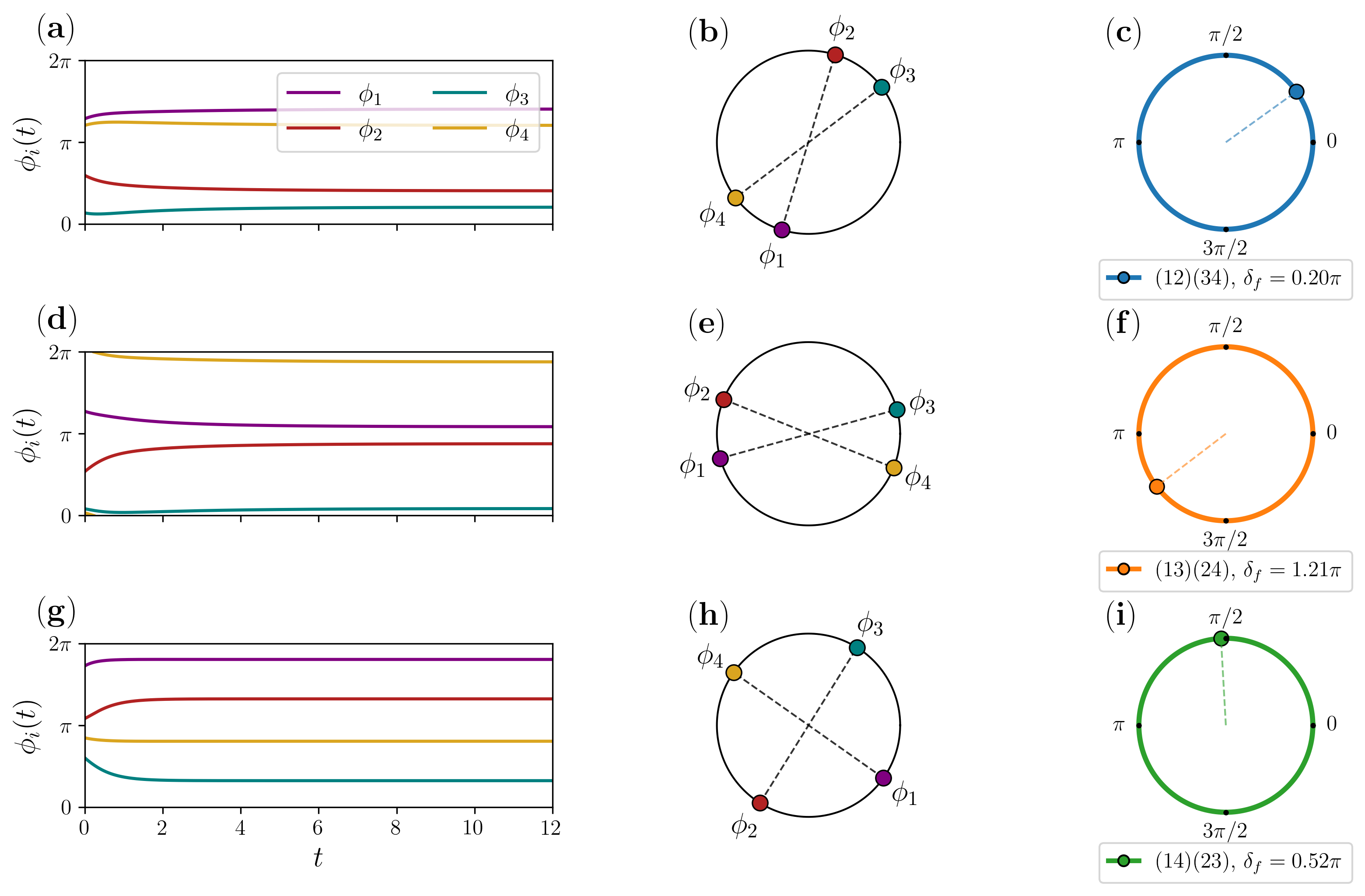}
    \caption{Representative trajectories for the tetrahedral motif, where each row shows relaxation from a different nonsymmetric initial condition. (a), (d), and (g) show the phase variables $\phi_i(t)$; (b), (e), and (h) show the corresponding final configurations on the unit circle, with dashed chords marking the dynamically selected antiphase pairs. The three rows end on the three possible pairing branches: $(12)(34)$, $(13)(24)$, and $(14)(23)$, respectively. (c), (f), and (i) show the corresponding locations on the $S^1$ branches of the reduced ground-state manifold, with $\delta_f$ denoting the final relative angle between the two antiphase pairs.}
    \label{fig:tetrahedron_trajectories}
\end{figure*}

\subsubsection{Ground-state manifold structure}

Equation~\eqref{eq:tet_zero_sum_condition} gives the tetrahedral ground-state condition $\sum_{i=1}^4\mb{S}_i = 0$, which forces the four planar unit vectors to form two antipodal pairs. To see this, define $\mb{A} = \mb{S}_1 + \mb{S}_2$. Then, Eq.~\eqref{eq:tet_zero_sum_condition} implies $\mb{S}_3 + \mb{S}_4 = -\mb{A}$. If $\mb{A}=0$, then $\mb{S}_2 = -\mb{S}_1$ and $\mb{S}_4 = -\mb{S}_3$, so the two antipodal pairs are $(\mb{S}_1,\mb{S}_2)$ and $(\mb{S}_3,\mb{S}_4)$. We denote this antipodal pairing by $(12)(34)$. If $\mb{A}\neq 0$, then the set of unit vectors $\{\mb{S}_1,\mb{S}_2\}$ is symmetric about the direction of $\mb{A}$. The set of unit vectors $\{\mb{S}_3,\mb{S}_4\}$ with sum $-\mb{A}$ can then be obtained by rotating this pair by $\pi$. If $\mb{S}_1$ rotates into $\mb{S}_3$, then the antipodal pairing is $(13)(24)$. Otherwise, $\mb{S}_1$ rotates into $\mb{S}_4$, so the antipodal pairing is $(14)(23)$. Therefore, every tetrahedral ground state consists of two opposite-phase pairs.

We have also shown how the four oscillators can be paired antipodally, namely, that there are three possible perfect matchings of four labeled sites:
\begin{equation}
    \M_4 = \{(12)(34),(13)(24),(14)(23)\}.
\end{equation}
We note that $\M_4$ is just a three-element set of pairing sectors, not a permutation group. For each pairing sector, the ground states have one neutral global phase and one internal relative-pair angle. We choose a convention in which $\phi_4 = \Phi$ and the vertex paired with vertex 4 has phase $\Phi+\pi$. Then, the three sectors can be parameterized as
\begin{equation}
\begin{split}
    (12)(34)&:\: (\phi_1,\phi_2,\phi_3,\phi_4) = (\Phi + \delta,\Phi+\delta+\pi,\Phi+\pi,\Phi), \\
    (13)(24)&:\: (\phi_1,\phi_2,\phi_3,\phi_4) = (\Phi + \delta,\Phi+\pi,\Phi+\delta+\pi,\Phi), \\
    (14)(23)&:\: (\phi_1,\phi_2,\phi_3,\phi_4) = (\Phi + \pi,\Phi+\delta,\Phi+\delta+\pi,\Phi),
\end{split}
\label{eq:tet_gs_sector_parameterization}
\end{equation}
where all phases are understood modulo $2\pi$. The angle $\Phi\in [0,2\pi)$ parameterizes the global $U(1)$ phase associated with the continuous symmetry of the energy, while $\delta\in[0,2\pi)$ parametrizes the remaining $S^1$ degree of freedom corresponding to the relative orientation of the two antipodal pairs. Although $U(1)\cong S^1$, it is useful to keep the notation distinct, as $\Phi$ is a symmetry-protected flat direction associated with uniform rotation of all phases, while $\delta$ is an internal flat direction that changes the relative timing pattern within the ground-state manifold.

For a fixed perfect matching $P\in\M_4$, the corresponding ground-state sector has the generic form
\begin{equation}
    \G_P\cong U(1)\times S^1.
\end{equation}
The full tetrahedral manifold is therefore the union of the three pairing sectors:
\begin{equation}
    \G_{\rm tet} = \bigcup_{P\in\M_4}\G_P.
\end{equation}
Schematically, we may write
\begin{equation}
    \G_{\rm tet}\simeq U(1)\times S^1\times\M_4,
\end{equation}
but this product form should not be interpreted as a completely disjoint decomposition since the pairing sectors meet at special collinear configurations where two different perfect matchings describe the same physical phase pattern. We observe that the tetrahedral motif has a continuous ground-state degeneracy, in contrast with the triangular motif, where quotienting by the global $U(1)$ phase leaves only two isolated chiral minima.

\begin{figure}[tbp!]
    \centering
    \includegraphics[width=\linewidth]{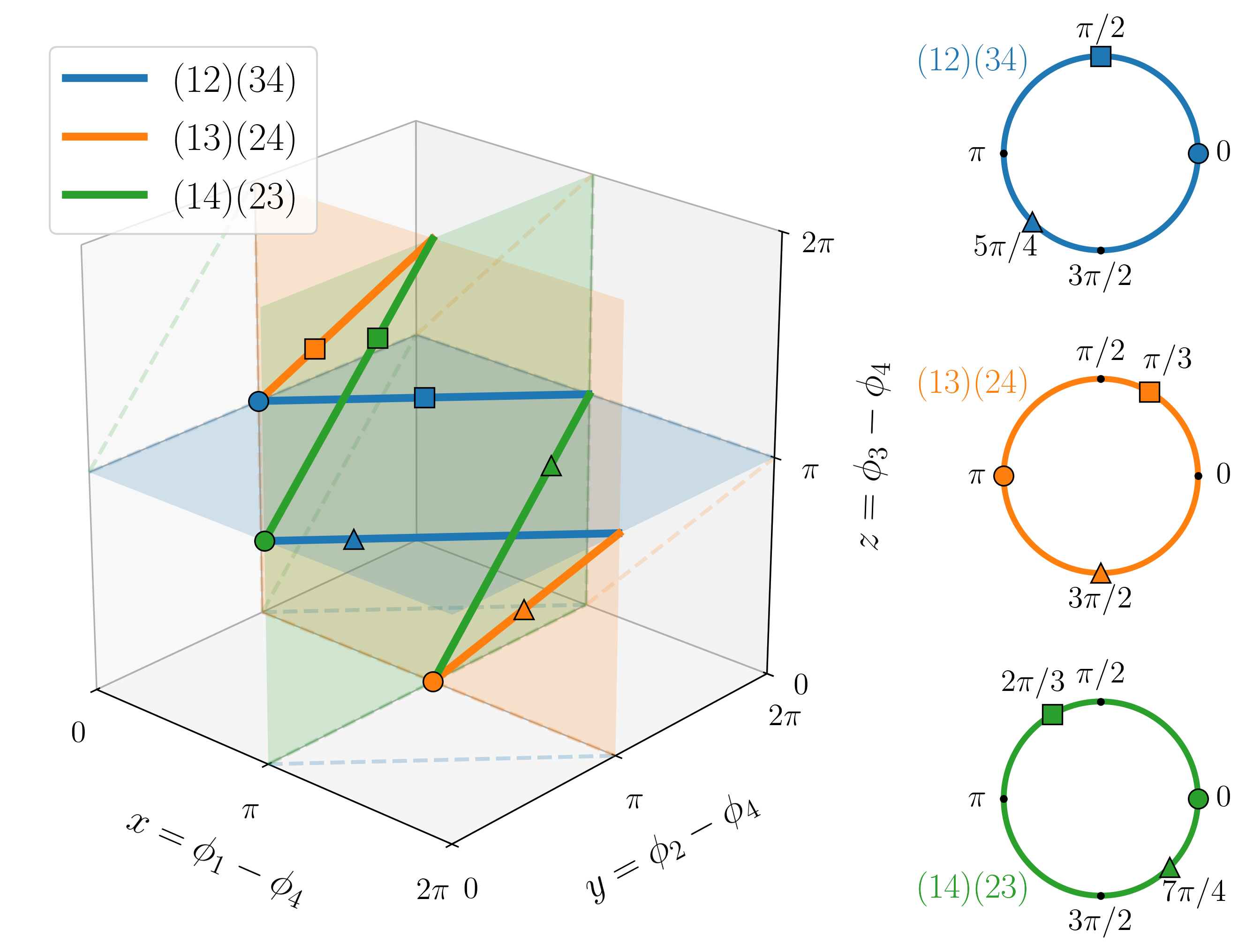}
    \caption{Reduced ground-state manifold of the tetrahedral motif. The left plot shows the three ground-state branches in the reduced phase space $(x,y,z) = (\phi_1-\phi_4,\phi_2-\phi_4,\phi_3-\phi_4)$, displayed in the fundamental domain $[0,2\pi)^3$. Each colored curve corresponds to one perfect matching of the four oscillators into two antiphase pairs: $(12)(34)$, $(13)(24)$, and $(14)(23)$. Because the reduced phase space is a three-torus $T^3$, each branch is a closed $S^1$ even though they appear as line segments in the fundamental cube; dashed projections indicate the periodic identifications across the cube boundaries. The translucent planes show the phase constraints associated with each pairing branch. On the right, the same three branches are shown as circles parameterized by the internal angle $\delta$ between the two antiphase pairs. The marked points illustrate corresponding locations on the three-dimensional representation and the associated $S^1$ branch.}
    \label{fig:tetrahedron_ground_state_manifold}
\end{figure}

The parameterizations in Eq.~\eqref{eq:tet_gs_sector_parameterization} also clarify what should be expected dynamically. In the triangular motif, a generic trajectory relaxes to a unique isolated phase-locked state. On the contrary, in the tetrahedral motif, a generic trajectory relaxes onto one of the three pairing sectors and retains a final value of the internal angle $\delta$. This behavior is illustrated in Fig.~\ref{fig:tetrahedron_trajectories}. The left column shows representative phase trajectories $\phi_i(t)$ from three nonsymmetric initial conditions. In each case, the phases approach a configuration consisting of two pairs, with the phases within each pair separated by $\pi$. The middle column shows the same final configurations as points on the unit circle, with the dashed chords making the antipodal-pair structure explicit. The three rows of Fig.~\ref{fig:tetrahedron_trajectories} were chosen to end on the three perfect possible matchings $(12)(34)$, $(13)(24)$, and $(14)(23)$, resulting in a dynamical realization of the decomposition $\G_{\rm tet} = \bigcup_{P\in\M_4}\G_P$. The right column then removes the global phase and records the final angle $\delta_f$ on the corresponding $S^1$ branch.

To describe this manifold after removing the global $U(1)$ phase, we use the reduced coordinates $x,y,z$ defined in Eq.~\eqref{eq:tet_reduced_coords}. In these coordinates, the three pairing sectors become
\begin{equation}
\begin{split}
    (12)(34):\:(x,y,z) &= (\delta,\delta+\pi,\pi), \\
    (13)(24):\:(x,y,z) &= (\delta,\pi,\delta+\pi), \\
    (14)(23):\:(x,y,z) &= (\pi,\delta,\delta+\pi),
\end{split}
\end{equation}
again with all coordinates understood modulo $2\pi$. Each of these pairing sectors is a closed circle $\G_P/U(1)\cong S^1_P$ in the reduced-coordinate three-torus $T^3$. Therefore, the reduced tetrahedral ground-state manifold is
\begin{equation}
    \G_{\rm tet}/U(1) = \bigcup_{P\in\M_4}S^1_P,
\end{equation}
or schematically,
\begin{equation}
    \G_{\rm tet}/U(1)\simeq S^1\times\M_4,
\end{equation}
with the same caveat that the three $S^1$ branches intersect at special collinear states. For example, the branches $(12)(34)$ and $(13)(24)$ meet at $(x,y,z)=(0,\pi,\pi)$. These intersection points will be discussed in more detail later in this subsection.

Figure~\ref{fig:tetrahedron_ground_state_manifold} shows these three branches inside the cut-open fundamental domain $[0,2\pi)^3$. The plotted line segments should therefore be interpreted as pieces of the closed $S^1$ branches whose endpoints are identified across opposite faces of the cube, shown by the dashed projections. The three colored branches in Fig.~\ref{fig:tetrahedron_ground_state_manifold} correspond directly to the three perfect matchings in $\M_4$, and the translucent planes indicate the simple phase constraints defining each branch. For example, the $(12)(34)$ branch satisfies
\begin{equation}
    z=\pi,\quad y=x+\pi\pmod{2\pi},
    \label{eq:tet_(12)(34)}
\end{equation}
which is the reduced coordinate form of $\phi_2 = \phi_1+\pi$ and $\phi_3 = \phi_4 + \pi$. The circular plots on the right of Fig.~\ref{fig:tetrahedron_ground_state_manifold} provide a complementary view of the same geometry, where each branch is drawn as its own $S^1$ parameterized by the internal angle $\delta$. The marked points show how selected locations in the reduced three-dimensional representation correspond to points on the appropriate $S^1$ branch. 

\begin{figure*}[tbp!]
    \centering
    \includegraphics[width=\linewidth]{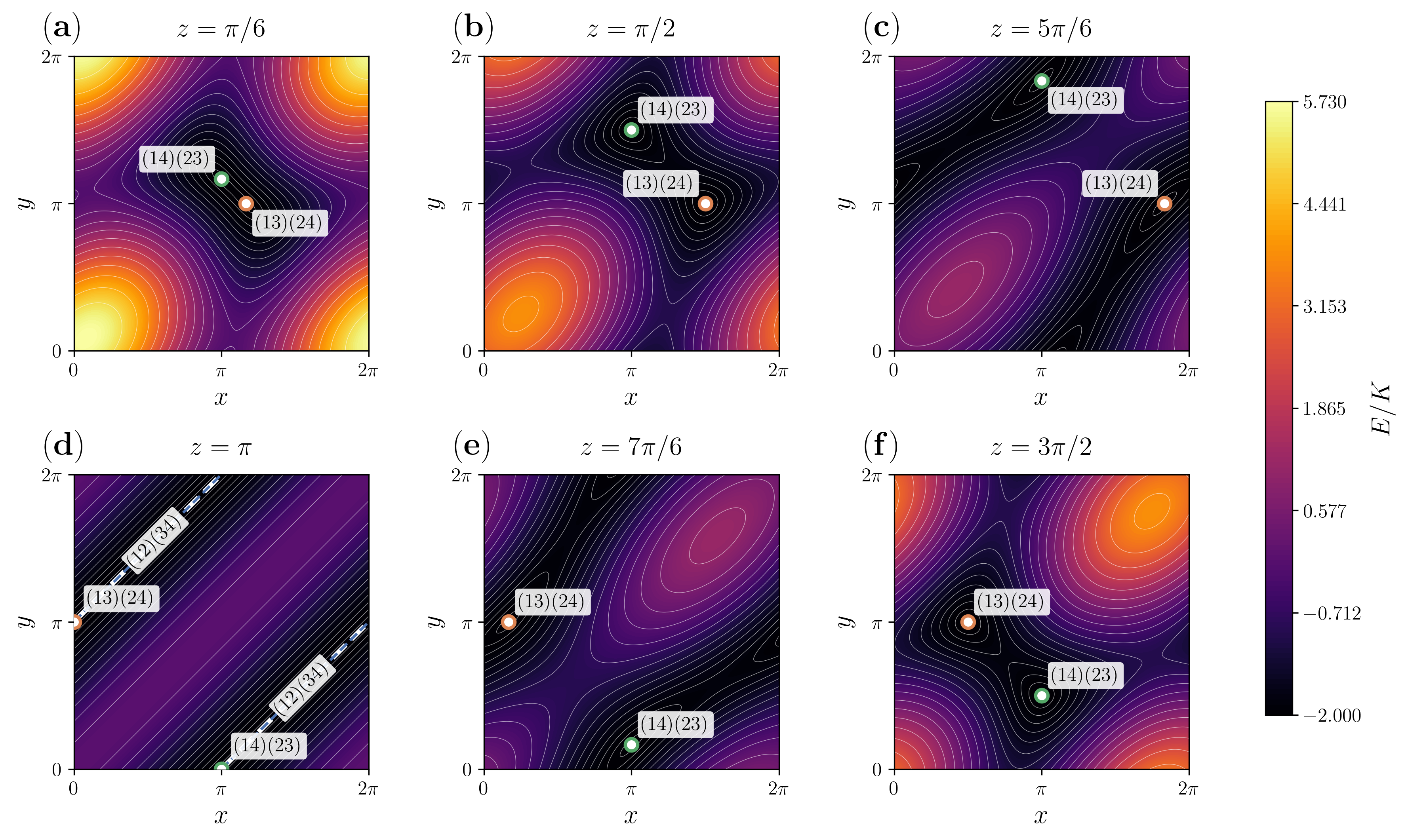}
    \caption{Energy landscape of the tetrahedral motif in reduced coordinates $(x,y,z)=(\phi_1-\phi_4,\phi_2-\phi_4,\phi_3-\phi_4)$. Each panel shows a two-dimensional slice of the reduced energy $E(x,y,z)/K$ at fixed $z$, with color indicating the energy and white contours showing lines of constant energy. The marked points indicate where the reduced ground-state branches intersect the corresponding slice, labeled by their antiphase pairing sectors $(12)(34)$, $(13)(24)$, and $(14)(23)$. In (d), the slice $z=\pi$ contains the full $(12)(34)$ ground-state $S^1$ branch, shown as a dashed line. The remaining slices cut the ground-state manifold at isolated points.}
    \label{fig:tetrahedron_energy_landscape}
\end{figure*}

\subsubsection{Energy landscape and phase-space geometry}

Having identified the reduced ground-state manifold geometrically, we now examine how these branches appear as minima of the full tetrahedral energy landscape. The reduced energy landscape $E(x,y,z)$ given in Eq.~\eqref{eq:tet_energy_landscape} is a function on the three-torus $(x,y,z)\in T^3$, so it is most naturally visualized through two-dimensional slices at fixed values of one reduced coordinate. On such a slice,
\begin{equation}
\begin{split}
    E(x,y&;z_0) = K[\cos(x-y)+\cos(x-z_0) \\ &\quad +\cos(y-z_0) +\cos x+\cos y+\cos z_0].
\end{split}
\end{equation}
The plotted domain $(x,y)\in[0,2\pi)^2$ should again be understood as a cut-open representation of a torus, so points on opposite edges are periodically identified.

The location of the ground-state manifold within these slices follows directly from the branch equations derived above. For a generic slice $z=z_0$, the slice intersects the $(13)(24)$ and $(14)(23)$ branches at the isolated points
\begin{equation}
    (x,y)=(z_0-\pi,\pi),\quad (x,y)=(\pi,z_0-\pi)\pmod{2\pi},
    \label{eq:tet_generic_slice_intersections}
\end{equation}
respectively. These are true ground-state points of the full tetrahedral landscape, with $E = E_{\rm min} = -2K$. By contrast, the $(12)(34)$ branch satisfies Eq.~\eqref{eq:tet_(12)(34)}, so it only appears in the special slice $z=\pi$. In that slice, the ground-state set is an entire one-dimensional valley:
\begin{equation}
    x-y = \pi\pmod{2\pi}.
\end{equation}
This difference between generic and special slices is clear visually. Most constant-$z$ cuts show isolated intersections with the ground-state manifold, while the slice $z=\pi$ contains one full reduced ground-state circle.

Figure~\ref{fig:tetrahedron_energy_landscape} shows several such slices. The dark regions mark low energy, while the lighter regions correspond to higher energy configurations in which the four phases are more nearly synchronized. The labeled markers indicate the intersections of the plotted slice with the reduced ground-state branches. For example, in the generic slices $z=\pi/6,\pi/2,5\pi/6,7\pi/6,3\pi/2$, the $(13)(24)$ and $(14)(23)$ branches appear as isolated minimum-energy points whose positions move through the $(x,y)$-plane according to Eq.~\eqref{eq:tet_generic_slice_intersections} as $z_0$ is varied. In the slice $z=\pi$, the full $(12)(34)$ branch lies in the plane and appears as a diagonal minimum-energy valley.

These slices make the relation between the energy landscape and ground-state manifold explicit. A two-dimensional slice usually cuts the three continuous $S^1$ branches of $\G_{\rm tet}/U(1)$, but when the slice is aligned with one of the branch constraints, it reveals an extended flat direction. The apparent point minima in most panels should therefore be interpreted as slice-dependent views of the continuous minimum-energy set in $T^3$, not isolated minima of the full reduced system.

\begin{figure*}[tbp!]
    \centering
    \includegraphics[width=\linewidth]{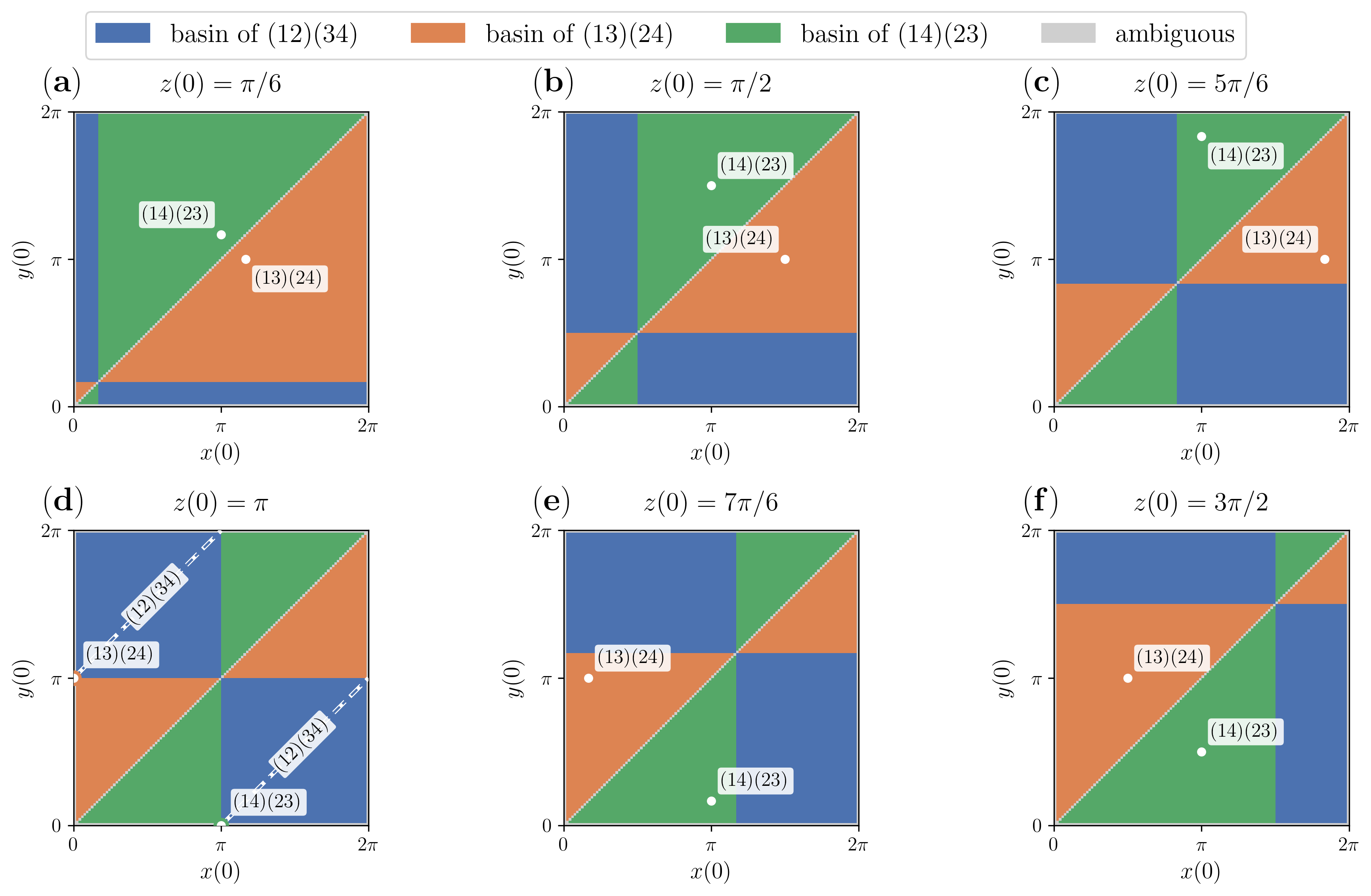}
    \caption{Basin structure of the tetrahedral motif in reduced coordinates $x,y,z$. Each slice fixes the initial value of one reduced coordinate, $z(0)$, and varies the initial condition over the $(x(0),y(0))$ plane. Colors indicate the pairing branch reached after relaxation: blue for $(12)(34)$, orange for $(13)(24)$, and green for $(14)(23)$. White markers indicate intersections of the corresponding ground-state branches with the plotted slice.}
    \label{fig:tetrahedron_basins}
\end{figure*}

The energy slices show where the ground-state branches lie, but they do not by themselves indicate which branch is selected dynamically from a given initial condition. To examine this branch-selection problem, we integrate the full reduced dynamics from many initial conditions and classify the final state by the antipodal pairing it approaches. In contrast to the energy-landscape panels above, the basin plots should be interpreted as slices through the space of initial conditions. Specifically, we choose to present $xy$-slices by fixing $z(0)=z_0$ and varying $(x(0),y(0))$, while allowing the subsequent trajectory to evolve in the full three-dimensional reduced phase space.

For classifying which branch a trajectory approaches, it is useful to define the pairing mismatch functions
\begin{equation}
\begin{split}
    D_{(12)(34)} &= |\mb{S}_1+\mb{S}_2|^2 + |\mb{S}_3+\mb{S}_4|^2, \\
    D_{(13)(24)} &= |\mb{S}_1+\mb{S}_3|^2 + |\mb{S}_2+\mb{S}_4|^2, \\
    D_{(14)(23)} &= |\mb{S}_1+\mb{S}_4|^2 + |\mb{S}_2+\mb{S}_3|^2.
\end{split}
\label{eq:mismatch}
\end{equation}
Each quantity measures the failure of a given perfect matching to form two antipodal pairs. At a generic point on the corresponding ground-state branch, exactly one of these mismatch functions vanishes. For example, $D_{(12)(34)}=0$ on a generic point of the $(12)(34)$ branch. The final pairing label of a trajectory can therefore be assigned by comparing the final state at $t_f$ to the three pairing manifolds and choosing the one with the smallest mismatch:
\begin{equation}
    P_f = \argmin_{P\in\mathcal M_4}D_P(t_f).
\end{equation}
Although this classification is unique for generic final states, two mismatch functions vanish simultaneously at the collinear intersection points of the ground-state branches. These Ising-like points are therefore intrinsically ambiguous from the standpoint of pairing labels, reflecting the fact that the same phase configuration admits more than one antipodal-pairing description.

Figure~\ref{fig:tetrahedron_basins} shows the resulting basin structure in the same fixed-$z(0)$ slices used to visualize the energy landscape. Each point in a panel is an initial condition, colored by the pairing branch reached after relaxation. The white markers indicate where the corresponding ground-state branches intersect the plotted initial-condition slice. Again, in generic slices, these intersections appear as isolated points, and in the special slice $z(0)=\pi$, the full $(12)(34)$ ground-state branch lies in the slice. 

The basin plots show that the local minima visible in an energy slice are only part of the dynamical story. Initial conditions need not all relax to the nearest visible ground-state intersection in that particular slice since the trajectory generally moves out of the initial slice as it descends through the full three-dimensional landscape. Basin membership is therefore organized by the geometry of the three ground-state branches and their stable manifolds in the full phase space. Regions of the initial-condition plane select different antiphase pairings, while the boundaries between colors mark basin boundaries. In the gradient-flow interpretation, these boundaries correspond to stable manifolds of unstable stationary sets separating the different branches.

\begin{figure*}[tbp!]
    \centering
    \includegraphics[width=\linewidth]{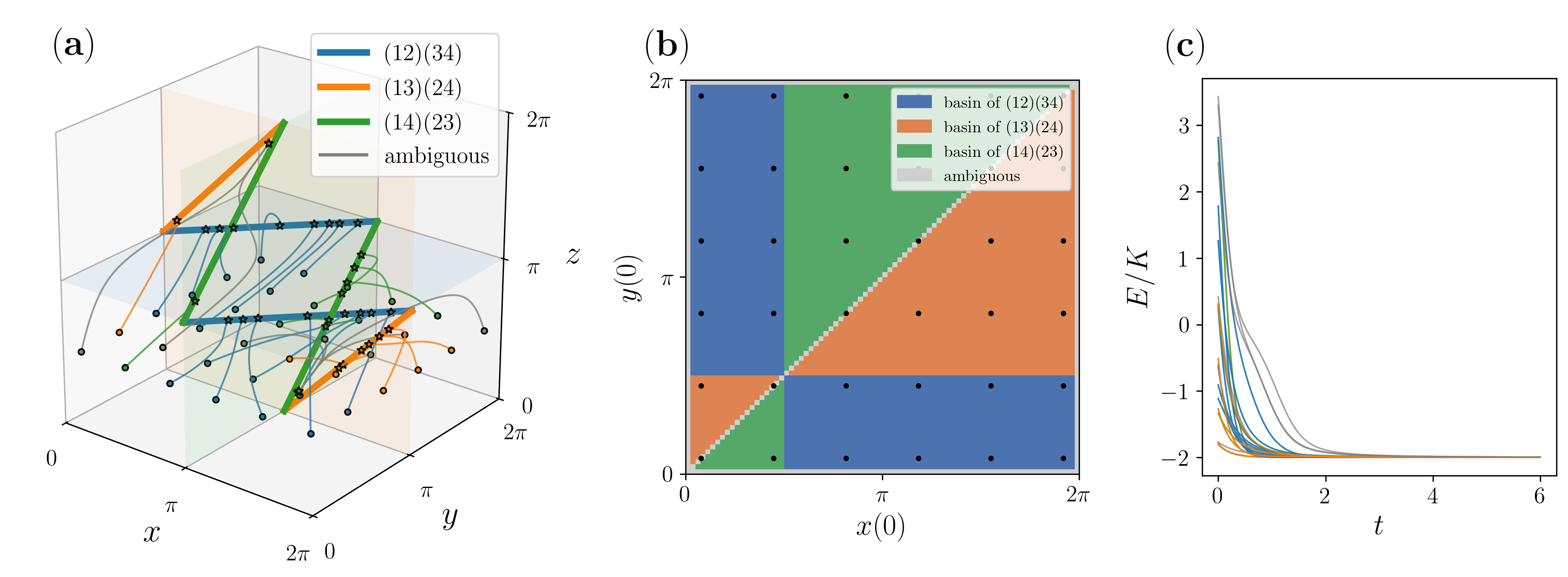}
    \caption{Gradient-flow dynamics for the tetrahedral motif in the reduced phase space. (a)~Representative trajectories in the three-dimensional reduced coordinates with the three reduced ground-state branches. Trajectories are colored by the pairing branch reached after relaxation, while gray trajectories indicate ambiguous cases near branch intersections or basin boundaries. (b)~The corresponding basin partition in the slice $z(0) = \pi/2$ used for the initial conditions, with black dots marking the initial conditions whose trajectories are shown in (a). (c)~The energy along the same trajectories. For all initial conditions, the energy decreases monotonically and approaches the ground-state value $E_{\rm min}=-2K$, confirming that the reduced dynamics relaxes onto $\mathcal G_{\rm tet}/U(1)$.}
    \label{fig:tetrahedron_gradient_flow}
\end{figure*}

To make this dynamical organization explicit, we next examine the reduced gradient flow itself. The basin slices give a static classification of initial conditions, whereas the gradient-flow trajectories show how those initial conditions descend through the reduced three-dimensional landscape and relax onto particular branches of $\mathcal G_{\rm tet}/U(1)$. As in the triangular motif, the phase equations are gradient flow equations for the energy in the unreduced phase variables: $\dot{\phi}_i = -\partial E/\partial\phi_i$. Passing to the relative coordinates, define $\mb{q} = (x,y,z)^\intercal$. After quotienting by the global phase, the reduced dynamics are again a metric-weighted gradient flow:
\begin{equation}
    \dot{\mb{q}} = -\mathsf{M}\nabla_{\mb{q}}E(\mb{q}),
    \label{eq:tet_gradient_flow}
\end{equation}
where
\begin{equation}
    \mathsf{M} = \mqty(2 & 1 & 1 \\ 1 & 2 & 1 \\ 1 & 1 & 2)
\end{equation}
and
\begin{equation}
    \frac{1}{K}\nabla_{(x,y,z)}E = \mqty(-\sin(x-y)-\sin(x-z)-\sin x \\ \sin(x-y)-\sin(y-z)-\sin y \\ \sin(x-z)+\sin(y-z)-\sin z).
\end{equation}
Similar to the triangular case, the reduced trajectories descend the energy monotonically, and the geometry of the flow is determined solely by the reduced energy landscape and the constant mobility matrix $\mathsf{M}$.

Figure~\ref{fig:tetrahedron_gradient_flow} illustrates this gradient-flow organization for initial conditions chosen from the slice $z(0)=\pi/2$. The basin plot in Fig.~\ref{fig:tetrahedron_gradient_flow}(b) identifies which pairing branch each initial condition ultimately selects, while the three-dimensional trajectories in Fig.~\ref{fig:tetrahedron_gradient_flow}(a) show how these initial conditions leave the slice and descend through the full reduced landscape. The same color convention is used in both panels, so the basin partition can be read directly as a map of branch selection under the reduced flow. The energy plots in Fig.~\ref{fig:tetrahedron_gradient_flow}(c) provide a useful check on the interpretation. For every trajectory shown, $E(t)$ decreases monotonically and approaches the ground-state value $E_{\min}=-2K$. The different colors therefore indicate $S^1_P$ branch selection within the single degenerate minimum set $\G_{\rm tet}/U(1)$, not relaxation to energetically distinct attractors.

The gradient-flow formulation also clarifies which stationary structures organize the basin boundaries. Since
\begin{equation}
    E = K\sum_{1\leq i<j\leq 4}\mb{S}_i\cdot\mb{S}_j = \frac{K}{2}\qty(\qty|\sum_{i=1}^4\mb{S}_i|^2 - 4),
\end{equation}
the stationary condition can be written directly in terms of the total spin
\begin{equation}
    \mb{S}_{\rm tot} = \sum_{i=1}^4\mb{S}_i.
\end{equation}
The derivative of the energy with respect to $\phi_i$ vanishes when
\begin{equation}
    \imaginary m\qty(e^{i\phi_i}\sum_{j=1}^4e^{-i\phi_j}) = 0.
\end{equation}
Equivalently, either the total spin $\mb{S}_{\rm tot}$ vanishes or every spin is parallel or antiparallel to $\mb{S}_{\rm tot}$. Therefore, the stationary sets fall into one of two classes: the zero-resultant ground-state manifold $\G_{\rm tet}$ discussed above or a finite set of Ising-like collinear non-ground stationary points.

In the reduced coordinates, the Ising-like stationary points are as follows. The fully synchronized state is
\begin{equation}
    (x,y,z) = (0,0,0),\quad E = 6K,
\end{equation}
which is the maximum of the energy landscape. The four ``three-against-one'' collinear states are
\begin{equation}
    (x,y,z) = (\pi,\pi,\pi),(\pi,0,0),(0,\pi,0),(0,0,\pi),\quad E = 0,
\end{equation}
which are saddle-type stationary points of the reduced energy landscape. Finally, the ``two-against-two'' collinear states are
\begin{equation}
    (x,y,z) = (0,\pi,\pi),(\pi,0,\pi),(\pi,\pi,0),\quad E = -2K,
\end{equation}
which are the aforementioned Ising-like ground states. Their special role is that they lie at intersections of the reduced ground-state branches, so the same phase configuration can be described by more than one antipodal pairing label.

The distinction between these stationary states is reflected in the local quadratic form of the reduced energy. Define the reduced Hessian by
\begin{equation}
    \mathsf{H}_{ab} = \pdv{E}{q_a}{q_b}.
\end{equation}
Since the reduced dynamics are a metric-weighted gradient flow $\dot{\mb{q}} = -\mathsf{M}\nabla_{\mb{q}}E$, the linearized dynamics near a stationary point are governed by
\begin{equation}
    \delta\dot{\mb{q}} = -\mathsf{M}\mathsf{H}\,\delta\mb{q}.
\end{equation}
Because the mobility matrix $\mathsf{M}$ is positive definite, the signs of the Hessian eigenvalues still determine the stable, unstable, and neutral directions of the flow.

At a generic point on a reduced ground-state branch, the Hessian has two positive directions and one zero direction. The positive directions attract trajectories transversely onto the branch, while the zero direction is tangent to the branch and corresponds to neutral motion in the internal angle $\delta$. At a collinear branch-intersection ground state, the quadratic form becomes more degenerate in the sense that, while there are still no negative directions, there is more than one zero direction since multiple ground-state branches meet.

The non-ground collinear stationary points have a different local structure. The fully synchronized point is a maximum of the reduced energy, while the three-against-one collinear states are saddle-type stationary points. For example, at $(x,y,z)=(\pi,\pi,\pi)$, the reduced Hessian has one positive and two negative directions. Therefore, this configuration has one stable direction and two unstable directions under the gradient flow. The other three-against-one states are related by permutations of the oscillator labels and play the same qualitative role, namely, they are unstable stationary configurations separating different downhill routes through the reduced landscape.

Because the ground states form intersecting continuous branches rather than isolated point attractors, the basin structure should be interpreted with some care. For a given pairing branch $S^1_P$, one may define the corresponding stable set
\begin{equation}
    W^s(S^1_P) = \qty{\mb{q}_0\in T^3\ \big|\ d\qty(\mb{q}(t;\mb{q}_0),S^1_P)\rightarrow 0\ \text{as}\ t\rightarrow\infty},
\end{equation}
where $d$ is the distance function and $\mb{q}(t;\mb{q}_0)$ is the trajectory that starts from the initial condition $\mb{q}_0$. The colored regions in the basin slices (Fig.~\ref{fig:tetrahedron_basins}) are two-dimensional cuts through these branch-selection sets, with the final branch label assigned using the pairing mismatch functions. We again emphasize that since the branches intersect, these are not basins of three completely disjoint attractors. The three-against-one saddle-type states also have lower-dimensional stable manifolds, such as the invariant collinear lines leading into $(\pi,0,0)$, $(0,\pi,0)$, $(0,0,\pi)$, and $(\pi,\pi,\pi)$, but these should be regarded as organizing structures rather than as the full basin boundaries. The boundaries between colored regions should instead be interpreted as separatrix-like basin boundaries for the branch-selection problem. They reflect both the unstable stationary structure of the reduced flow and the fact that the pairing label becomes nonunique near collinear branch intersections.

\begin{figure*}
    \centering
    \includegraphics[width=\linewidth]{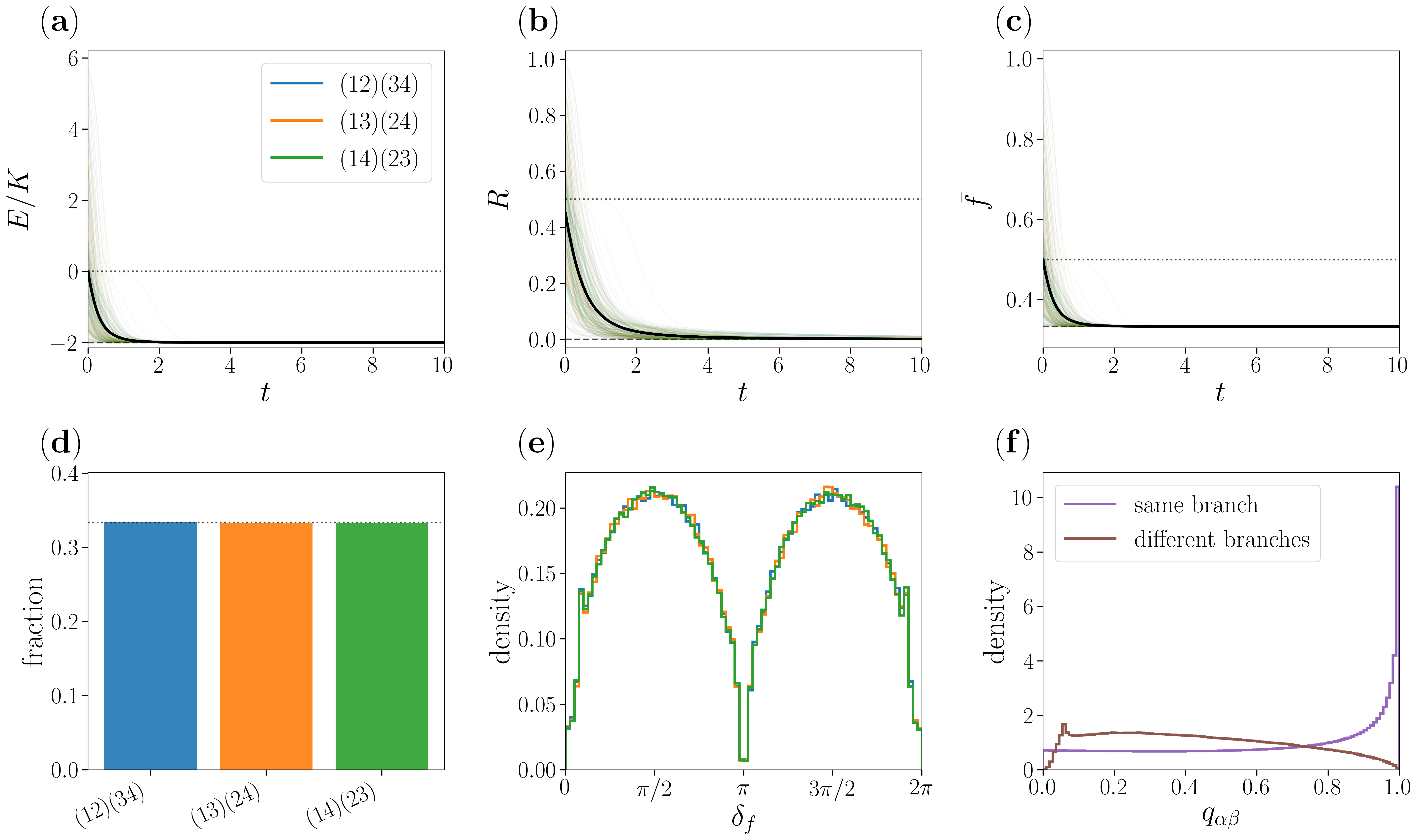}
    \caption{Ensemble diagnostics for the tetrahedral phase motif, computed from $N_{\rm init} = 10^6$ independent initial conditions sampled uniformly from $[0,2\pi)^4$. (a)~Relaxation of the energy $E/K$ toward the ground-state value $E_{\min}/K=-2$. (b)~Decay of the global synchrony order parameter $R$, showing relaxation toward the zero-total-spin condition $R_{\min}=0$. (c)~Relaxation of the mean bond frustration $\bar f$ toward its tetrahedral ground-state value $\bar f_{\min}=1/3$. In (a)--(c), faint colored curves show representative individual trajectories colored by final pairing branch, while solid black curves show branch-averaged relaxation. Dotted lines mark the three-against-one collinear reference values. \\
    (d) Final branch-selection probabilities for the three antipodal pairings. The three branches are selected with equal probability, up to sampling fluctuations, consistent with the permutation symmetry of $K_4$, and the dotted line marks $1/3$. (e) Distribution of the final internal angle $\delta_f$ along each selected $S^1$ branch. Although $\delta$ is a continuous ground-state coordinate, the relaxation dynamics induce a nonuniform final-state measure along it. (f) Distribution of final-state overlaps for pairs of trajectories ending on the same branch or on different branches.}
    \label{fig:tetrahedron_diagnostics}
\end{figure*}

\subsubsection{Dynamical content and ensemble diagnostics}

The preceding figures describe the geometry of the tetrahedral ground-state manifold and the organization of its basins. We now quantify the same behavior over a large unbiased ensemble of initial conditions. The initial phases are sampled independently and uniformly from $[0,2\pi)^4$, and each trajectory is evolved under the tetrahedral gradient flow. To make Fig.~\ref{fig:tetrahedron_diagnostics}, we use an ensemble containing $10^6$ initial conditions. This large sample is useful because the three pairing branches are symmetry-related, so any residual imbalance in the final branch probabilities should be attributable only to finite-sampling fluctuations.

For each trajectory, we record the energy $E$, the global order parameter $R$, and the mean bond frustration $\bar f$. These quantities are not independent in the tetrahedral motif. Using the total-spin identity,
\begin{equation}
    \frac{E}{K}=8R^2-2,
    \label{eq:tet_ER}
\end{equation}
we also have
\begin{equation}
    \bar f=\frac{1}{3}+\frac{2}{3}R^2.
    \label{eq:tet_fR}
\end{equation}
Thus, the ground-state limit is equivalently characterized by
\begin{equation}
    \frac{E}{K}\to -2,\qquad
    R\to 0,\qquad
    \bar f\to \frac{1}{3}.
\end{equation}
The three-against-one collinear stationary states provide another useful reference scale,
\begin{equation}
    \frac{E}{K}=0,\qquad
    R=\frac{1}{2},\qquad
    \bar f=\frac{1}{2}.
\end{equation}
These values are marked by the dotted horizontal lines in Fig.~\ref{fig:tetrahedron_diagnostics}.

The final pairing branch is assigned using the mismatch functions defined in Eq.~\eqref{eq:mismatch}. In this statistical discussion, we label a pairing branch by
\begin{equation}
    P\in\M_4 = \{(12)(34),(13)(24),(14)(23)\}.
\end{equation}
The basin probability associated with branch $P$, defined generally in
Sec.~\ref{sec:minimal_phase_theory}, is estimated as
\begin{equation}
    p_P = \frac{N_P}{N_{\rm init}},
\end{equation}
where $N_P$ is the number of trajectories classified as ending on branch $P$, and $N_{\rm init}$ is the total number of initial conditions. For an unbiased ensemble on the fully symmetric tetrahedral graph, permutation symmetry predicts
\begin{equation}
    p_{(12)(34)} = p_{(13)(24)} = p_{(14)(23)} = \frac{1}{3},
\end{equation}
up to finite-sampling fluctuations.

Figures~\ref{fig:tetrahedron_diagnostics}(a)--(c) show the relaxation of $E/K$, $R$, and $\bar f$, respectively. The colored faint curves are representative individual trajectories, colored by the final pairing branch, while the solid black curves show the branch-averaged behavior. The branch averages lie on top of each other, as expected from the symmetry of $K_4$. Most trajectories relax rapidly toward the ground-state manifold, with only a small tail of slower trajectories remaining near the collinear reference values for an extended time. These slow relaxations correspond to initial conditions passing near unstable collinear structures or symmetry-related basin boundaries before settling onto $\mathcal G_{\rm tet}/U(1)$.

The spread in Fig.~\ref{fig:tetrahedron_diagnostics}(b) is visibly broader than in Figs.~\ref{fig:tetrahedron_diagnostics}(a) and Fig.~\ref{fig:tetrahedron_diagnostics}(c). This is because $R$ measures the residual total spin linearly, whereas both the excess energy and the excess mean frustration are quadratic in $R$, as shown in Eqs.~\eqref{eq:tet_ER} and \eqref{eq:tet_fR}. Consequently, trajectories with noticeably different small values of $R$ can already be nearly indistinguishable in energy and mean frustration. The order parameter $R$ is therefore the more sensitive diagnostic of late-time approach to the ground-state manifold.

The ensemble statistics confirm that the discrete part of the final state is unbiased. As shown in Fig.~\ref{fig:tetrahedron_diagnostics}(d), the measured basin probabilities $p_\alpha$ are essentially equal and lie very close to the reference value $1/3$. Hence, although each trajectory selects a particular pairing branch, the unbiased ensemble does not prefer any one of the three branches. This is consistent with the full permutation symmetry of the tetrahedral motif.

The branch label, however, does not fully specify the final state. After a trajectory has selected a branch, it still lands at some position along the corresponding $S^1$ ground-state circle. Figure~\ref{fig:tetrahedron_diagnostics}(e) shows the resulting distribution of the final internal angle $\delta_f$. The three histograms nearly coincide, again reflecting the symmetry between pairing sectors, but the distribution along each branch is not uniform, showing that different intervals of $\delta_f$ have different basin volumes in the full reduced phase space.

The suppressed weight near $\delta_f=0,\pi,2\pi$ indicates that relatively few random initial conditions relax to final states close to the Ising-like branch-intersection configurations. These points are special because the pairing label is not unique, and as discussed in the local stability analysis, the quadratic form also has additional degeneracies at these intersections compared with a generic point on a branch. Consequently, the landscape near an intersection has extra soft directions and does not funnel nearby trajectories towards a single branch as strongly as a generic branch segment does. Nearby trajectories can be routed toward different outgoing branches, and only a relatively small number of initial conditions land close to the intersection itself. By contrast, the larger weight near $\delta_f=\pi/2$ and $3\pi/2$ shows that the largest basin volume is associated with noncollinear two-pair configurations far from the branch intersections.

The phase-overlap diagnostic $q_{\alpha\beta}$ defined in Sec.~\ref{sec:minimal_phase_theory} gives a complementary view of the mixed discrete-continuous structure. For the tetrahedral motif, the general definition reduces to
\begin{equation}
    q_{\alpha\beta} = \qty|\frac{1}{4}\sum_{j=1}^{4}e^{i\qty(\phi_j^{(\alpha)}-\phi_j^{(\beta)})}|,
\end{equation}
where $\alpha$ and $\beta$ label final states in the ensemble, not pairing sectors. Because the absolute value removes global phase
differences, $q_{\alpha\beta}$ compares the relative phase patterns of
the two final configurations.

For two final states on the same pairing branch, the overlap depends only on their separation along the internal $S^1$ coordinate. Specifically, for two relaxed configurations on the same branch,
\begin{equation}
    q_{\alpha\beta} = \qty|\cos(\frac{\delta_\alpha-\delta_\beta}{2})|.
\end{equation}
Therefore, same-branch pairs with nearby values of $\delta_f$ contribute near $q_{\alpha\beta}=1$, while pairs separated by roughly $\pi$ along the same ground-state circle contribute near small overlap. The strong peak at $q_{\alpha\beta}=1$ arises because the relaxation dynamics concentrate final states within high-density regions of the $\delta_f$ distribution shown in Fig.~\ref{fig:tetrahedron_diagnostics}(e). Many trajectories assigned to the same $S^1$ branch therefore land near one another in the branch coordinate, producing same-branch final-state pairs with nearly identical phase configurations.

Pairs ending on different branches probe a different structure. Their overlap is no longer controlled only by distance along one $S^1$ branch because the two states also differ in which sites are paired antipodally. Consequently, different-branch pairs produce a broader distribution concentrated at intermediate values of $q_{\alpha\beta}$. This broad distribution reflects the fact that distinct pairing sectors are neither completely unrelated nor generically identical; they are different embeddings of the same two-antiphase-pair structure in the four labeled sites, and they meet only at the special Ising-like branch-intersection configurations.

The utility of Fig.~\ref{fig:tetrahedron_diagnostics}(f) is that it packages the information in Figs.~\ref{fig:tetrahedron_diagnostics}(d) and (e) into a pairwise diagnostic. Same-branch pairs are enhanced near $q_{\alpha\beta}=1$ when the two trajectories land at nearby values of $\delta_f$, while different-branch pairs form a broader distribution because the antipodal pairing of the labeled sites differs. The overlap distribution therefore provides a compact summary of both branch selection and residual motion along the selected $S^1$ branch.

Taken together, the tetrahedral motif provides the first example in which the frustrated phase dynamics produce a continuous reduced ground-state structure. The triangle resolves frustration by selecting one of two isolated chiral minima after quotienting by the global phase. By contrast, the tetrahedron relaxes to a zero-total-spin manifold consisting of three intersecting $S^1$ branches, one for each antipodal pairing of the four labeled oscillators. Therefore, the final state cannot be specified by a single discrete order parameter. It contains both a discrete component, the selected pairing branch $P\in\M_4$, and a continuous component, the internal angle $\delta_f$ along that branch.

This mixed discrete-continuous structure is reflected dynamically in several ways. The reduced gradient flow drives trajectories monotonically toward $E_{\min}=-2K$, suppressing the global order parameter to $R=0$ while leaving a residual flat direction within the ground-state manifold. Basin slices reveal how different regions of initial phase space select different pairing branches, while the ensemble diagnostics show that the three branches are selected with equal probability for unbiased initial conditions. At the same time, the nonuniform distributions of the final internal angle $\delta_f$ and the final-state overlap $q_{\alpha\beta}$ show that the flow induces a nontrivial measure on the degenerate ground-state manifold. The tetrahedral motif therefore goes beyond the triangle by combining frustration, basin selection, branch intersections, and continuous degeneracy in the smallest complete graph for which such structure appears.

\subsection{Small kagome lattice}
\label{subsec:kagome}

We now transition from isolated frustrated motifs to an extended two-dimensional frustrated network. The kagome lattice, one of the most famous frustrated systems in condensed matter physics, is a lattice of corner-sharing triangles. It therefore repeats the same local incompatibility found in the triangular motif while allowing the local $120^\circ$ constraints to propagate through a many-site system. The result is a constrained coloring sector, spatial chirality patterns, and a landscape containing both ground-state colorings and non-ground torque-balanced final states.

\begin{figure}[tbp!]
    \centering
    \includegraphics[width=0.8\linewidth]{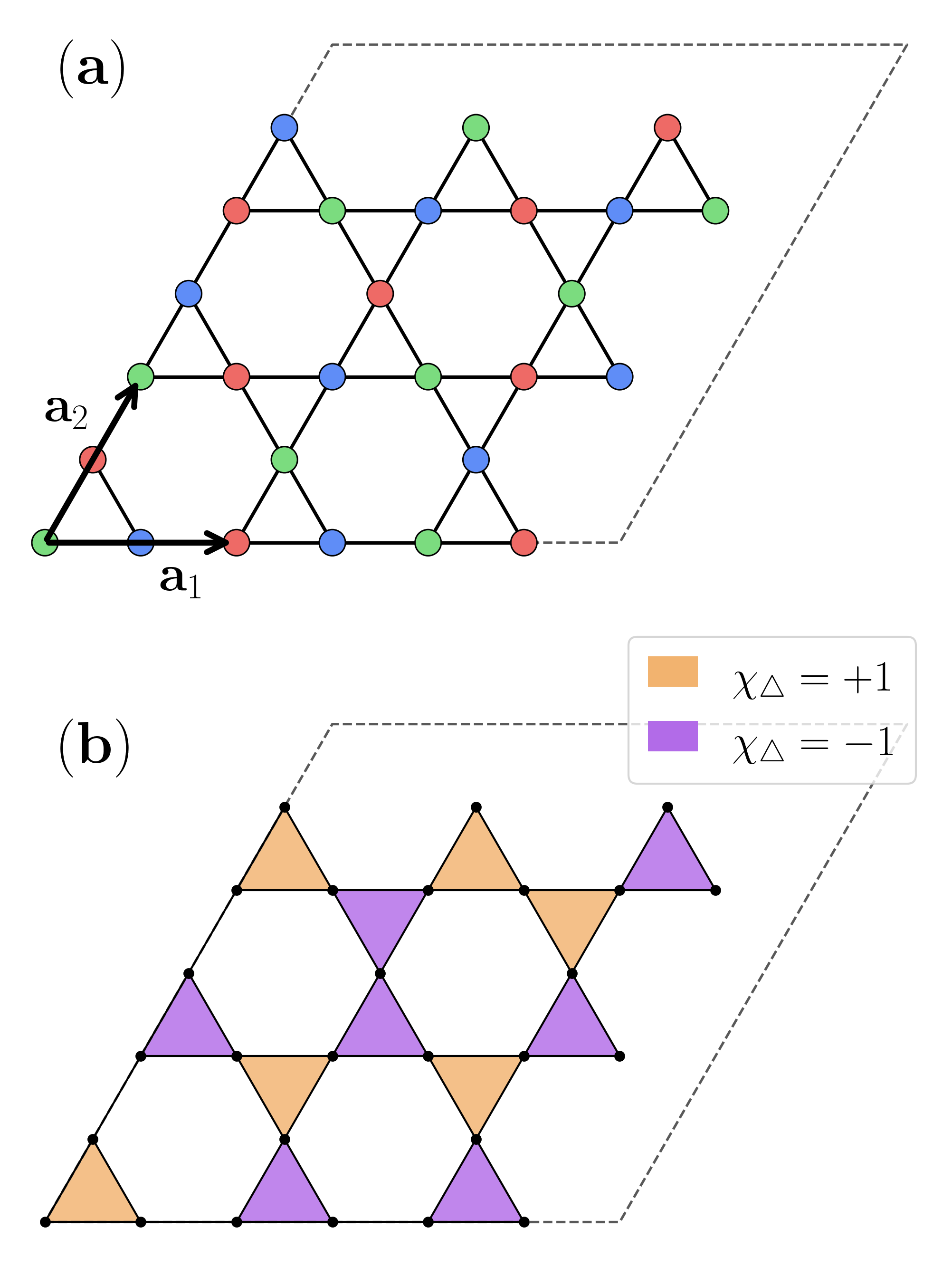}
    \caption{Kagome lattice geometry, three-coloring constraint, and local chirality structure. (a)~Representative $3\times 3$ kagome lattice with primitive Bravais vectors $\mb{a}_1$ and $\mb{a}_2$ and periodic boundary conditions indicated by the dashed parallelogram.  The three site colors denote phases separated by $2\pi/3$, corresponding to $\phi_i=\Phi+2\pi c_i/3$ with $c_i\in\mathbb Z_3$.  A ground-state coloring requires every elementary triangle to contain all three colors so that each triangle satisfies the local $120^\circ$ constraint. (b)~The corresponding chirality pattern for the same coloring.  For a fixed orientation convention, each elementary triangle has $\chi_\triangle=+1$ or $\chi_\triangle=-1$ depending on the cyclic ordering of the three colors.}
    \label{fig:kagome_coloring_chirality}
\end{figure}

We take the kagome lattice to be a triangular Bravais lattice with a three-site basis, the structure of which is displayed in Fig.~\ref{fig:kagome_coloring_chirality}(a). The primitive vectors are 
\begin{equation}
    \mb{a}_1 = (1,0),\quad\mb{a}_2=\qty(\frac{1}{2},\frac{\sqrt{3}}{2}),
\end{equation}
and the basis vectors are
\begin{equation}
    \bs{\delta}_A = (0,0),\quad\bs{\delta}_B = \qty(\frac{1}{2},0),\quad\bs{\delta}_C = \qty(\frac{1}{4},\frac{\sqrt{3}}{4}).
    \label{eq:kagome_basis}
\end{equation}
A lattice site is labeled by a Bravais-cell coordinate $\mb{R}_{\mb{n}} = n_1\mb{a}_1 + n_2\mb{a}_2$ and a sublattice index $\mu\in\{A,B,C\}$:
\begin{equation}
    \mb{r}_{\mb{n},\mu} = \mb{R}_{\mb{n}} + \bs{\delta}_\mu.
\end{equation}
In this work, we consider $L_x\times L_y$ kagome clusters with periodic boundary conditions, which identify Bravais indices as
\begin{equation}
    (n_1,n_2)\equiv(n_1+L_x,n_2)\equiv(n_1,n_2+L_y).
\end{equation}
The total number of sites is
\begin{equation}
    N = 3L_xL_y.
\end{equation}
In the numerical examples below, we use the smallest periodic cluster large enough to display nontrivial coloring, chirality, and metastable basin structure, namely $L_x=L_y=3$, so $N=27$. Nearest-neighbor bonds are the edges of the elementary triangles. With the basis in Eq.~\eqref{eq:kagome_basis}, one upward-pointing triangle with edges $(A_{\mb{n}},B_{\mb{n}}),(B_{\mb{n}},C_{\mb{n}}),(C_{\mb{n}},A_{\mb{n}})$ is contained within each unit cell. The neighboring downward-pointing triangle sharing the site $A_{\mb{n}}$ has edges $(A_{\mb{n}},B_{\mb{n}-\mb{1}}),(B_{\mb{n}-\mb{1}},C_{\mb{n}-\mb{2}}),(C_{\mb{n}-\mb{2}},A_{\mb{n}})$, where $\mb{1}$ and $\mb{2}$ denote translations by one unit cell along $\mb{a}_1$ and $\mb{a}_2$, respectively. With periodic boundary conditions, each site has coordination number $z=4$, so the number of nearest-neighbor bonds is $N_b = 2N$.

On the kagome lattice, we use the same repulsively coupled phase dynamics with the antiferromagnetic $XY$ energy
\begin{equation}
    E = K\sum_{\ev{ij}}\cos(\phi_i - \phi_j),
\end{equation}
where $\ev{ij}$ denotes nearest-neighbor bonds. The local constraint is inherited directly from the triangular motif. On any elementary triangle $\triangle = (i,j,k)$, the energy is minimized when the three phases form a $120^\circ$ pattern:
\begin{equation}
    \{\phi_i,\phi_j,\phi_k\} = \qty{\Phi,\Phi+\frac{2\pi}{3},\Phi+\frac{4\pi}{3}}\bmod{2\pi},
\end{equation}
where $\Phi$ represents the global $U(1)$ phase. The kagome ground-state problem can therefore be phrased as a constrained three-coloring problem: assign a color $c_i\in\Z_3$ with $\phi_i = \Phi + 2\pi c_i/3$ such that every elementary triangle contains all three colors. Figure~\ref{fig:kagome_coloring_chirality}(a) illustrates one such coloring, where the red, green, and blue sites are visual representatives of the three elements of $\Z_3$. This local coloring constraint is the starting point for characterizing the ground-state manifold.

\subsubsection{Ground-state coloring manifold structure}

The local constraint gives a useful exact description of the kagome ground states. Let $\T$ denote the set of elementary up- and down-pointing triangles, let $\Lambda_{\rm kag}$ denote the set of sites of the finite periodic kagome cluster, and let $c_i\in\Z_3$ be the color assigned to site $i$. We define the kagome coloring sector by
\begin{equation}
    \C = \qty{c:\Lambda_{\rm kag}\to\Z_3\,\big|\,\{c_i,c_j,c_k\} = \Z_3\,\forall\triangle = (i,j,k)\in\T},
\end{equation}
so $\C$ is the set of three-colorings for which the corresponding phase configurations are ground states. Given any coloring $c\in\C$, a corresponding configuration is obtained by choosing a global phase $\Phi\in [0,2\pi)$ and setting 
\begin{equation}
    \phi_i(\Phi,c) = \Phi + \frac{2\pi c_i}{3}.
    \label{eq:kagome_phase_coloring_map}
\end{equation}
Every nearest-neighbor bond in such a configuration connects two different colors, so $\phi_i-\phi_j = \pm2\pi/3\pmod{2\pi}$ and $\cos(\phi_i-\phi_j) = -1/2$. Therefore, all bonds carry the same residual frustration cost. For a kagome cluster with $N = 3L_xL_y$ sites and $N_b = 2N$ bonds, the ground-state energy is
\begin{equation}
    E_{\min} = -\frac{K}{2}N_b = -KN.
\end{equation}
Equivalently, the average bond frustration is $\bar f_{\min} = 1/4$. As in the triangular motif, ground states distribute frustration uniformly over every bond in the lattice.

The parameterization in Eq.~\eqref{eq:kagome_phase_coloring_map} is not one-to-one, as different choices of the global angle $\Phi$ and the color labels $c_i$ can represent the same physical phase configuration. The three phase colors can be permuted by
\begin{equation}
    S_3\cong\Z_3\rtimes\Z_2,
\end{equation}
where $\rtimes$ is the semidirect product, but only the cyclic subgroup $\Z_3$ is a redundancy of the parameterization. A global phase rotation by $2\pi/3$ can be absorbed into a cyclic relabeling of the three colors:
\begin{equation}
    \Phi\mapsto\Phi+\frac{2\pi m}{3},\quad c_i\mapsto c_i-m,\quad m\in\Z_3,
\end{equation}
which leaves every physical phase $\phi_i$ unchanged. Hence, the kagome ground-state manifold has the structure
\begin{equation}
    \G_{\rm kag} \cong \frac{U(1)\times\C}{\Z_3}.
\end{equation}
The quotient removes the redundant cyclic color relabeling associated with the global $U(1)$ rotation, but it does not remove the remaining $\Z_2$ in $S_3$, which corresponds to color reversal $c_i\mapsto -c_i$. This reversal is therefore a symmetry of the energy, but not a gauge redundancy of the representation in Eq.~\eqref{eq:kagome_phase_coloring_map}.

The physical meaning of this remaining $\Z_2$ is local chirality. For a general phase configuration on an oriented elementary triangle $\triangle=(i,j,k)$,
\begin{equation}
\begin{split}
    \chi_\triangle(\phi) = \frac{2}{3\sqrt{3}}&\big[\sin(\phi_j-\phi_i) \\ &\quad + \sin(\phi_k-\phi_j) + \sin(\phi_i-\phi_k)\big].
\end{split}
\end{equation}
For a ground-state coloring, this reduces to the coloring expression
\begin{equation}
\begin{split}
    \chi_\triangle&(c) = \frac{2}{3\sqrt{3}}\bigg[\sin\qty(\frac{2\pi}{3}(c_j-c_i)) \\ &+ \sin\qty(\frac{2\pi}{3}(c_k-c_j)) + \sin\qty(\frac{2\pi}{3}(c_i-c_k))\bigg].
\end{split}
\end{equation}
For a valid coloring, this takes the values $\chi_\triangle=\pm 1$, with the sign distinguishing the two cyclic orderings of the three colors on the chosen oriented triangle. Under color reversal $c_i\mapsto -c_i$, we have $\chi_\triangle\mapsto -\chi_\triangle$. Thus, chirality measures precisely the part of the color-permutation structure that survives the $\Z_3$ quotient; it belongs to the discrete coloring sector $\C$, not to the overall $U(1)$ phase.

A representative chirality pattern is shown in Fig.~\ref{fig:kagome_coloring_chirality}(b). Each triangle inherits a sign from the cyclic ordering of the colors in Fig.~\ref{fig:kagome_coloring_chirality}(a). However, these signs are not independent Ising variables. Neighboring triangles share sites, so choosing the coloring on one triangle constrains the allowed color orderings on adjacent triangles. Therefore, only certain assignments of $\chi_\triangle=\pm1$ can be lifted to a globally consistent three-coloring, so it is useful to define the set of allowed chirality patterns as
\begin{equation}
    \X_{\rm kag} = \qty{\{\chi_\triangle(c)\}_{\triangle\in\T}\ \big|\ c\in\C}\subset\{\pm 1\}^{|\T|}.
\end{equation}
The inclusion is generally strict because an arbitrary pattern of $\pm1$ signs on triangles need not correspond to any valid kagome coloring. For a connected cluster, once a compatible chirality pattern and the colors on one reference triangle are fixed, the colors on the rest of the lattice are determined by propagation across shared edges. Consistency around closed loops is precisely the condition that the chirality pattern belongs to $\mathcal X_{\rm kag}$.

The structure of the ground-state coloring manifold should be distinguished from the finite motifs discussed earlier in this section. After quotienting by the global $U(1)$, the triangle has only two chiral ground states, while the tetrahedron has a continuous ground-state degeneracy associated with antipodal pairings. The kagome lattice instead supports an extended coloring sector $\C$, resulting from the requirement that the local $120^\circ$ rule must be satisfied consistently across a periodic network of corner-sharing triangles. As a result, the discrete part of the kagome ground-state structure contains more information than a single chirality label. The lattice can support spatially varying arrangements of positive and negative triangle chiralities while still satisfying the local constraint everywhere. Thus, while each elementary triangle locally resembles the frustrated triangular ground state, the global ground-state sector consists of many compatible coloring and chirality patterns across the periodic lattice.

\subsubsection{Zero-temperature quench dynamics and metastable torque-balanced states}

We now ask how the system reaches, or fails to reach, the coloring manifold $\G_{\rm kag}$ dynamically. As in the finite motifs above, we sample random initial phases and evolve them under the deterministic gradient-flow dynamics with no noise. In the language of condensed matter physics, this procedure is a zero-temperature quench, where the random initial condition represents a disordered high-temperature state and the subsequent deterministic evolution represents relaxation after instantaneous cooling to zero temperature. The word ``quench'' therefore emphasizes the analogy to suddenly cooling a frustrated magnet and observing which ground-state configuration or metastable local minimum is selected by the relaxation dynamics. In the present phase-oscillator model, this analogy is implemented as an initial-value protocol, with phases drawn from a disordered distribution and then allowed to relax downhill according to the fixed zero-temperature kagome energy.

This protocol is useful because it probes the basin structure of the landscape. The system does not search globally for the lowest energy state; rather, each random initial condition flows downhill until it reaches the stationary state in whose basin it started. Since the dynamics are zero-temperature gradient flow, the energy is non-increasing $\mathrm{d}E/\mathrm{d}t\leq 0$. Therefore, once a trajectory enters a basin separated from the coloring manifold by an energy barrier, it cannot escape by thermal activation or noise-induced hopping.

Specializing the stationarity condition $\partial E/\partial \phi_i = 0$ in Eq.~\eqref{eq:stationary_condition} to the kagome graph, it is useful to write the gradient-flow force on site $i$ as a local torque:
\begin{equation}
    \tau_i = -\pdv{E}{\phi_i} = K\sum_{j=1}^NA_{ij}\sin(\phi_i-\phi_j),
    \label{eq:kagome_site_torque}
\end{equation}
where $A_{ij}$ is the adjacency matrix of the kagome graph. It follows that stationarity is the torque-balance condition
\begin{equation}
    \tau_i = 0
\end{equation}
for all $i$. We call these states torque-balanced because each bond contributes a phase torque
\begin{equation}
    \tau_{ij} = KA_{ij}\sin(\phi_i-\phi_j)
\end{equation}
to oscillator $i$, and the total torque $\tau_i = \sum_j\tau_{ij}$ vanishes at each site.

\begin{figure*}[tbp!]
    \centering
    \includegraphics[width=\linewidth]{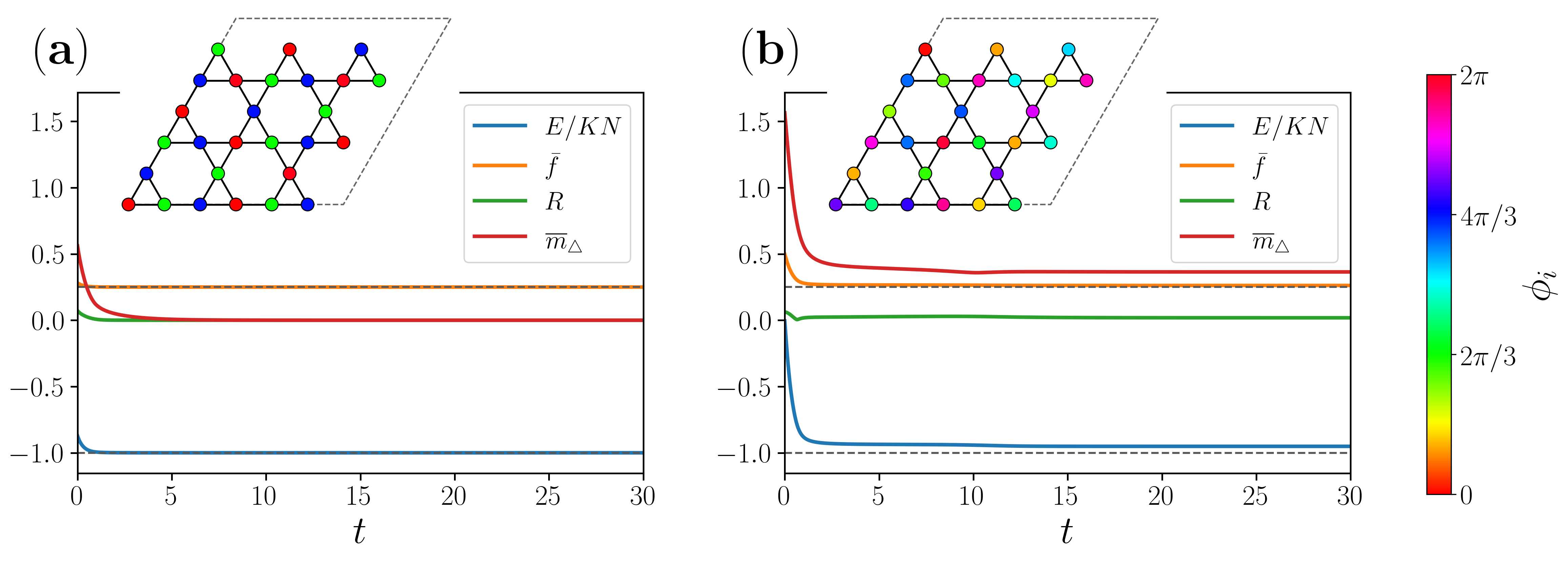}
    \caption{Representative zero-temperature relaxation trajectories on the kagome lattice. The upper insets show the final phase configuration on the periodic kagome cluster, with site colors indicating the phase $\phi_i$. The time traces show the energy density $E/KN$, mean bond frustration $\bar f$, global phase coherence $R$, and mean triangle imbalance $\overline{m}_\triangle$. Dashed horizontal lines mark the kagome ground-state values $E_{\min}/KN=-1$ and $\bar f_{\min}=1/4$. (a)~A trajectory that relaxes into the kagome coloring manifold. The final state satisfies the local $120^\circ$ constraint on every triangle, has vanishing global coherence, and reaches the ground-state energy. (b)~A trajectory that relaxes instead to a torque-balanced metastable state outside the ground-state coloring manifold. Although the final state remains nearly incoherent globally, it retains nonzero local triangle imbalance and an energy above the ground-state value.}
    \label{fig:kagome_trajectories}
\end{figure*}

In planar-spin language with $\mb{S}_i = (\cos\phi_i,\sin\phi_i)$, the neighboring spins define a local field
\begin{equation}
    \mb{h}_i = \sum_{j=1}^NA_{ij}\mb{S}_j.
\end{equation}
Using the sign convention above, the torque is then
\begin{equation}
    \tau_i = K(\mathbf h_i\times \mathbf S_i)_z.
\end{equation}
Torque balance means that the net local tendency for the phase to rotate vanishes. A ground-state coloring is one example of a torque-balanced state, but it is not the only one.

This distinction is essential on the kagome lattice. A ground-state coloring satisfies the stronger local condition that every elementary triangle forms an exact $120^\circ$ pattern, but torque balance only requires the sum of the bond torques at each site to vanish. Because each kagome site belongs to two adjacent triangles, deviations from the $120^\circ$ condition on one triangle can be compensated by deviations in the neighboring triangle. The result is a stationary configuration in which no individual phase wants to move, even though the configuration is not a ground state.

Simple torque-balanced states include the previously discussed highly symmetric unstable examples, such as the fully synchronized state or collinear $0/\pi$ configurations, for which every bond torque vanishes individually. However, the physically important nonground torque-balanced states in the kagome problem are metastable states reached by generic zero-temperature quenches. In these states, the phases are spread over many angles, and the elementary triangles do not satisfy the exact $120^\circ$ coloring constraint. Nevertheless, the final configuration is stationary because the nonzero bond torques incident on each site cancel in the sum, giving $\tau_i=0$ to numerical precision. Metastability on the kagome lattice reflects the extended frustrated constraint structure, where local triangle errors can be mutually compensated across the network to produce stable nonground attractors.

The kagome lattice is therefore dynamically different from the isolated triangle. For one triangle, the stable zero-temperature endpoints are the two chiral $120^\circ$ ground states up to global $U(1)$ rotation. On the kagome lattice, the exact coloring manifold $\G_{\rm kag}$ coexists with many additional torque-balanced states outside that manifold. Random initial conditions often fall into these nonground basins before the system can organize into a globally compatible three-coloring. In this sense, the quench dynamics sample the full attractor landscape of the frustrated lattice, not just its ground-state manifold.

To diagnose the relaxation, we monitor the normalized energy density $E/KN$, the average bond frustration $\bar f$, the global coherence $R$, and the average local triangle moment $\ov{m}_\triangle$ defined in Eq.~\eqref{eq:mean_triangle_moment}. Here, $\ov{m}_\triangle=0$ if and only if every elementary triangle satisfies the local \(120^\circ\) constraint. Then, for a ground-state coloring,
\begin{equation}
    \frac{E_{\min}}{KN} = -1,\quad \bar f_{\min} = \frac{1}{4},\quad R_{\min} = 0,\quad \ov{m}_{\triangle,\min} = 0.
    \label{eq:kagome_minimum_values}
\end{equation}
Since $N_b = 2N$, the general relation between $E$ and $\bar f$ from Eq.~\eqref{eq:fbar_energy_relation_general} becomes
\begin{equation}
    \bar f = \frac{1}{2} + \frac{1}{4}\frac{E}{KN},
    \label{eq:kagome_fE}
\end{equation}
but we still plot both because $\bar f$ has the direct interpretation of a mean local bond cost.

Representative quenches are shown in Fig.~\ref{fig:kagome_trajectories}. Figure~\ref{fig:kagome_trajectories}(a) displays a trajectory that reaches the ground-state coloring manifold $\G_{\rm kag}$, showing relaxation to the minimum values in Eq.~\eqref{eq:kagome_minimum_values}. The inset illustrates the corresponding final three-coloring. Figure~\ref{fig:kagome_trajectories}(b) shows the more typical nonground outcome. The energy decreases rapidly, and the global coherence $R$ remains small, so the system still has suppressed global synchrony. However, the final energy remains above $E_{\min} = -KN$, the bond frustration remains above $\bar f_{\min} = 1/4$, and $\ov{m}_\triangle$ remains nonzero. The final phases shown in the inset are spread over many angles rather than collapsing onto the three ideal coloring values. The state is therefore not in $\G_{\rm kag}$, but it is still stationary under the deterministic dynamics. In terms of the site torque defined in Eq.~\eqref{eq:kagome_site_torque}, the final state satisfies $\max_i|\tau_i|<10^{-12}$, showing that the nonground state is indeed torque-balanced.

This illustrates the warning from Sec.~\ref{sec:minimal_phase_theory} that low $R$ does not by itself imply local frustrated order. Both the exact coloring state and the metastable state in Fig.~\ref{fig:kagome_trajectories} have small $R$, so both are globally desynchronized, but they still differ locally. In the ground state, every triangle satisfies the $120^\circ$ constraint, but in the metastable state, the system has suppressed global synchrony without satisfying all local kagome constraints. The finite value of $\ov{m}_\triangle$ detects precisely this local failure.

\begin{figure*}[tbp!]
    \centering
    \includegraphics[width=\linewidth]{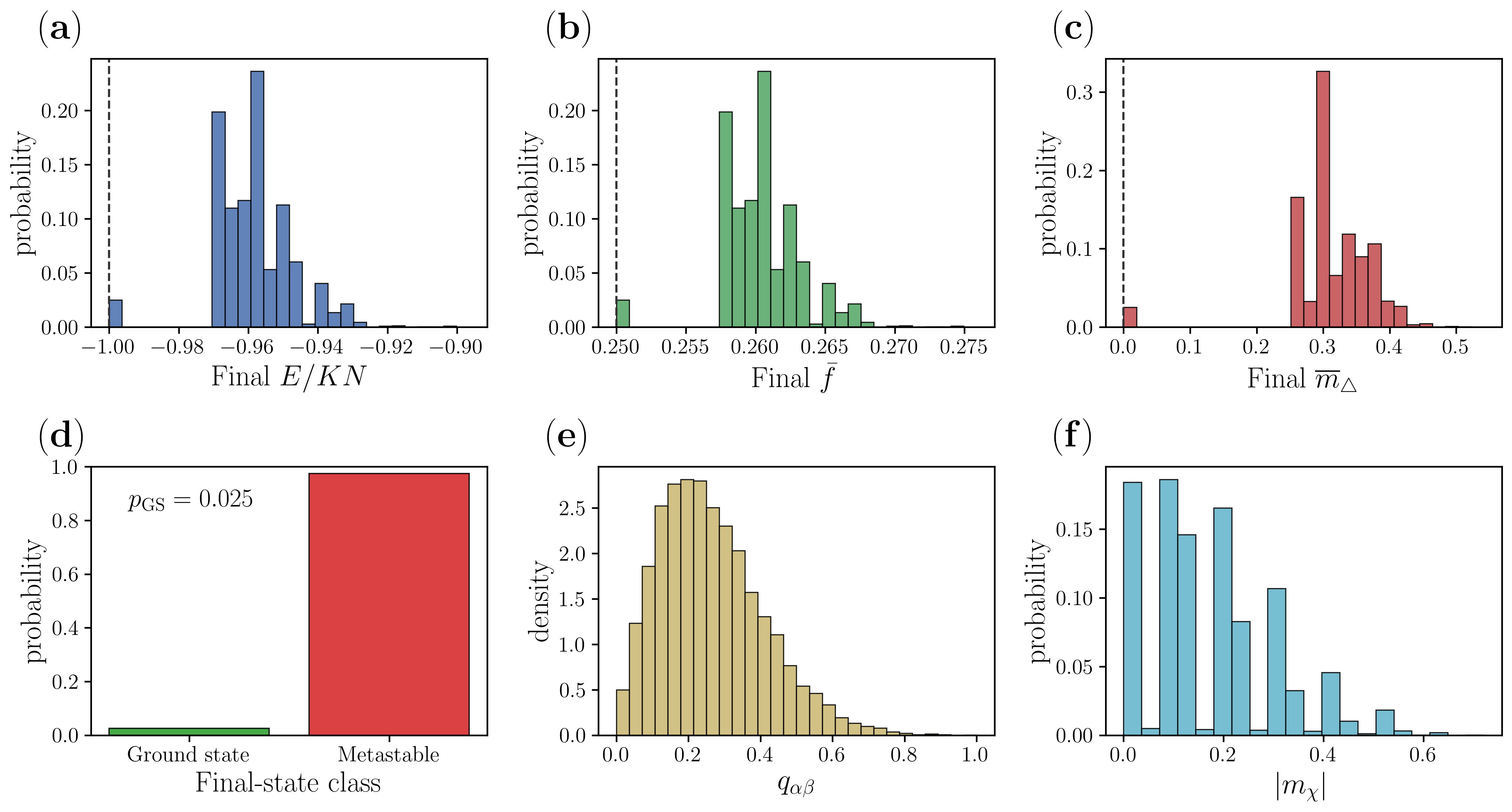}
    \caption{Ensemble statistics for zero-temperature quenches on a $3\times 3$ periodic kagome lattice, computed from $N_{\rm init} = 10^6$ independent initial conditions. Each trajectory is initialized from random phases and evolved under deterministic gradient flow until it reaches a torque-balanced final state. Dashed vertical lines in (a)--(c) mark the exact ground-state coloring values. (a)~Final energy density $E/KN$, with the ground-state value $E_{\min}/(KN)=-1$. (b)~Final average bond frustration $\bar f$, with the ground-state value $\bar f_{\min}=1/4$. (c)~Final average triangle moment $\ov{m}_\triangle$, which vanishes when every elementary triangle satisfies the local $120^\circ$ constraint. (d)~Basin probabilities for ground-state and metastable final-state classes, showing the small ground-state basin weight $p_{\rm GS}\approx 0.025$. (e)~Pairwise final-state overlap distribution. (f)~Distribution of the net chirality magnitude $|m_\chi|$. Together, these diagnostics show that most random quenches terminate in metastable torque-balanced states outside the exact kagome coloring manifold.}
    \label{fig:kagome_ensemble_statistics}
\end{figure*}

The key point is that the kagome quench samples a larger attractor landscape than the ground-state manifold alone. $\G_{\rm kag}$ is characterized by exact three-colorings, but random zero-temperature quenches often terminate in metastable torque-balanced states outside this manifold. In these states, distortions of neighboring triangles compensate so that the net torque on every site vanishes, even though the local $120^\circ$ constraint is not satisfied everywhere. Once the trajectory enters such a basin, the zero-temperature gradient flow cannot cross the intervening barriers. Thus, the nonground outcomes are emergent dynamical consequences of the extended frustrated landscape.

\subsubsection{Ensemble diagnostics}

The trajectory-level diagnostics show that individual quenches can either reach the exact coloring manifold or become trapped in nonground torque-balanced states. We now ask which outcome is typical over an ensemble of random initial conditions. Figure~\ref{fig:kagome_ensemble_statistics} summarizes the final states obtained by initializing an ensemble of $10^6$ random kagome states and evolving by zero-temperature gradient flow until a stationary torque-balanced configuration is reached. The dashed vertical lines in Figs.~\ref{fig:kagome_ensemble_statistics}(a)--(c) mark the exact coloring-manifold values in Eq.~\eqref{eq:kagome_minimum_values}.

Figures~\ref{fig:kagome_ensemble_statistics}(a) and (b) show that the final states are typically low-energy but nonground.  The energy distribution has a small peak at the coloring value $E_{\min}/KN=-1$, but most of the probability weight lies slightly above it.  The same behavior appears in the distribution of $\bar f$, whose ground-state value is $\bar f_{\min} = 1/4$. Since $\bar f$ is linearly related to $E/KN$ by Eq.~\eqref{eq:kagome_fE}, Fig.~\ref{fig:kagome_ensemble_statistics}(b) rewrites the energetic information as an average local bond cost. This is useful physically because it shows that the metastable states are not highly excited configurations, usually having only modest excess frustration per bond, even though they have failed to enter the exact coloring manifold.

Figure~\ref{fig:kagome_ensemble_statistics}(c) provides a clearer local distinction. For most quenches, the distribution of $\ov{m}_\triangle$ is concentrated away from zero, indicating that the final states are not merely distorted three-colorings, as they often retain significant local violations of the $120^\circ$ constraint. This is the essential separation produced by the extended kagome network, namely, that torque balance at every site does not imply that every elementary triangle is individually minimized.

The basin probabilities in Fig.~\ref{fig:kagome_ensemble_statistics}(d) convert this observation into a dynamical statement. Only a fraction
\begin{equation}
    p_{\rm GS}\approx 0.025
\end{equation}
of random initial conditions reach the exact ground-state coloring manifold, and the dominant outcome is a metastable torque-balanced state. This result is best interpreted in terms of basin volume. The kagome lattice has many compatible colorings, but under deterministic zero-temperature relaxation, their combined basin of attraction occupies only a small portion of the sampled phase space. Most random initial conditions instead flow into nonground local minima or stable stationary states before a globally compatible coloring can be assembled.

The remaining panels show that the metastable sector is not a single generic ``other'' outcome. The overlap distribution in Fig.~\ref{fig:kagome_ensemble_statistics}(e) is broad, with substantial weight at low and intermediate $q_{\alpha\beta}$, so two independently quenched final states are typically not related by a global phase rotation or by a small local distortion. Since the ground-state basin weight is so small, this broad overlap structure is dominated by metastable torque-balanced states rather than by exact colorings. The result is therefore not merely that most trajectories miss $\G_{\rm kag}$, but that they miss it in many different ways. We conclude that the kagome attractor landscape contains a large nonground sector with diverse spatial phase patterns and nontrivial basin structure.

Finally, Fig.~\ref{fig:kagome_ensemble_statistics}(f) characterizes the chirality content of these final states using the net chirality magnitude $|m_\chi|$ defined in Eq.~\eqref{eq:net_chirality_magnitude}. A small value of $|m_\chi|$ indicates that positive and negative local chiralities largely compensate, while a larger value indicates a stronger imbalance between the two handednesses. The broad distribution in Fig.~\ref{fig:kagome_ensemble_statistics}(f) therefore shows that the final-state ensemble contains a variety of chirality organizations, rather than a unique chiral ordering pattern. The distribution is not expected to be smooth for the finite $3\times 3$ cluster studied here, since $|m_\chi|$ is built from a finite number of triangle chiralities and the allowed values are constrained by the kagome connectivity and attractor structure. This means that the jagged appearance of the histogram reflects finite-size effects, not a failure of the diagnostic. For exact colorings, $|m_\chi|$ probes the compatible chirality sector $\mathcal{X}_{\rm kag}$ associated with $\mathcal C$; for metastable states, it remains a useful coarse measure of how locally triangular phase order is arranged across the lattice.

The kagome lattice completes the transition from isolated frustrated motifs to an extended frustrated network. Geometrically, it repeats the triangular incompatibility on a periodic array of corner-sharing triangles, so the local ground-state condition is still the familiar $120^\circ$ phase constraint. Globally, however, satisfying this constraint on every triangle produces the coloring manifold $\G_{\rm kag} = (U(1)\times \C)/\Z_3$, where $\C$ encodes the compatible three-colorings and their associated chirality patterns. The zero-temperature dynamics shows that this geometrically well-defined ground-state manifold is not the same as the dynamically accessible attractor set. Random quenches typically suppress global synchrony and relax to low-energy torque-balanced states, but most do not reach an exact coloring. In these metastable final states, residual triangle distortions remain even though the net torque on every site vanishes. The ensemble diagnostics make this separation explicit by distinguishing energy, local satisfaction of the $120^\circ$ constraint, basin weight of $\G_{\rm kag}$, overlap structure, and chirality organization. Because the present calculation uses a small periodic $3\times3$ cluster, the numerical values of the basin probabilities and histogram shapes should be interpreted as finite-size diagnostics rather than thermodynamic-limit estimates. The qualitative point is the distinction between the exact coloring manifold and the larger metastable torque-balanced attractor set. The kagome system therefore provides the first extended-network realization of the frustrated-neuron mechanism in this paper. Local antiphase incompatibility suppresses global synchrony while preserving structured local order, and the extended constraint network produces a many-state metastable landscape substantially larger than the exact ground-state coloring manifold. This distinction between global coherence, local constraint satisfaction, and metastable basin structure is the main result carried into the biophysical interpretation of Sec.~\ref{sec:biophys_real}.

\section{Towards biophysical realization}
\label{sec:biophys_real}

The preceding sections established the frustrated-neuron mechanism in its cleanest mathematical form. By reducing each rhythmic unit to a phase and imposing a uniform repulsive interaction on each edge, we isolated the geometrical content of the problem. The purpose of this section is to explain how this phase-level mechanism should be carried back toward more biophysical neuron models.

Roughly, this is the reverse direction from the reduction discussed in the introduction and Subsec.~\ref{subsec:neural_timing_frustrated_phase}, where phase oscillators were motivated as the simplest timing-level description of rhythmic neural systems. Now, the phase theory functions as a target, identifying which timing structures a more detailed model must reproduce in order to realize geometrical frustration. While a biophysical realization will not be literally equivalent to an antiferromagnetic $XY$ magnet, it must possess rhythmic degrees of freedom with effective interactions that prefer a nonzero phase difference and a network architecture that prevents those preferred relations from being simultaneously satisfied. Under these conditions, the diagnostics developed above provide a concrete language for analyzing neural timing beyond the usual dichotomy between synchrony and desynchrony.

\subsection{Embedding phase-level frustration in biophysical variables}

A biophysical realization of the frustration mechanism must embed the target mechanism in variables such as membrane voltage, recovery currents, ionic gates, synaptic activation, electrical coupling, and noise. Therefore, the first question we must consider is how these biophysical variables can collectively generate stable timing preferences on individual edges and incompatible timing constraints on closed motifs.

Following the standard conductance-based current-balance framework augmented by chemical synaptic currents, electric gap-junction coupling, and stochastic drive, a schematic biophysical network can be written as~\cite{hodgkin1952quantitative, destexhe1994efficient, koch1998methods, connors2004electrical, faisal2008noise}
\begin{equation}
\begin{split}
    C_i\dot{V}_i &= -I_i^{\rm ion}(V_i,\mb{w}_i) + I_i^{\rm app} \\ &\qquad - \sum_jA_{ij}I_{ij}^{\rm chem} - \sum_jA_{ij}I_{ij}^{\rm gap} + \sigma_i\xi_i(t),
    \label{eq:schem_biophys_network}
\end{split}
\end{equation}
where $V_i$ is the membrane voltage of unit $i$, $C_i$ is its capacitance, $I_i^{\rm app}$ is an applied or tonic drive, and $\xi_i(t)$ represents unresolved fluctuations. The intrinsic ionic current is represented in conductance-based form as
\begin{equation}
    I_i^{\rm ion}(V_i,\mb{w}_i) = \sum_\ell \bar{g}_{\ell i}M_{\ell i}(\mb{w}_i)(V_i-E_{\ell i}),
\end{equation}
where $\ell$ labels ionic current types, $\bar g_{\ell i}$ is a maximal conductance, $E_{\ell i}$ is the corresponding reversal potential, and $M_{\ell i}$ is a product of activation and inactivation variables. In Hodgkin--Huxley-type notation, for example, these products include factors such as $m^3h$ for sodium activation and inactivation or $n^4$ for potassium activation~\cite{hodgkin1952quantitative}. The associated gating variables may be written in relaxation form
\begin{equation}
    \dot{w}_{\ell a,i} = \frac{w_{\ell a,\infty}(V_i) - w_{\ell a,i}}{\tau_{\ell a}(V_i)}
\end{equation}
or, equivalently, in terms of voltage-dependent transition rates. This is the level at which a Hodgkin--Huxley~\cite{hodgkin1952quantitative} or Morris--Lecar~\cite{morris1981voltage} realization would be specified.

The coupling currents determine whether the biophysical system can implement the timing preferences of the phase model. A standard conductance-based chemical synapse from unit $j$ to unit $i$ can be written as~\cite{destexhe1994efficient, koch1998methods, ermentrout2010mathematical}
\begin{equation}
    I_{ij}^{\rm chem} = g_{ij}^{\rm chem}s_{ij}(t)(V_i - E_{ij}^{\rm syn}),
    \label{eq:chemical_synapse_current}
\end{equation}
where $g_{ij}^{\rm chem}$ is the maximal synaptic conductance, $s_{ij}(t)$ is the synaptic activation variable, and $E_{ij}^{\rm syn}$ is the synaptic reversal potential. With the sign convention used in Eq.~\eqref{eq:schem_biophys_network}, $I_{ij}^{\rm chem}$ is an outward current. Whether the synapse is effectively excitatory or inhibitory depends on the location of $E_{ij}^{\rm syn}$ relative to the membrane voltage trajectory. A common kinetic description takes $s_{ij}$ to obey first-order activation and decay dynamics:
\begin{equation}
    \dot{s}_{ij} = \alpha_{ij}S_{ij}\qty(V_j(t-\tau_{ij}))(1-s_{ij}) - \beta_{ij}s_{ij}.
\end{equation}
Here, $\alpha_{ij}$ and $\beta_{ij}$ are activation and decay rates, $\tau_{ij}$ is a conduction or synaptic delay, and $S_{ij}(V_j)$ is a smooth approximation to transmitter release. A typical phenomenological choice is a sigmoid threshold function
\begin{equation}
    S_{ij}(V_j) = \frac{1}{1+\exp[-(V_j-\theta_{ij}^{\rm syn})/\kappa_{ij}]},
\end{equation}
where $\theta_{ij}^{\rm syn}$ sets the release threshold and $\kappa_{ij}$ controls the sharpness of activation. These parameters affect the amplitude of the synaptic current as well as the phase at which the postsynaptic oscillator is perturbed, so they contribute directly to the induced interaction $H_{ij}^{\rm eff}$. 

Electrical coupling through gap junctions is usually represented by an Ohmic current proportional to the voltage difference between coupled cells~\cite{connors2004electrical, bennett2004electrical}:
\begin{equation}
    I_{ij}^{\rm gap} = g_{ij}^{\rm gap}\qty(V_i - V_j(t-\tau_{ij}^{\rm gap})),
    \label{eq:gap_junction_current}
\end{equation}
where $\tau_{ij}^{\rm gap} = 0$ in the simplest instantaneous approximation. Although gap junctions often promote synchrony, their effective phase-level role can depend on waveform shape, recovery dynamics, coupling strength, and delay. More generally, neither chemical nor electrical coupling imposes frustration by itself. Conductances, reversal potentials, synaptic time scales, delays, electrical coupling strengths, and intrinsic oscillator dynamics determine the induced interaction $H_{ij}^{\rm eff}$, and hence the preferred phase lag $\psi_{ij}^{\ast}$ on each edge. Frustration appears only when these local preferences are embedded in a network whose closed motifs make them mutually incompatible.

To make the connection with the phase theory explicit, suppose that the uncoupled $i$th unit has a stable rhythmic solution
\begin{equation}
    \mb{X}_i^0(t) = \mb{X}_i^0(t+T_i),\quad\omega_i = \frac{2\pi}{T_i},
\end{equation}
where $\mb{X}_i = \qty(V_i,\mb{w}_i,\{s_{ij}\},\hdots)$ denotes the full state of the unit and its local dynamical variables. A phase function $\Theta_i(\mb{X}_i)$ assigns a phase $\theta_i = \Theta_i(\mb{X}_i)$ along the cycle, and the infinitesimal phase response curve is
\begin{equation}
    \mb{Z}_i(\theta) = \nabla_{\mb{X}_i}\Theta_i\Big|_{\mb{X}_i = \mb{X}_i^0(\theta)}.
\end{equation}
For weak coupling, the detailed interaction from unit $j$ to unit $i$ induces an effective phase interaction $H_{ij}^{\rm eff}$, defined schematically by the averaged expression
\begin{equation}
\begin{split}
    H_{ij}^{\rm eff}(\varphi) = \frac{\epsilon}{2\pi}\int_0^{2\pi}\mb{Z}_i(\vartheta)\cdot\mb{P}_{ij}&\big[\mb{X}_i^0(\vartheta), \\ \mb{X}_j^0(\vartheta+\varphi&-\omega_j\tau_{ij});\mb{p}_{ij}\big]\dd{\vartheta}.
    \label{eq:effective_phase_interaction}
\end{split}
\end{equation}
Here, $\varphi$ denotes a generic phase difference, $\epsilon$ denotes the effective coupling scale$, \mb{P}_{ij}$ denotes the perturbation to oscillator $i$ generated by the coupling from $j$, and $\mb{p}_{ij}$ collects biophysical parameters such as $g_{ij}^{\rm chem}$, $E_{ij}^{\rm syn}$, $\alpha_{ij}$, $\beta_{ij}$, $g_{ij}^{\rm gap}$, and delay times. Equation~\eqref{eq:effective_phase_interaction} defines the effective phase interaction generated by a particular biophysical coupling. It is obtained by averaging the voltage-, gating-, and synapse-level perturbation against the phase-response curve of the postsynaptic oscillator, so it summarizes how the underlying biophysical variables shift the relative timing of the two cycles~\cite{ermentrout1991multiple, hoppensteadt1997weakly, brown2004phase, smeal2010phase, ashwin2016mathematical, park2017utility}.

In this notation, the corresponding phase dynamics take the form
\begin{equation}
    \dot{\theta}_i = \omega_i + \sum_jA_{ij}H_{ij}^{\rm eff}(\theta_j - \theta_i) + \mathcal{O}(\epsilon^2).
\end{equation}
The minimal model studied in Sec.~\ref{sec:minimal_phase_theory} corresponds to the special target
\begin{equation}
    H_{ij}^{\rm eff}(\varphi) \simeq -K_{ij}\sin\varphi,\quad K_{ij}>0.
\end{equation}
Defining the relative phase $\psi_{12} = \theta_1 - \theta_2$, two identical units with symmetric coupling give
\begin{equation}
    \dot{\psi}_{12} = H^{\rm eff}(-\psi_{12}) - H^{\rm eff}(\psi_{12}) = 2K\sin\psi_{12},
\end{equation}
so $\psi_{12} = 0$ is unstable and $\psi_{12} = \pi$ is stable. This is the phase-level signature that a biophysical pair must reproduce in order to behave as an antiferromagnetic timing bond.

More generally, a biophysical edge need not realize a pure sinusoid or exact antiphase locking. For a pair of units $i$
 and $j$, define the oriented relative phase
\begin{equation}
    \psi_{ij} = \theta_i - \theta_j.
\end{equation}
Then, the two-unit relative-phase dynamics can be written as
\begin{equation}
    \dot{\psi}_{ij} = G_{ij}(\psi_{ij})
\end{equation}
with
\begin{equation}
    G_{ij}(\psi) = \Delta\omega_{ij} + H_{ij}^{\rm eff}(-\psi) - H_{ji}^{\rm eff}(\psi),\quad\Delta\omega_{ij} = \omega_i - \omega_j.
\end{equation}
A stable preferred phase lag $\psi_{ij}^{\ast}$ is then defined by
\begin{equation}
    G_{ij}(\psi_{ij}^\ast) = 0,\quad G_{ij}'(\psi_{ij}^\ast)<0.
    \label{eq:G_stable_lag}
\end{equation}
The idealized repulsive Kuramoto model is the symmetric case with $\Delta\omega_{ij} = 0$ and $\psi_{ij}^\ast = \pi$. In a conductance-based realization, however, $\psi_{ij}^\ast$ may be shifted by synaptic delay, slow inhibition, gap-junctional coupling, intrinsic adaptation, or higher harmonics in $\H_{ij}$. More generally, a Fourier expansion of $\H_{ij}$ can shift the preferred lag or generate multiple phase lags. For reciprocal coupling with an odd phase-dependent interaction, the sine harmonics can still be integrated to give a generalized bond potential, while nonconstant even components generally destroy the gradient structure. Even in the latter case, the stable roots of Eq.~\eqref{eq:G_stable_lag} define the local timing preferences, and geometrical timing frustration requires these preferred lags to be mutually incompatible around closed loops. A useful loop-level criterion is
\begin{equation}
    \sum_{\ev{ij}\in C}\sigma_{ij}^{(C)}\psi_{ij}^\ast\neq 0\pmod{2\pi},
\end{equation}
where $C$ is a closed cycle and $\sigma_{ij}^{(C)} = \pm 1$ records whether the oriented edge is traversed with or against the convention used to define $\psi_{ij}^\ast$. The triangular motif from Subsec.~\ref{subsec:triangle} is the most transparent case:
\begin{equation}
    \psi_{12}^\ast + \psi_{23}^\ast + \psi_{31}^\ast = 3\pi\neq 0\pmod{2\pi}.
\end{equation}

The biophysical variables introduced above provide several routes to such nonzero preferred lags. Because the effective interaction $\H_{ij}$ is obtained by averaging the coupling perturbation against the postsynaptic phase-response curve, the same microscopic connection can induce different timing effects in different dynamical regimes. Furthermore, the sign, phase shift, and harmonic content of $\H_{ij}$ depend jointly on the voltage waveform, recovery dynamics, synaptic time course, reversal potential, delay, and operating regime of the oscillator~\cite{smeal2010phase, park2017utility}. Thus, inhibition should not be identified automatically with antiphase coupling. Depending on the phase-response curve and synaptic waveform, inhibitory coupling can promote synchrony, desynchrony, antiphase locking, or an intermediate phase lag.

The clearest biological route to the antiphase bond is reciprocal inhibition in a half-center oscillator~\cite{marder1996principles, marder2001central, daun2009control}. In such circuits, two rhythmic units or populations inhibit one another and often alternate between active and suppressed phases, providing a natural biophysical analog of the satisfiable two-oscillator baseline discussed in Subsec.~\ref{subsec:two_oscillator}. More generally, inhibition is central to cortical and hippocampal rhythm generation, but it does not imply a universal antiphase rule since inhibitory circuits can support synchronizing or lagged timing relations depending on cellular and network context~\cite{whittington2000inhibition, buzsaki2012mechanisms}. Beyond this special case, the synaptic activation variables, reversal potentials, electrical conductances, recovery variables, and delays appearing in Eqs.~\eqref{eq:chemical_synapse_current}--\eqref{eq:gap_junction_current} provide a broader route to nonzero preferred lags by shifting when a presynaptic signal affects the postsynaptic cycle.

This motivates the effective-interaction viewpoint used here. One should not infer timing frustration directly from labels such as excitatory, inhibitory, or electrical; instead, one should determine how the biophysical variables shape $\H_{ij}$, identify the stable lag $\psi_{ij}^\ast$ on each edge, and then ask whether those lags are compatible around the network. Recent micropatterned neuron-network experiments make this perspective especially relevant, since controlled biological networks can exhibit frustration-like suppression of synchronization when connectivity and gap-junction-mediated communication compete with collective timing organization~\cite{li2025collective}. Together with the phase-level results above, this motivates reduced biophysical tests of the same mechanism, in which the voltage, recovery, coupling, and synaptic variables can be varied systematically.

A natural starting point is a FitzHugh--Nagumo-type realization~\cite{fitzhugh1961impulses,nagumo1962active}, which could take the schematic form
\begin{equation}
\begin{split}
    \dot v_i &= v_i-\frac{v_i^3}{3} - w_i + I_i - \sum_j A_{ij}g_{ij}^{\rm chem} s_{ij}(v_i-E_{ij}^{\rm syn}) \\ 
    &\quad\ - \sum_j A_{ij}g_{ij}^{\rm gap}(v_i-v_j) + \sigma_i\xi_i(t), \\
    \dot w_i &= \epsilon_i(v_i+a_i-b_iw_i), \\
    \dot s_{ij} &= \alpha_{ij}S_{ij}(v_j(t-\tau_{ij}))(1-s_{ij}) - \beta_{ij}s_{ij}.
\end{split}
\end{equation}
Here, the voltage-recovery dynamics generate an oscillation from which a phase can be extracted, and the coupling terms are tuned or measured to determine whether the induced $H_{ij}^{\rm eff}$ has the required preferred phase lag. This makes FitzHugh--Nagumo networks a natural next step, since they are simple enough to analyze systematically, but detailed enough to test whether frustrated timing survives when amplitude dynamics, excitability, and synaptic variables are restored. 

The same embedding problem applies to Morris--Lecar, Hodgkin--Huxley, Izhikevich, Rulkov, and neural-mass realizations. The variables and coupling rules differ from model to model, but the test is the same: determine whether the detailed dynamics generate effective edge preferences $\psi_{ij}^{\ast}$, whether those preferences are geometrically incompatible, and whether the resulting attractors retain the phase-theory signatures of suppressed global coherence, local timing order, basin-dependent relaxation, and metastability.

This perspective also clarifies the role of the energy landscape. The symmetric phase model in Sec.~\ref{sec:minimal_phase_theory} has an exact Lyapunov function and can be viewed as zero-temperature relaxation in an antiferromagnetic $XY$ landscape. A conductance-based or reduced spiking network generally does not need to possess such a scalar energy, in which case the magnetic language should be understood dynamically. Ground states become low-cost timing patterns, metastable states become stable or long-lived attractors, and torque balance becomes a vanishing effective phase drive.

\begin{table*}[t]
    \caption{\label{tab:biophysical_diagnostics} Biophysical interpretation of the phase-level diagnostics used in this paper, after phases have been extracted from voltage, spike, burst, or population activity.}
    \renewcommand{\arraystretch}{1.3}
    \begin{ruledtabular}
    \begin{tabular}{cc}
    \textbf{Phase-level quantity} & \textbf{Biophysical interpretation} \\
    \hline
    $R$ & Strength of global voltage, spike, burst, or population synchrony \\
    $f_{ij}$ & Pairwise timing error relative to the preferred lag $\psi_{ij}^{\ast}$ \\
    $\bar f$ & Mean residual pairwise timing conflict across the network \\
    $m_\triangle$ & Deviation of one local circuit from triphasic timing \\
    $\overline m_\triangle$ & Average local triphasic-timing distortion across motifs \\
    $\chi_\triangle$ & Direction of a three-step local activation sequence \\
    $|m_\chi|$ & Net bias toward one local sequence direction over the other \\
    $\tau_i^{\rm eff}$ & Inferred residual phase torque/timing drive on unit $i$ \\
    $p_A$ & Trial probability of selecting timing class $A$ \\
    $q_{\alpha\beta}$ & Similarity between timing patterns reached in two trials \\
    $T$ & Effective temperature/noise strength in the phase model \\
    $\sigma_i$ & Biophysical noise amplitude in the voltage equation \\
    \end{tabular}
    \end{ruledtabular}
\end{table*}

\subsection{Biophysical interpretation of diagnostics}

Having formulated the sense in which a phase-level frustration mechanism can be embedded in higher-dimensional biophysical dynamics, we turn to the question of how the motifs and observables of the phase theory can be interpreted in a neural system. The natural common variable is the phase. If the phase function is known, then the phase of unit $i$ is
\begin{equation}
    \theta_i = \Theta_i(\mb{X}_i(t)),
\end{equation}
with $\mb{X}_i$ denoting the full biophysical state of the neuron. In simulations or experiments where the full asymptotic phase function is not available, $\Theta_i$ can be replaced by a practical phase estimator. For example, a simple event-based phase can be obtained from spiking or bursting dynamics by interpolating between consecutive spike times or burst onsets:
\begin{equation}
    \theta_i(t) = 2\pi\frac{t-t_{i,n}}{t_{i,n+1} - t_{i,n}}\pmod{2\pi},\quad t_{i,n}\leq t<t_{i,n+1},
\end{equation}
where $t_{i,n}$ and $t_{i,n+1}$ are successive events for unit $i$. Regardless of the implementation, the purpose is to place the detailed biophysical dynamics in phase coordinates so that the diagnostics of Sec.~\ref{sec:minimal_phase_theory} can be interpreted as neural timing observables. This allows us to distinguish disordered desynchronization from structured frustrated timing, in which weak global coherence coexists with local phase order, sequence handedness, basin dependence, and metastable final states.

The motif hierarchy in Sec.~\ref{sec:oscillator_dyn_energy_land} gives these diagnostics their biological interpretation. A stable nonzero lag between two units corresponds to a reproducible timing offset between voltage, spike, burst, or population rhythms. On triangular motifs, local incompatibility of such lags is expressed as a three-step sequential activation, the direction of which is measured by the chirality. In larger networks, the same local organization can coexist with weak global coherence, so the global order parameter $R$ must be interpreted together with local and ensemble diagnostics.

For a trianglar motif $\triangle = (i,j,k)$, the triangle moment
\begin{equation}
    m_\triangle(t) = \qty|e^{i\theta_i(t)} + e^{i\theta_j(t)} + e^{i\theta_k(t)}|
\end{equation}
measures the deviation from local triphasic timing. Biophysically, a small $m_\triangle$ means that the three units participate in a coherent three-step firing, bursting, or activity sequence, even if the surrounding network is not globally synchronized. The associated chirality
\begin{equation}
    \chi_\triangle(t) = \frac{2}{3\sqrt{3}}\qty[\sin(\theta_j-\theta_i) + \sin(\theta_k-\theta_j) + \sin(\theta_i-\theta_k)]
\end{equation}
records the handedness of that local sequence. $\chi_\triangle$ can then be interpreted as the direction of local sequential activation, beyond purely magnetic terminology. In an extended network,
\begin{equation}
    \ov{m}_\triangle(t) = \frac{1}{N_\triangle}\sum_\triangle m_\triangle(t),\quad m_\chi(t) = \frac{1}{N_\triangle}\sum_\triangle \chi_\triangle(t)
\end{equation}
measure, respectively, the average local violation of triphasic timing
and the net imbalance between opposite sequence directions.

The edge-level frustration diagnostic has a similarly direct timing
interpretation. If the effective interaction on edge $ij$ has a stable
preferred lag $\psi_{ij}^{\ast}$ and $\psi_{ij}(t) = \theta_i(t) - \theta_j(t)$ is the relative phase, then
\begin{equation}
    f_{ij}(t) = \frac{1}{2}\qty[1 - \cos\qty(\psi_{ij}(t) - \psi_{ij}^\ast)]
\end{equation}
measures the timing error relative to that preferred lag. For an ideal
antiphase bond, $\psi_{ij}^{\ast}=\pi$, so $f_{ij}$ vanishes for
alternating activity and is maximal for synchrony. The network average
$\bar f$ therefore measures residual pairwise timing conflict. Similarly, the effective phase torque
\begin{equation}
    \tau_i^{\rm eff} = \omega_i + \sum_j A_{ij} H_{ij}^{\rm eff}(-\psi_{ij}) - \Omega
\end{equation}
measures the residual drift of unit $i$ in a frame rotating with collective frequency $\Omega$. This is the effective-interaction analog of the phase torque $\tau_i$ used in Sec.~\ref{sec:oscillator_dyn_energy_land}; in the special case $H_{ij}^{\rm eff}(\varphi)=-K\sin\varphi$, it reduces to the repulsive Kuramoto torque up to the common rotating-frame subtraction. In a biophysical neural model, $\tau_i^{\rm eff}$ should be understood as an inferred phase-level drive rather than as a literal mechanical torque. A state can therefore be timing-balanced, in the sense of small $\tau_i^{\rm eff}$, while still retaining local timing distortions measured by nonzero $m_\triangle$.

The ensemble diagnostics translate the landscape language into trial-level timing observables. The basin weight $p_\mathcal{A}$ becomes the fraction of repeated trials that select timing class $\mathcal{A}$ after a common preparation, perturbation, or initialization protocol. The final-state overlap
\begin{equation}
    q_{\alpha\beta} = \qty|\frac{1}{N}\sum_{j=1}^Ne^{i(\theta_j^{(\alpha)} - \theta_j^{(\beta)})}|
\end{equation}
compares the phase patterns reached in two trials $\alpha$ and $\beta$ after quotienting out a common phase rotation. A narrow overlap distribution indicates reliable selection of the same timing pattern, whereas a broad distribution suggests that trial-to-trial variability reflects access to multiple structured timing states.

The biophysical interpretation of the diagnostics is summarized in Table~\ref{tab:biophysical_diagnostics}. Taken together, they provide a compact biological interpretation of the phase theory. A detailed FitzHugh--Nagumo, Morris--Lecar, Hodgkin--Huxley, Izhikevich, Rulkov, or neural-mass network need not possess an exact antiferromagnetic $XY$ energy, but nevertheless, after phases have been extracted through $\Theta_i$ or an empirical phase estimator, the same diagnostics can determine whether the biophysical dynamics exhibit the organizing signatures of frustrated timing.

\subsection{Scope and limitations of the minimal theory}

The deterministic simulations in Sec.~\ref{sec:oscillator_dyn_energy_land} correspond to the zero-temperature limit of overdamped Langevin relaxation on the phase landscape introduced in Sec.~\ref{sec:minimal_phase_theory}. This limit is useful because each initial condition relaxes downhill to the attractor whose basin contains it, exposing the basin geometry as cleanly as possible. Real neural systems, however, are noisy and heterogeneous. Channel fluctuations, synaptic release variability, background network input, unresolved degrees of freedom, and trial-to-trial changes in internal state all contribute to stochastic timing variability~\cite{faisal2008noise}. Thus, the temperature parameter $T$ in the phase theory should be read biophysically as an effective phase-noise strength. It is the phase-level counterpart of voltage-, synapse-, or channel-level fluctuations such as the $\sigma_i\xi_i(t)$ term in Eq.~\eqref{eq:schem_biophys_network}, after those fluctuations are projected onto the phase coordinate.

The same qualification applies to the coupling used in the minimal model. In a more detailed realization, intrinsic-frequency detuning~\cite{burton2012intrinsic, lowet2022tuning}, heterogeneous coupling functions~\cite{skardal2015erosion, petkoski2018phase}, conduction or synaptic delays~\cite{panchuk2013synchronization, petkoski2016heterogeneity}, spike-frequency adaptation~\cite{ladenbauer2012impact, augustin2013adaptation}, and synaptic or electrical asymmetries~\cite{smeal2010phase, park2017utility, le2025asymmetric, madadi2023delay} generally modify the effective interaction $H_{ij}^{\rm eff}$ and shift the preferred lags $\psi_{ij}^{\ast}$. The ideal limit studied in this paper corresponds to identical oscillators, zero noise, symmetric coupling, and the repulsive sinusoidal interaction $H_{ij}^{\rm eff}(\varphi)=-K\sin\varphi$, but away from this limit, exact degeneracies, equal basin weights, and symmetry-related selection probabilities need not survive unchanged.

Among the departures from this reference limit, noise and heterogeneity fundamentally reshape the dynamical landscape. Weak noise broadens deterministic attractors into fluctuating timing states, while stronger noise can drive chirality switching, diffusion along nearly degenerate manifolds, or transitions between metastable timing patterns. If the noise is large enough, local timing order is eventually lost. Heterogeneity plays a complementary role by lifting degeneracies and biasing basin weights, as unequal frequencies, delays, exitabilities, or coupling strengths can favor one chirality, pairing branch, or metastable class over another. Although these effects change how the landscape is explored, the underlying criterion for frustration is the same.

This is also where the magnetic analogy must be interpreted at the appropriate level. In the symmetric phase model, the analogy is literal since the dynamics possess an exact Lyapunov function, but in more detailed biophysical models, delays, adaptation, asymmetric coupling, strong interactions, and amplitude dynamics need not preserve a scalar energy. The minimal theory is therefore a controlled reference case in which the geometrical source of timing frustration is isolated before these additional effects are restored. Beyond this limit, the energy landscape may be replaced by a more general attractor landscape, and changing the interaction topology may alter the ground-state structure, metastable states, and their basins of attraction. These changes limit which specific landscape properties of the minimal model survive, but not the defining test: geometrical timing frustration occurs when the induced preferred lags cannot be made mutually consistent around closed motifs.

\section{Conclusions and outlook}
\label{sec:conclusions}

Weak global coherence is often interpreted as a failure of synchronization, a loss of order, or evidence that collective activity has become disorganized. The results in this paper support a different interpretation. In a frustrated timing network, the absence of global synchrony can be the macroscopic signature of local organization. Specifically, the phases may fail to align globally not because the system lacks structure, but because local timing preferences are mutually incompatible across the network. Geometrical frustration provides the language for making this distinction precise. In the framework we presented, a neural timing pattern is organized by the preferred phase relations imposed on individual edges and by the way those relations propagate through the interaction graph. When the network geometry prevents the local preferences from being realized simultaneously, relaxation occurs in a frustrated landscape whose stable states reflect the compromises required to balance incompatible timing constraints.

The purpose of the minimal theory developed in this work was to isolate this mechanism in a setting where the landscape can be defined without ambiguity. For identical phase oscillators with symmetric and repulsive sinusoidal coupling, the rotating-frame dynamics possess an exact Lyapunov function and are equivalent to zero-temperature relaxation in an antiferromagnetic $XY$ energy landscape. In this limit, the magnetic analogy gives a concrete diagnostic framework for neural timing; ground states, local constraint violations, chirality, metastable attractors, basin probabilities, and overlaps become measurable properties of phase-locked timing patterns. 

This framework also underscores why synchrony alone is an incomplete observable. The same small value of the Kuramoto order parameter can conceal very different phase organizations, ranging from an exact frustrated ground state to a low-energy metastable compromise or even an incoherent configuration with little local structure. Distinguishing these cases requires local and ensemble diagnostics. The central role of the phase theory is therefore to move beyond synchrony as a single diagnostic and give a landscape-based description of how local timing order, global coherence, and relaxation dynamics are related.

The hierarchy of analyzed geometries shows how a single local rule can generate progressively richer energy landscapes. The two-oscillator bond establishes the reference case in which a local antiphase preference has a unique stable realization. The triangle then shows what changes when the same preference is placed on the smallest incompatible motif. The system cannot satisfy each edge independently, so the ground states become collective $120^\circ$ compromises with chirality as an emergent discrete choice. This example also shows that different diagnostics respond to different aspects of the landscape, with chirality detecting motion away from the Ising-like saddle before the bond frustration or global coherence change appreciably. The tetrahedron demonstrates that frustration can produce a ground-state set with complex internal geometry. Its intersecting antipodal-pairing branches show that relaxation selects both a discrete pairing sector and a continuous coordinate along that sector. An important point is that this continuous degeneracy is not sampled uniformly by the dynamics. The final internal angle along each branch carries a nontrivial basin-induced measure, separating the static geometry of the ground-state manifold from the dynamical measure produced by relaxation.

The kagome lattice is the point in the hierarchy where the frustrated-neuron mechanism becomes an extended-network phenomenon. Each elementary triangle still imposes the same $120^\circ$ constraint found in the isolated motif, but extending that constraint across a lattice of corner-sharing triangles turns the local frustrated compromise into a constrained three-coloring manifold. The critical result is that this ground-state manifold does not determine the typical relaxation outcome. Random zero-temperature quenches usually settle instead into low-energy metastable states outside the ground coloring sector. These states have suppressed global coherence and vanishing site torques, making them dynamically stable timing patterns, yet their significant triangle distortions show that they do not satisfy the exact local $120^\circ$ rule. The kagome analysis therefore separates exact local constraint satisfaction from dynamical stationarity, showing that the dominant attractors of an extended frustrated timing network can remain organized by local incompatibility while lying outside the exact ground-state manifold.

The biophysical interpretation follows from this separation between the exact phase landscape and the attractor structure selected by relaxation. The phase theory should be used as an effective-interaction target, identifying the timing constraints and dynamical signatures that more biophysically detailed neural models must reproduce in order to realize geometrical frustration. In a conductance-based, spiking map-based, or neural-mass realization, the relevant question is how the voltage, recovery, synaptic, electrical, delay, and noise variables shape the induced phase interaction on each edge. If those interactions generate stable preferred lags that become incompatible around closed motifs, then the biophysical system realizes the same frustrated-timing mechanism, even though its dynamical landscape may no longer be an exact antiferromagnetic $XY$ energy landscape. This is the value of the frustrated-magnet analogy beyond the minimal model: it provides a dictionary for identifying the effective constraints, measuring their violations, distinguishing local timing order from global coherence, and interpreting metastable neural activity as part of a structured attractor landscape.

A natural continuation of this work is therefore to ask how much of the phase-level landscape survives when the idealized timing variables are embedded back into neural dynamics. Reduced excitable-cell models, such as FitzHugh--Nagumo or Morris--Lecar networks, provide a first setting for this test because they restore voltage and recovery variables while preserving enough simplicity to identify the induced phase interactions. More detailed spiking, bursting, map-based, and population-level realizations, such as Hodgkin--Huxley, Izhikevich, Rulkov, and neural-mass models, can then determine how the same mechanism is reshaped by amplitude dynamics, synaptic kinetics, adaptation, delays, and coupling asymmetries. Across these formulations, the important task is to follow the effective interaction by determining which preferred phase lags are produced by the underlying biophysical variables, asking whether those lags remain compatible around closed motifs, and measuring whether the resulting attractors retain the signatures of frustrated timing~\cite{paper2}.

Noise and heterogeneity enter this program as tests of robustness. The zero-temperature model studied in this paper exposes the basin geometry in its simplest form, but more realistic neural systems will blur and reconnect that landscape through stochastic timing variability and nonidentical local dynamics. Future work should therefore ask how much of the frustrated organization survives this deformation, namely, whether local incompatibility still suppresses global coherence, whether metastable timing states remain distinct, and whether relaxation continues to carry a memory of the basin structure~\cite{paper3}.

The same framework also opens a path towards richer condensed-matter analogies. If preferred lags, delays, or effective coupling signs become heterogeneous, the frustrated timing landscape begins to resemble a spin-glass problem, where overlaps, preparation history, and the distribution of metastable states become central. If local timing rules define an allowed manifold with defect-like violations, the analogy moves closer to ice physics, with constrained dynamics, residual entropy, and propagating timing defects. In larger architectures, one may also look for spin-liquid-like regimes in which the system does not freeze into a single timing pattern, but continues to explore a constrained manifold of locally compatible phase configurations. The natural diagnostics would then be correlation functions and structure-factor-like measures that characterize order within the fluctuating frustrated manifold~\cite{paper3}.

Together, these directions frame a broader series of frustrated-neurons studies. The present paper establishes the minimal phase-level foundation by isolating the geometrical mechanism, deriving the exact antiferromagnetic $XY$ mapping in the symmetric limit, and developing the diagnostic dictionary needed to compare global coherence, local constraint satisfaction, basin structure, and metastability. Subsequent papers in this series will determine how this landscape changes when biophysical detail, noise, disorder, and larger architectures are restored. Across these extensions, the organizing criterion remains the same: frustrated neural timing arises when locally preferred phase relations cannot be made mutually consistent across the network, producing timing landscapes in which global coherence alone cannot reveal the local order selected by relaxation.

\begin{acknowledgements}
    I thank Dima Watkins and Ishaan U. Patel for beneficial discussions.
\end{acknowledgements}

\section*{Data availability}

There are no publicly available research data or software supporting this paper. Requests for further information or data should be sent to the author.

\bibliography{refs}

\end{document}